\documentclass[twocolumn]{aastex631}

\usepackage{multirow}
\usepackage{graphicx}
\usepackage{subfigure}
\usepackage[normalem]{ulem}

\begin{document}

\title{Barium Stars Across the Milky Way: Probing Their Origins via the GALAH Survey}

\author[0009-0008-0347-4347]{Jaden Levine}
\affiliation{Department of Astronomy, San Diego State University, San Diego, CA 92182 USA}
\affiliation{Department of Astronomy, The University of Texas at Austin, Austin, TX 78712, USA}
\altaffiliation{Corresponding author email: jt.levine212@gmail.com}

\author[0000-0002-0900-6076]{Catherine Manea}
\affiliation{Department of Physics \& Astronomy, University of Utah, Salt Lake City, UT 84112, USA}

\author[0000-0002-1423-2174]{Keith Hawkins}
\affiliation{Department of Astronomy, The University of Texas at Austin, Austin, TX 78712, USA}

\author[0000-0001-6873-8501]{Kendall Sullivan}
\affiliation{Department of Astronomy \& Astrophysics, University of California Santa Cruz, CA 95064, USA}
\affiliation{Mullard Space Science Laboratory, University College London, Holmbury St Mary, Dorking, Surrey RH5 6NT, UK}

\author[0000-0001-6248-1864]{Kate H. R. Rubin}
\affiliation{Department of Astronomy, San Diego State University, San Diego, CA 92182 USA}

\author[0000-0002-0475-3662]{Zachary Maas}
\affiliation{Department of Astronomy, Indiana University, Bloomington, IN 47405, US}

\author[0000-0002-6478-0611]{Andrew C. Nine}
\affiliation{Department of Physics, University of Wisconsin-Whitewater, Whitewater, WI 53190, USA}

\begin{abstract}
Barium stars are unusually enriched in barium ([Ba/Fe] $\ge1.0$ dex) and not predicted by current Galactic chemical evolution models. Previous observations of barium stars have found evidence that they form through mass transfer from a companion asymptotic giant branch (AGB) star or through radiative levitation. The chemical abundance and kinematic information of barium stars may help constrain AGB stellar nucleosynthesis, binary star evolution, and internal evolutionary processes that affect surface abundances. Using $\sim$450,000 stars from the GALactic Archaeology with Hermes (GALAH) survey, we identify {nearly 3000 new} barium-rich stars and separate them into hot ($ T_{\text{eff}}>6000$ K) and cool ($ T_{\text{eff}} < 6000$ K) populations. Cross-matching with Gaia DR3, we find that 47.7\% of our barium stars {within 1 kpc} have elevated re-normalized unit weight error (RUWE $\ge1.4$), compared to $16.3\%$ of a comparable sample of the GALAH field, suggesting multiplicity plays an important role in the formation of both populations of barium stars. A subset of hot barium stars exhibit low RUWE (RUWE $<1.2$) and [$\alpha$/Fe] $<-0.2$, supporting radiative levitation as an origin as well. We determine Galactic memberships using both kinematics and chemistry and find that barium stars exist in the thin disk, thick disk, and halo though they are slightly more prevalent at lower metallicities. Overall, we show evidence for barium stars produced by mass transfer and for those produced by radiative levitation, with both formation mechanisms occurring ubiquitously across the Galaxy.

\end{abstract}


\section{Introduction} \label{sec:intro}
Barium (Ba) is an element created primarily through the slow neutron-capture process ($s$-process) in the interiors of asymptotic giant branch (AGB) stars \citep[e.g.,][]{Karakas2016}. Observations of disk stars and theoretical Galactic chemical evolution (GCE) models show that typical field stars have [Ba/Fe] ratios ranging from -0.25 to +0.25 dex depending on their metallicity, age, and position in the Galaxy \citep[e.g.,][]{Bensby2014, Delgado2017, Bedell2018, Molero2023, Storm2025}. However, observations show there exist stars with unusually strong barium lines and high inferred barium abundances ([Ba/Fe] $\ge$ 1) that do not match disk stars at similar metallicities. Barium stars, which appear to have over ten times the barium abundance of the Sun, were first discovered in the mid-twentieth century by \citet{firstba}. Numerous studies since have investigated these objects to understand the origins of their extreme compositions {\citep[e.g.,][and references throughout this manuscript]{Burbidge1957, Warner1965, McClure1980, Boffin1988, McClure1990, Jorissen1992, North1994, North2000, Jorissen1998, deCastro2016, Merle2016, Jorissen2019}}. 

Two leading theories have surfaced explaining the extreme compositions of these stars. First posited by \cite{McClure1980}, one theory suggests that mass transfer from a companion AGB star is responsible for the barium enhancement. AGB stars are the precursor to the final stage of stellar evolution for low to intermediate mass (M $\lesssim$ 8-10 M$_\odot$) stars. This stage typically lasts around $\sim100,000$ years, during which the $^{13}$C($\alpha$, n)$^{16}$O reaction in the $^{13}$C pocket of the AGB star's convective envelope (or, for M \textgreater 4 M$_\odot$ stars, the $^{22}$Ne($\alpha$, n)$^{25}$Mg reaction) creates free neutrons that enable the $s$-process \citep[e.g.,][]{Lugaro2003, Karakas2006, Cristallo2009, Kappeler2011, dawesreview, Buntain2017}. Successive neutron-captures onto seed nuclei (typically Fe) followed by beta decays create progressively heavier elements. The products of this nucleosynthesis are subsequently convected to the surface of the star and, for single AGB stars, expelled into the interstellar medium at a rate of up to $10^{-4} \ \text{M}_\odot \ \text{yr}^{-1}$ via a stellar wind \citep[e.g.,][]{Goldman2017}. For AGB stars with companions, a fraction of these products (parametrized by the so-called dilution factor, e.g., \citealp{Cseh2022, Vilagos2024}) can instead be transferred to the other star. Barium stars created via mass transfer are physically rich in the $s$-process elements newly synthesized by the AGB companion and are therefore effective tracers of AGB star nucleosynthesis \citep[e.g.,][]{Theuns1996,Vilagos2024}. Due to the relatively short lifespans of AGB stars, most companions of barium-rich stars are likely white dwarfs. Dedicated follow-up efforts including radial velocity monitoring and UV photometry have collected evidence for white dwarfs in numerous such systems \citep[e.g.,][]{McClure1980, binarynature, Jorissen1992, Jorissen1998, Gray2011, Kong2018, Jorissen2019, Escorza2019, Nine2023, Nine2024, Pal2024,Yamaguchi2025}. If barium stars are primarily made through mass transfer, then discoveries of barium stars in large spectroscopic surveys may provide new insights on multiplicity and mass transfer in different stellar populations. For example, barium stars in large surveys might be useful for identifying non-eclipsing, unresolved binary stars that are beyond distances at which Gaia RUWE can effectively trace multiplicity (see Section \ref{subsec:gaiaandruwe}). 

Barium stars may also arise from the effects of radiative levitation within a star's envelope \citep[e.g.,][]{radiativelev}. Radiative levitation, also termed radiative acceleration, is a complex mechanism  that is difficult to model \citep[e.g.,][]{atomicdifussionandturbulentmixing} and is a result of the transfer of momentum from emergent photons to atoms in the stellar envelope via bound-bound or bound-free atomic transitions \citep[e.g.,][]{Michaud1970}. In stars with significant radiative envelopes ($T_{\text{eff}}$ \textgreater 6000 K), this process lifts atoms to the surface of the star, competing with gravitational settling, where elements settle deeper in the star \cite[e.g.,][]{Michaud2015, Alecian2020}. Heavier elements with a greater number of atomic transitions and larger absorption cross sections (e.g., neutron capture elements: Ba, Y, Nd, etc.) are more susceptible to radiative acceleration than lighter elements (e.g., $\alpha$-elements: Ca, Si, and Mg). The heavy elements are preferentially lifted to the stellar surface, affecting the resulting spectrum. Heavier elements will have stronger absorption lines that are interpreted as over-{abundances} while the lighter elements appear relatively under-abundant when compared to the intrinsic composition of the star \citep[e.g.,][]{radiativelevmodeling}. Therefore, some hot ($T_{\text{eff}}$ \textgreater 6000 K) barium stars with radiative envelopes may have strong barium lines as a result of radiative levitation, and a lack of a significant convective envelope to offset the effect, rather than an enhanced total barium abundance \citep[e.g.,][]{Alecian2020}.

The recent decade has ushered in an era of large spectroscopic surveys collecting medium-to-high resolution optical spectra of millions of Milky Way stars \citep[e.g.,][]{GaiaESO, GALAH, LAMOSTsurvey}. This spectroscopic revolution offers unprecedented opportunities to identify and study barium stars at a population scale.  {\cite{lamost}} conducted one of the first spectroscopic survey-based population-level studies of barium stars using the Large Sky Area Multi-Object Fiber Spectroscopic Telescope (LAMOST, \citealp{LAMOSTsurvey}) survey. They found evidence for mass transfer by studying the abundances of carbon and nitrogen, two additional products of AGB star nucleosynthesis \citep{dawesreview}, and they also found evidence of radiative levitation by looking to $\alpha$-element abundances. Here, we leverage the GALactic Archaeology with HERMES (GALAH) spectroscopic survey with a resolving power ($\text{R}=\lambda/\Delta\lambda$) of R$\sim$28,000, a higher resolution survey than LAMOST (R$\sim$ 1800), to investigate the origins of barium stars. Similar to the methods of {\cite{lamost}}, we search for evidence of both mass transfer and radiative levitation in barium stars by studying their elemental abundances. However, unlike the aforementioned study, we also leverage astrometric data from Gaia to search for evidence of barium stars created through mass transfer. We also direct readers to Masegian et al, in prep., which also performs an investigation of the barium stars in GALAH, paying additional attention to the {statistics of companionship} of barium stars and implications for AGB star nucleosynthesis.

The questions this paper will answer are as follows: 1) Are there barium-rich stars in GALAH? 2) What information can large stellar surveys provide to constrain potential formation mechanisms of barium stars? 3) Do barium stars belong preferentially to certain Milky Way stellar populations?

This paper is organized as follows: In Section \ref{sec:methods}, we introduce the GALAH and Gaia data sets, and we define our sample of barium-rich ($\text{[Ba/Fe]} \ge 1$) stars. Section \ref{sec:results} presents our analysis of the available chemical and kinematic data for our sample to search for signs of multiplicity and characterize their Milky Way memberships. In Section \ref{sec:discussion} we interpret the results found in Section \ref{sec:results}. Section \ref{sec:conclusion} presents the conclusions. 

\section{The Data} \label{sec:methods}

\subsection{The GALAH Chemical Abundance Survey} \label{subsec:galah}
We require a large chemical abundance dataset to identify and chemically characterize barium stars across different stellar populations. The GALAH survey is ideal for this task \citep{GALAH}.  GALAH is a magnitude-limited (V\textless 14 mag) southern-hemisphere spectroscopic survey that uses the HERMES spectrograph on the 3.9 m Anglo-Australian Telescope. HERMES is a {high} resolution ($\text{R}\sim28,000$), optical (4710 \textless $\lambda$ \textless 7890 \AA), multi-object spectrograph capable of obtaining up to 392 simultaneous spectra of objects within a two degree field of view. GALAH provides stellar parameters and chemical abundances derived using Spectroscopy Made Easy {(SME, \citealt{oldsme}; \citealt{newsme})} and by adopting 1D MARCS model atmospheres (\citealt{1975MARCS}; \citealt{1976MARCS}; \citealt{2008MARCS}). We summarize some details regarding their parameter and abundance determination methods here, and we direct the reader to {\cite{GALAH}} for a deeper discussion.  SME is a radiative transfer and spectral synthesis code that determines stellar parameters and abundances by fitting synthetic spectra to observed spectra \citep{oldsme}. When solving for stellar parameters (e.g., $ T_{\text{eff}}$, log g, microturbulence, broadening velocity, macroturbulence, etc.), GALAH restricts the SME fits to spectral windows centered on Fe, Ti, and Sc, and select Balmer lines.  Upon constraining stellar parameters, GALAH measures abundances in up to 30 elements for nearly 600,000 stars. Local thermodynamic equilibrium (LTE) is assumed during abundance determination for most elements, though some elemental abundances (most relevant to this work, C, O, Mg, Si, Ca, and importantly, Ba) are determined under the non-LTE assumption using departure coefficient grids from \citet{Amarsi2020}. {GALAH also provides $1\sigma$ uncertainties on their stellar parameters and chemical abundances which we integrate into our Monte Carlo (MC) sampling to determine uncertainties throughout this work.}

{The validation of GALAH DR3 abundances is discussed extensively in \citet{GALAH}. DR3 was validated against multiple external high-resolution anchors (e.g., \citealp{Ramirez2011, Bedell2018, gbs,Hegedus2023}) to confirm accuracy and ensure representative precision. For example, comparisons between GALAH DR3 abundances and those of the Gaia Benchmark Stars sample \citep[e.g.,][]{gbs} returns offsets $0.00 <\Delta\rm[X/Fe] < 0.06$ which are captured by the associated abundance uncertainties.  Additional focused comparisons between DR3 abundances and those determined from higher resolution spectra further validate the accuracy and precision of GALAH abundances \citep[e.g.][]{Aguado2021, Lu2025wCatherine}. Throughout this work, we assume that GALAH abundances are accurate, their precisions are representative, and their quality flags are thorough.  However, as with any abundances determined through automated means, GALAH abundances may benefit from higher-resolution follow-up or boutique re-analysis of the spectra.  For higher precision constraints on detailed abundance patterns for, e.g., comparison to theoretical models, barium stars identified in this work would benefit from careful follow-up and reanalysis.}

\subsection{Gaia and RUWE} \label{subsec:gaiaandruwe}
In addition to chemical abundances from GALAH, this work implements astrometric data from Gaia \citep{GaiaDR3} to determine barium stars' {Galactic memberships and possibilities of binarity, the latter of which is relevant for identifying candidate post-mass transfer systems}. Gaia DR3 provides statistical indicators that measure the quality of the astrometric fit. {One such indicator} is the re-normalized unit weight error (RUWE) which is described in detail in \citet{Lindegren2018}. For well-behaved astrometric solutions (such as those for single stars), RUWE will be near 1. RUWE {can} be inflated for stars where the single star solution is a poor approximation such as stars with protoplanetary disks \citep[e.g.,][]{Fitton2022} and, most relevant to this work, multi-star systems. {In this work, we adopt the assumption that within 1 kpc, stars with RUWE $\ge 1.4$ likely host a companion star, and that those with RUWE $<1.2$ are likely single stars \citep[][]{Hernandez2023,Fitton2022,Belokurov2020,Lindegren2018}.  We discuss the implications of these assumptions below.}

{There is extensive literature exploring the sensitivity and accuracy of RUWE as a binarity indicator \citep[e.g.,][]{Lindegren2018, Pourbaix2019, Jorissen2019b, Belokurov2020, Fitton2022, Hernandez2023, Sullivan2024, Elbadry2025}.  Overall, RUWE has been widely demonstrated to be an effective indicator of binarity within 1 kpc \citep[e.g.,][]{Belokurov2020, Sullivan2024}. Beyond 1 kpc, RUWE values tend to fall below the astrometric noise level and binary stars can no longer be {reliably} detected. For this reason, we implement a 1 kpc distance cut when considering RUWE values {or discussing binarity} among our sample (e.g. {Sections} \ref{subsec:masstransfer} and \ref{subsec:where}). Within the 1 kpc cut, RUWE will identify fewer candidates at larger distances because the angular separation decreases for a given physical separation. Furthermore, RUWE is most sensitive to intermediate angular separation, intermediate mass ratio binaries, so our usage of RUWE biases our ``likely binary" sample to these types of systems \citep[e.g.][]{Sullivan2024}. }  

{To gauge the sensitivity and completeness of RUWE as a binarity indicator in the context of barium stars, we examine the RUWE values of a catalog of known binary barium stars from \cite{VanderSwaelmen2017} and \cite{Jorissen2019} (see figure in Appendix \ref{appendix_ruwe}).  $\sim$ 71\% of their confirmed barium binary stars within 1 kpc have RUWE $>1.4$ while $\sim$17\% have RUWE  $<1.2$, implying the contamination of false negatives is on the order of $\sim20-30\%$.  RUWE becomes unreliable as a binarity indicator for stars beyond 1 kpc and for those with longer orbital periods ($\gtrsim7,500$ days). However, there exists a star (HD 104979) with an orbital period of 19295.0 days \citep{Jorissen2019} and RUWE $>1.4$. This is likely due to the star's short distance from the Sun ($\sim50$ pc), illustrating the importance of distance when considering RUWE as a binary proxy. This analysis demonstrates the influence of orbital period and parallax on the effectiveness of RUWE as a binary tracer, and reaffirms RUWE's efficacy as a binary indicator within 1 kpc.}

{Though RUWE has been demonstrated to be a broadly effective binarity indicator within 1 kpc for intermediate mass ratio and angular separation binary stars, there are numerous other Gaia metrics for multiplicity we considered.  Radial-velocity (RV) variations due to a companion can be detected in stars with at least three Gaia RV measurements \citep[e.g.][]{Katz2023, Chance2025} and are most sensitive to close ($a\lesssim$ 3 AU for g $\sim$ 14) binaries.  However, only $\sim$40\% of our barium-rich sample has Gaia-reported RV information, whereas all but one barium-rich star have reported RUWE.  Furthermore, within 1 kpc, all but 35 barium stars that show significant RV variations (identified using the catalog and associated recommendations of \citet{Chance2025}) also show RUWE  $>1.4$.  We also consider including information from the associated Gaia Non-Single-Star (NSS) catalog \citep{NSS}. However, of the 402 barium-rich stars in our sample with NSS information, 314 of them have RUWE $\ge$ 1.4, and the remainder lie beyond 1 kpc.  In summary, RV variations and the NSS catalog would serve to expand our ``likely binary'' barium star group beyond 1 kpc.  However, for the purpose of studying general trends between barium stars and their barium-normal counterparts, the 1 kpc distance cut that we impose also aids in controlling for Galactic chemical evolution effects.  Therefore, the limited additional stars beyond 1 kpc that would join the ``likely binary'' group would complicate the interpretation of general trends.  We therefore do not consider other binarity indicators beyond RUWE in this work but encourage their investigation in future work.}

{We emphasize that the overall goal of this work is not to define pure samples of single-star or binary-star systems but rather to present barium star population trends as a whole. Therefore, we ultimately proceed by considering $d < 1~\rm kpc$ stars with RUWE $\ge1.4$ as our ``likely binary" group to probe for mass transfer and use those with RUWE $<1.2$ to probe for radiative levitation in likely single star systems, omitting other potential binarity indicators from this study. We acknowledge the incompleteness of RUWE as a standalone binary proxy but treat it on a population scale as an indicator of multiplicity to highlight trends between barium star and field populations and identify particularly interesting systems for higher resolution follow-up.  We leave the development of a pristine sample of confirmed binary barium stars in GALAH to a future work.}

{We note that GALAH provides users with Gaia astrometric data as part of the main DR3 catalog (GALAH\_DR3\_main\_allstar\_v2). However, the third data release of GALAH reports Gaia DR2 parameters. We update these values by crossmatching GALAH DR3 data with Gaia DR3 using the Gaia DR3 source IDs provided by GALAH before proceeding with our analysis. The GALAH team uses Gaia position, parallax, and proper motion along with their own radial velocity measurements to derive heliocentric UVW velocities for their stars. We transform the heliocentric velocities into local standard of rest velocities (LSR) assuming $\rm(U, V, W)_\odot=(11.1, 12.24,7.25)$ km/s adopted from \citealt{schonrich2010}.  Furthermore, for cases where we require estimates of barium star distance (e.g., when considering RUWE), we adopt the distances from {\cite{Bailerjones}} which are derived using a probabilistic approach that imposes a prior based on a three-dimensional Galactic model.  Basic parallax inversion for determining stellar distances is not ideal for stars with significant fractional parallax uncertainties \citep{parallaxfordistancebad}.}

\subsection{Defining the Sample} \label{subsec:using data}
We must begin with a subset of stars with reliable chemical abundances worthy of investigation. We make use of GALAH's flagging system \citep{GALAH} to identify this subset. Starting with a population of 588,464 stars, we define our sample with the following selections and motivate them in the subsequent paragraph: \begin{enumerate}
  \item $\text{flag}_{\text{sp}} \le 1$, stellar parameter quality flag \label{selection1}
  \item $\text{flag}_{\text{[Fe/H]}} = 0$, quality flag for overall iron abundance \label{selection2}
  \item $\text{flag}_{\text{[Ba/Fe]}} = 0$, quality flag for barium abundance \label{selection3}
  \item $\text{flag}_{\text{[X/Fe]}} = 0$, quality flag for [X/Fe] where ``X" represents the element being used \label{selection4}
  \item $4000 \text{ K} \le T_{\text{eff}} \le 7000 \text{ K}$, temperature range for optimal SME performance \label{selection5}
\end{enumerate}
A $\text{flag}_{\text{sp}} = 0$ indicates the GALAH team found no identifiable problems with the SME stellar parameter determination. A $\text{flag}_{\text{sp}} = 1$ is given to stars with Gaia DR2-reported RUWE $> 1.4$. We use $\text{flag}_{\text{sp}} \le 1$ in this case because we are intentionally studying the high-RUWE stars that are being flagged. Higher values of $\text{flag}_{\text{sp}}$ are used to indicate unreliable broadening, low signal to noise, reduction issues, etc., so we omit these unreliable stars (see \citealt{GALAH}, Table 4 {for the full} list of flag values). The quality flag for iron abundance, $\text{flag}_{\text{[Fe/H]}}$, is equal to zero when there are no identifiable issues with the [Fe/H] abundance determinations. We apply this to ensure reliable iron abundances. Similarly, the $\text{flag}_{\text{[X/Fe]}}$ value is equal to zero when there are no identifiable issues with the [X/Fe] abundance determination. Every star in our sample satisfies selection \ref{selection3}, while selection \ref{selection4} is only used when studying abundances other than [Ba/Fe] and [Fe/H] ({Sections} \ref{subsec:masstransfer} and \ref{subsec:where}). We also limit our sample to stars with 4000 K $\le T_{\text{eff}} \le$ 7000 K to avoid temperature regimes where SME tends to fail \citep[][]{newsme,2008MARCS}. After the selections are applied, our sample consists of 450,676 stars.

\section{Analysis} \label{sec:results}

\begin{figure*}[]
  \centering
  \includegraphics[width=\textwidth]{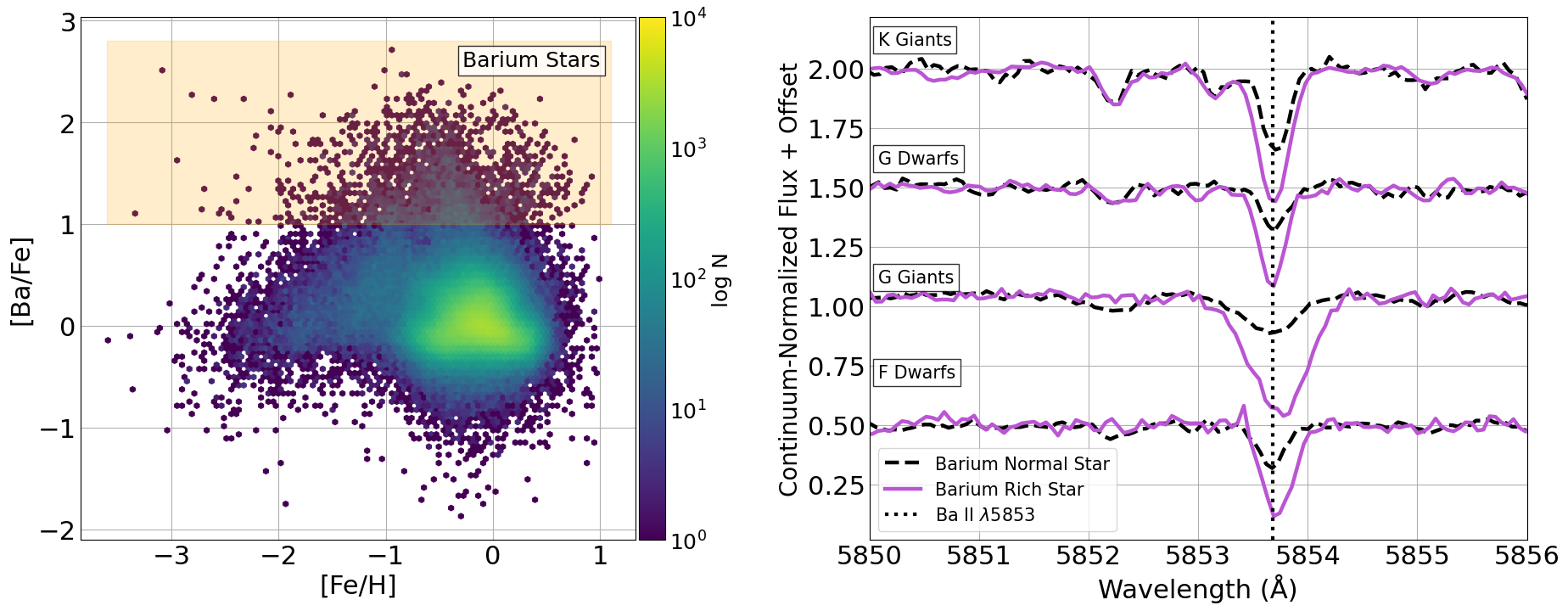}
  \caption{\textit{Left:} [Ba/Fe] versus [Fe/H] for our sample (see Section \ref{subsec:using data}) of the GALAH field. Stars in the shaded region are deemed barium stars and have $\text{[Ba/Fe]} \ge 1$. \textit{Right:} Spectral comparisons for barium-enhanced (solid line) and barium-normal (dashed line) stars for a K giant, G dwarf, G giant, and F dwarf from top to bottom respectively. The barium stars and the comparison GALAH stars are chosen such that they have similar temperatures, surface gravity, and metallicities. This figure confirms the existence of a population of stars in the GALAH survey that are barium-rich and qualitatively confirms their Ba enhancement spectroscopically.}
  \label{fig:numbersubplot}  
\end{figure*}

\subsection{Finding Barium Stars} \label{subsec:finding}
The primary goal of this work is to identify and characterize the barium stars within GALAH DR3. We are solely interested in strong barium stars \citep[e.g.,][]{Warner1965} and therefore adopt the definition that barium stars are those with [Ba/Fe] $\ge 1$. This selects stars that are over ten times as barium abundant as the Sun, exceeding what could be produced through standard Galactic chemical evolution alone \citep[e.g.,][]{Molero2023}. Figure \ref{fig:numbersubplot} presents the [Ba/Fe] vs. [Fe/H] distribution of our sample. Our selection criteria identifies 2810 stars with [Ba/Fe] $ \ge 1$, shown by the highlighted region in Figure \ref{fig:numbersubplot}. {Including GALAH-provided uncertainties on [Ba/Fe], MC sampling yields an effective barium star count of $3085\pm^{26}_{25}$, corresponding to $0.68\pm_{0.01}^{0.01}\%$ of the GALAH field stars with high quality abundances (see Section \ref{subsec:using data}) meeting the [Ba/Fe] $\ge1$ criterion.}

To qualitatively verify the barium enhancements of stars in our sample, we turn to their HERMES spectra. In Figure \ref{fig:numbersubplot}, we compare the strengths of the \ion{Ba}{2} $\lambda5853$ line of a randomly-selected barium-rich star to that of a comparison star with [Ba/Fe] $\sim$ 0 for one K giant (Gaia DR3 4422593353008932736), one G dwarf (Gaia DR3 6233918569409734400), one G giant (Gaia DR3 3038914422505373824) and one F dwarf (Gaia DR3 3790375232188743168). We seek visual confirmation of an exceptionally strong barium line with respect to the barium-normal reference star. This comparison is restricted to stars of the same spectral type, luminosity class, and metallicity to ensure that any differences in the \ion{Ba}{2} $\lambda5853$ line are due to differences in the stars' chemical composition as opposed to stellar parameter effects. In Appendix \ref{table}, we present the key information of the four barium stars {presented} in Figure \ref{fig:numbersubplot} as well as other noteworthy barium stars discussed throughout this text.

\begin{figure}
  \centering
  \includegraphics[width=\linewidth]{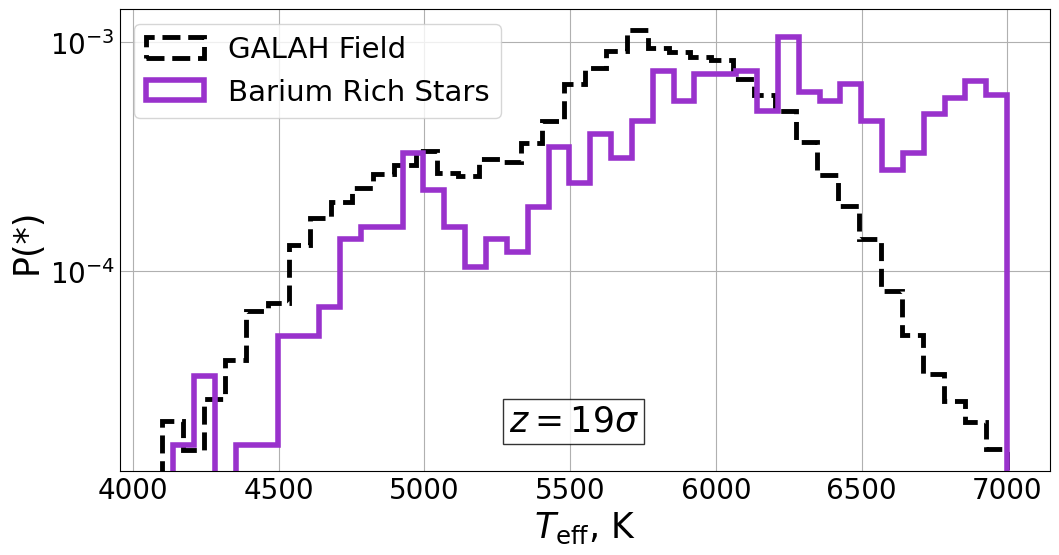}
  \caption{$T_{\text{eff}}$ distributions within 1 kpc for barium stars (purple) and the non-barium-enhanced GALAH field (dashed black). The histograms are normalized to integrate to unity. Barium stars tend to be hotter on average relative to their non-barium-enhanced counterparts. The underrepresentation of K and G type barium stars could be explained by mass transfer (See Section \ref{discuss:masstrans}).}
  \label{fig:temphist}
\end{figure}

 To begin our search for evidence of both theories of barium star formation, we separate barium stars into hot and cool categories, with $T_{\text{eff}}$ $<$ 6000 K defining our cool category and $T_{\text{eff}} >$ 6000 K defining our hot category. Stars with $T_{\text{eff}} > 6000 $ K have shallow convective envelopes and significant radiative envelopes that are conducive to the effects of radiative levitation.  Meanwhile, stars with lower temperatures are not subject to the effects of radiative levitation due to significant convective envelopes, so other mechanisms must explain their barium enhancement {\citep[e.g.,][]{radlevalphaelement}}. In Figure \ref{fig:temphist}, we present {the} temperature distribution of our barium-rich sample within 1 kpc. The distance cutoff is implemented because it better probes the local distribution of barium star evolutionary states. Without this cutoff, the $T_{\text{eff}}$ distribution is skewed towards cooler giants due to their brightness (and thus enhanced detectability at larger distances) compared to dwarfs. The distance of 1 kpc is chosen to remain consistent with future distance cuts and simultaneously ensures the sample is sensitive to RUWE as a tracer of multiplicity {(see discussion in Section \ref{subsec:gaiaandruwe})}. {To quantify the significance of the difference between the barium-rich and GALAH field $T_{\rm eff}$ distributions, we implement the two-sample Kolomogorov-Smirnov test and derive a p-value, which we transform into its Gaussian-equivalent significance (z-score) using the inverse standard normal survival function. For the two $T_{\rm eff}$ distributions in Figure \ref{fig:temphist}, we derive a z-score of $z=19.13\sigma$, meaning the KS test rejects the null hypothesis at the $19.13\sigma$ level. We apply this methodology to future distributions and present rounded z-scores in associated figures.}
 
 {In addition to studying their temperature distribution, we compare the spectral type distributions of barium-rich and non-barium-rich stars in Table \ref{spec_table}, employing MC sampling to determine uncertainties on percentages.}  We define K stars to have $4000\text{ K}<T_{\text{eff}}<5300\text{ K}$, G stars as those with $5300\text{ K}<T_{\text{eff}}<6000\text{ K} $, and  F stars as $6000\text{ K} < T_{\text{eff}}<7000 \text{ K}$ \citep{Pecaut2013}. {For ease of comparison, we also implement the use of fractional excess: $$f_{\rm excess}={ {f_{\rm Ba}-f_{\rm Field}}\over f_{\rm Field} }$$ where $f_{\rm Ba}$ is the fraction of the total barium stars that are in a given bin and $f_{\rm Field}$ is the equivalent fraction of non-barium-rich stars. Within this convention, $f_{\rm excess}=1.0$ equates to twice as many barium stars for a given classification relative to the field, and $f_{\rm excess}= -0.5$ describes half as many barium stars for a given classification compared to the field. When comparing the nearby ($< 1$ kpc) barium stars (807 stars) to the nearby, non-barium enhanced GALAH field (274,476 stars), we find that barium stars have fractional excesses of  $f_{\rm excess}=-0.40$ for K stars, $f_{\rm excess}= -0.43$ for G stars, and $f_{\rm excess}= 1.16$ for F stars. This analysis reveals that the barium stars in our sample tend to be hotter on average than their non-barium-enhanced counterparts, providing our first hint towards the mass transfer formation mechanism of barium stars (see interpretation in Section \ref{discuss:masstrans})}.

\begin{table}[] 

 \centering
 \caption{{Spectral type distributions of nearby barium stars (807 stars) and the nearby non-barium-enhanced GALAH field (279,476 stars) determined with MC sampling.}}
 \begin{tabular}{c|c|c}

 Spectral Type & Barium Stars & GALAH Field \\ 
 
  \hline
 \hline
K Type & $12.67\pm^{0.38}_{0.36}$\% & $21.07\pm^{0.03}_{0.03}\%$  \\
G Type & $29.51\pm^{0.83}_{0.79}$\% & $52.17\pm^{0.05}_{0.05}\%$\\ 
F Type & $57.70\pm^{0.75}_{0.74}\%$ & $26.77\pm^{0.04}_{0.04}\%$ \\ 
\hline
 \end{tabular}
 \label{spec_table}
\end{table}

\begin{table*}[] 

\centering
 \caption{{RUWE value distributions for cool and hot subgroups (see Section \ref{subsec:finding}) of our barium-rich and non-barium-rich (field) stars within 1 kpc determined with bootstrapping. $\rm RUWE \ge1.4$ defines our ``likely binary" group while RUWE $<1.2$ defines our ``likely single star" group (see discussion in Section \ref{subsec:gaiaandruwe}).}}
 \begin{tabular}{c|c|c|c|c}

 RUWE & Cool Barium Stars & Cool Field & Hot Barium Stars & Hot Field \\ 
 
  \hline
 \hline
$\rm RUWE \ge1.4$     & $50.45\pm^{2.70}_{2.70}$\% & $15.97\pm^{0.08}_{0.08}\%$ & $45.78\pm^{2.32}_{2.32}\%$ & $17.00\pm^{0.14}_{0.13}\%$ \\
$1.2\le \rm RUWE <1.4$& $6.91\pm^{1.20}_{1.50}$\% & $5.60\pm^{0.05}_{0.05}\%$ & $8.65\pm^{1.27}_{1.27}\%$ & $6.57\pm^{0.09}_{0.09}\%$\\ 
$\rm RUWE <1.2$        & $42.64\pm^{2.75}_{2.70}\%$ & $78.43\pm^{0.09}_{0.09}\%$ & $45.57\pm^{2.11}_{2.32}\%$ & $76.44\pm^{0.15}_{0.16}\%$\\ 
 \hline
 \end{tabular}
  \label{ruwe_table}
\end{table*}

\begin{figure}
  \centering
  \includegraphics[width=\linewidth]{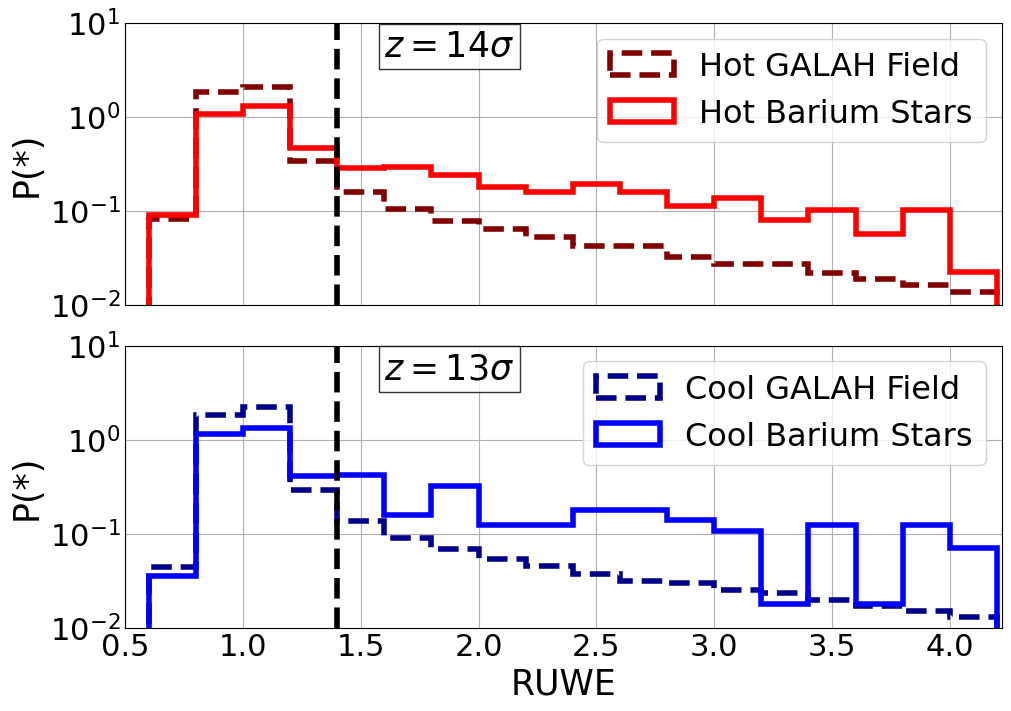}
  \caption{Normalized RUWE distributions for nearby ($\le$ 1 kpc) ``hot" (top panel, $T_{\text{eff}}$ \textgreater 6000 K) and ``cool" (bottom panel, $ T_{\text{eff}}$ $<$ 6000 K) barium stars with a dashed black line marking $\text{RUWE} = 1.4$, the value above which stars tend to be in binary systems (see Section \ref{subsec:gaiaandruwe}). The GALAH field subsamples in both panels do not include barium-rich stars and are selected to have the same $T_{\text{eff}}$ ranges as the barium stars. Both panels show an excess of barium stars with RUWE values greater than 1.4 relative to the field. This supports the mass transfer formation theory for both hot and cool barium star populations. }
  \label{fig:ruwehist}
\end{figure}

\subsection{Exploring Formation Mechanisms of Barium Stars} \label{subsec:masstransfer}
Barium stars in multi-star systems allow for the possibility of past mass transfer, one of the proposed mechanisms for their creation. As mentioned in Section \ref{subsec:gaiaandruwe}, RUWE {can be an} effective tracer of multiplicity for stars within 1 kpc. Figure \ref{fig:ruwehist} presents the RUWE distributions for barium stars (solid line) and our sample (see Section \ref{subsec:using data}) of the GALAH field void of barium stars (dashed line) within 1 kpc of the Sun (to limit the sample to distances that are sensitive to the RUWE multiplicity tracer). {We find that cool, nearby barium stars (333 stars) have $f_{\rm excess}= 2.16$ for RUWE $\ge 1.4$, $f_{\rm excess}=0.23$ for $1.2\le \rm RUWE < 1.4$, and $f_{\rm excess}=-0.46$ for RUWE $<1.2$ relative to the cool, nearby non-barium-enhanced GALAH field (206,427 stars). Similarly, when comparing the hot, nearby barium-rich (474 stars) and non-barium-rich stars (72,719 stars), we find barium stars have fractional excesses of $f_{\rm excess}=1.69$ for RUWE $\ge1.4$, $f_{\rm excess}=0.32$ for $1.2\le\rm RUWE <1.4$, and $f_{\rm excess}=-0.40$ for RUWE $<1.2$. Table \ref{ruwe_table} provides the equivalent percentages, with uncertainties derived from bootstrapping. With RUWE $\ge 1.4$ being an indicator of multiplicity (Section \ref{subsec:gaiaandruwe}), this is evidence that barium stars experience higher rates of multiplicity relative to the barium-normal field, regardless of surface temperature.}

\begin{figure*}
  \centering
  \includegraphics[width=\textwidth]{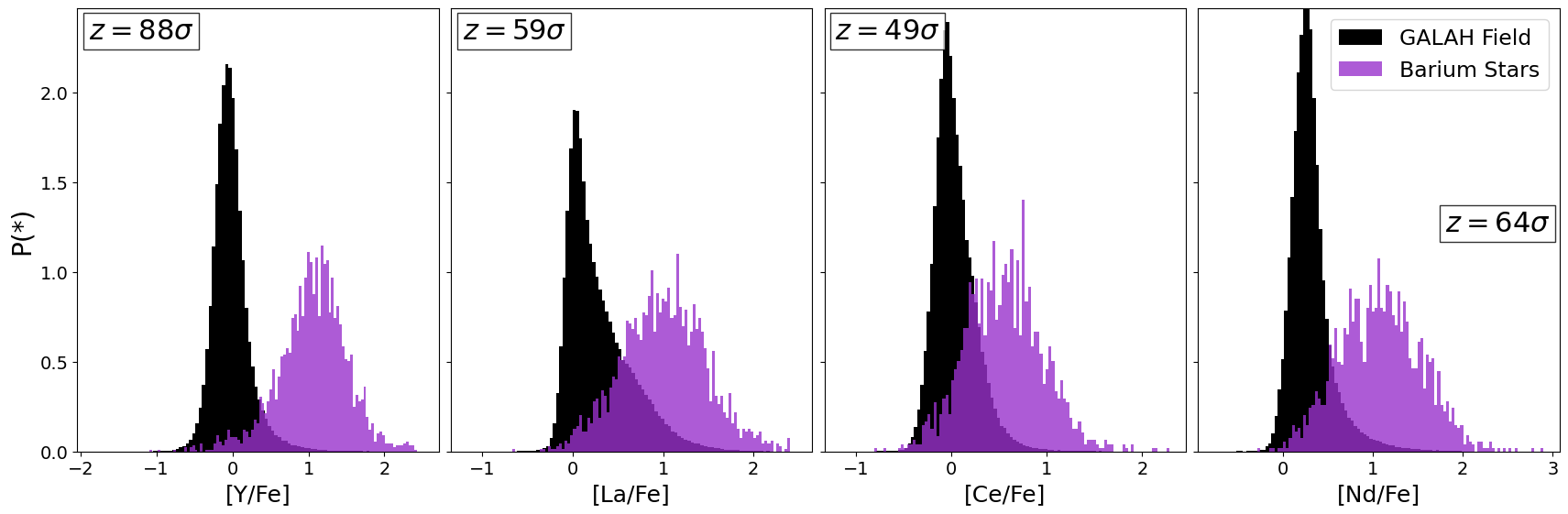}
  \caption{[X/Fe] distributions of $s$-process elements for barium-rich stars (purple) and the GALAH field with barium-rich stars removed (black). Barium stars tend to be richer in $s$-process elements than the background field. This implies that processes which foster barium enhancement also lead to enrichment in the other $s$-process elements. Y shows the strongest and most distinct correlation to barium abundance among the $s$-process elements. This motivates our exploration of Ba and Y abundances in Fig. \ref{fig:ba_y_ruwe} and Fig. \ref{fig:meshgrid}.}
  \label{fig:sprocessspectra}
\end{figure*}

\begin{figure}
  \centering
  \includegraphics[width=\linewidth]{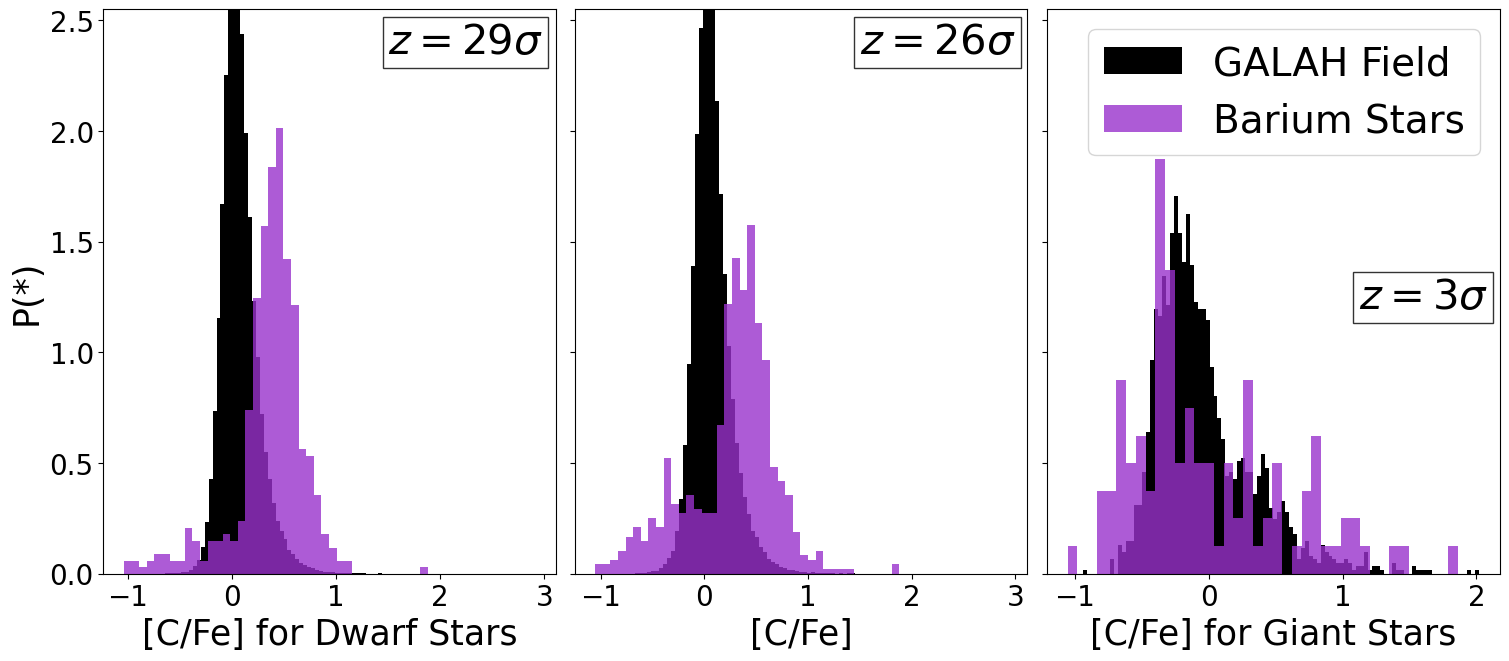}
  \caption{Equivalent distributions as in Figure \ref{fig:sprocessspectra} but for [C/Fe] in dwarfs (left), all stars (middle), and giants (right). We note that barium dwarfs show significant enhancements in [C/Fe] {($+0.29\pm0.01$ dex on average) relative to the field sample while barium giants do not ($+0.02\pm0.01$ dex)}.}
  \label{fig:cspectra}
\end{figure}

Beyond Gaia RUWE, we can return to GALAH chemical abundances to help distinguish between different barium star formation mechanisms. For example, barium stars created through radiative levitation should show a depletion in the $\alpha$-elements \citep[e.g.,][]{2010Vick, atomicdifussionandturbulentmixing}, and those that are created via mass transfer should be $\alpha$-element normal \citep[e.g.,][]{Karakas2016}. In both cases, we expect the barium stars to be rich in all $s$-process elements. In Figure \ref{fig:sprocessspectra}, we compare the [Y/Fe], [La/Fe], [Ce/Fe], and [Nd/Fe] abundance distributions of barium stars to those of the barium star-free GALAH field. {Using MC sampling with GALAH's chemical abundance uncertainties, we find the barium star sample has average abundance in excesses of $\rm \Delta\langle{[Y/Fe]}\rangle=+1.08\pm0.004$, $\rm \Delta\langle{[La/Fe]}\rangle =+0.72\pm0.002$, $\rm \Delta\langle{[Ce/Fe]}\rangle =+0.51\pm0.003$ and $\rm \Delta\langle{[Nd/Fe]}\rangle =+0.77\pm0.003$} relative to the average abundances of the GALAH field void of barium stars. These results demonstrate the correlated $s$-process enrichment: stars that are rich in barium tend to be rich in other $s$-process elements as well, and we find that abundance ratios of [Y/Fe] have the strongest correlation with barium enhancement in our sample.  In Figure \ref{fig:cspectra}, we also investigate the [C/Fe] abundance distributions of our sample, another element affected by AGB star nucleosynthesis \citep[e.g.,][]{dawesreview, Karakas2016}.  Giants are known to alter their surface C abundances due to mixing along the red giant branch \citep[e.g.]{Spoo2022}, so we separate our populations into dwarfs (left panel) and giants (right panel) in addition to examining all evolutionary states together (middle panel).  We find that dwarf barium stars in our sample show average [C/Fe] enhancements of $+0.29\pm0.01$ dex relative to barium-normal field dwarfs. Barium giants, however, do not show such enhancements {, exhibiting $\Delta\langle\rm[C/Fe]\rangle$ of $+0.02\pm0.01$ dex}.

\begin{figure}
  \centering
  \includegraphics[width=\linewidth]{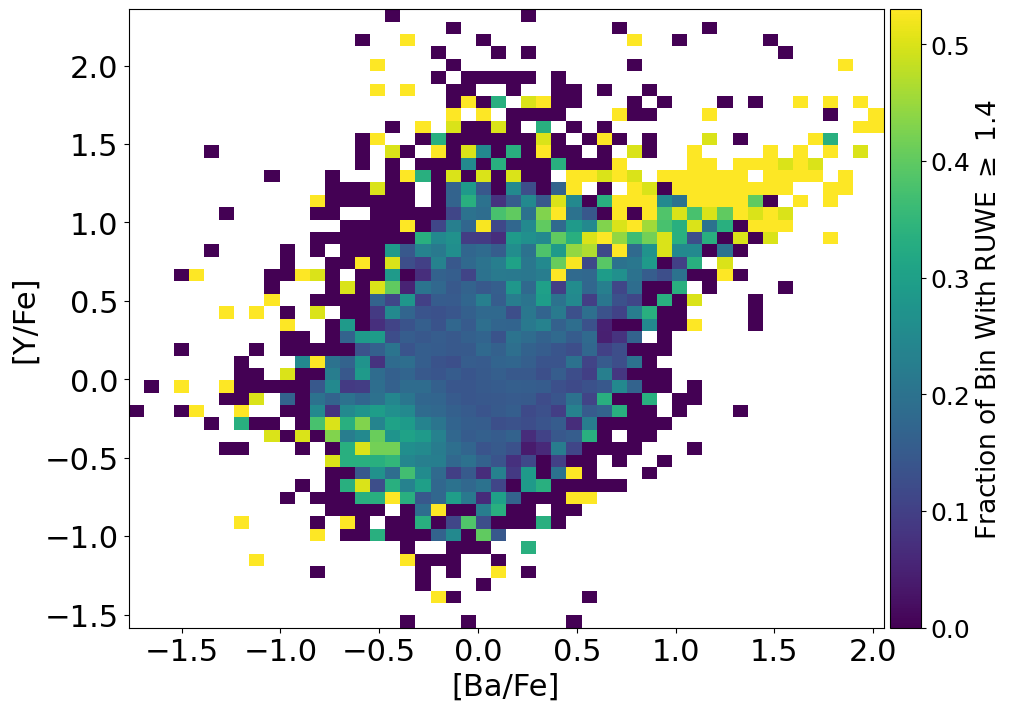}
  \caption{[Y/Fe] vs [Ba/Fe] for nearby ($\le$ 1 kpc) cool stars in the GALAH field. The colors represent the fraction of the bin that has RUWE $\ge 1.4$. The stars that are [Ba/Fe] and [Y/Fe] rich show higher rates of enhanced RUWE relative to the field. This is evidence for mass transfer theory for cool barium stars.}
  \label{fig:ba_y_ruwe}
\end{figure}

When searching for barium stars created through mass transfer, it is {more straightforward} to focus on the cooler stars, as this will ensure that the [Ba/Fe] {enrichment} is not due to radiative levitation \citep[e.g.,][]{radlevalphaelement}. Moving forward, we {use} the cool ($T_{\text{eff}} <6000$ K) stars to probe for mass transfer origins. So far, we have found that barium stars appear to have higher rates of multiplicity (as traced by RUWE) and also tend to be enhanced in other $s$-process elements beyond barium. Next, we combine these results by considering the overlap between these two characteristics. Figure \ref{fig:ba_y_ruwe} isolates the nearby ($\le$ 1 kpc) cool ($T_{\rm eff}$ \textless~6000 K) stars in our GALAH sample and places them in the [Y/Fe] vs [Ba/Fe] plane. Each bin in this plane is colored by the fraction of stars that have RUWE $\ge 1.4$. We {find cool, nearby stars that} are rich in barium and yttrium (196 stars, top right region) {have $f_{\rm excess}=3.28$ for RUWE $\ge$ 1.4 ($68.37\pm^{3.57}_{3.06}\%$ vs. $15.97\pm^{0.08}_{0.08}\%$) relative to the cool, nearby, non-barium or yttrium enhanced field (206,671 stars).} These results {may} suggest that $s$-process {enhanced stars in general have a higher incidence of multiplicity and that multiplicity is correlated to barium star formation in our sample}.

\begin{figure*}
  \centering
  \includegraphics[width=\textwidth]{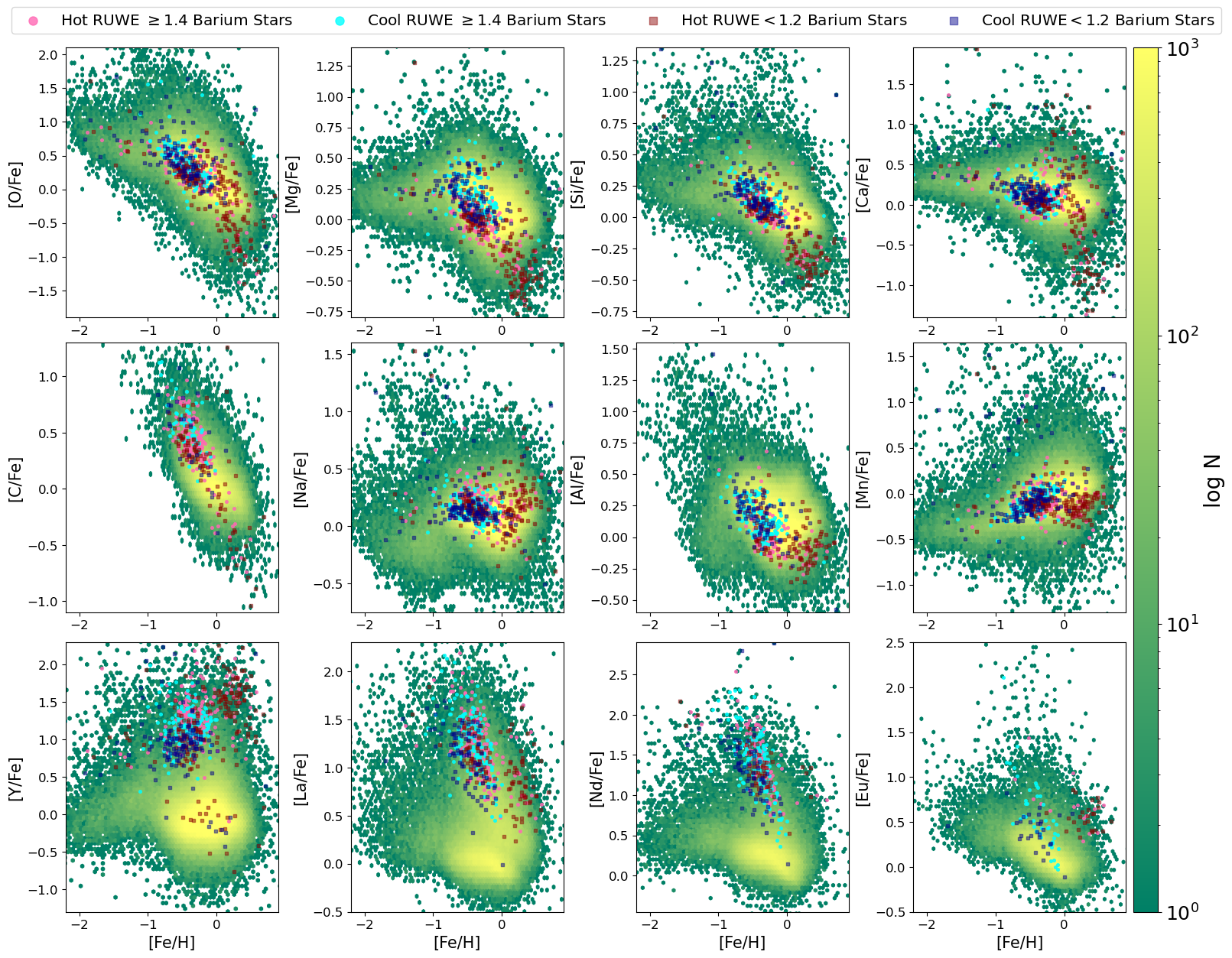}
  \caption{[X/Fe] vs. [Fe/H] distributions within 1 kpc for hot, RUWE $<1.4$ (red), hot RUWE $\ge1.4$ (pink), cool RUWE $\ge 1.4$ (light blue), and cool RUWE $<1.4$ (dark blue) barium stars against those of the GALAH field (background density distribution). Hot barium stars show depletions of $\alpha$-elements (top row) relative to the field while the cool stars mirror the field. Both hot and cool barium stars have distributions similar to the GALAH field in carbon (also created in AGB stars) and various elements not related to the $\alpha$ or $s$-processes (middle row). Finally, both the cool and hot stars are enhanced in the $s$-process elements Y, La, and Nd (bottom row) relative to the background field. The final panel shows the $r-$process element Eu which is discussed in Section \ref{subsec:europiumdiscussion}. }
  \label{fig:bigplot}
\end{figure*}

The positions of barium stars in the [X/Fe] vs. [Fe/H] plane can further inform the formation mechanisms of barium stars by highlighting other potential chemical anomalies (or lack thereof). Stars engaging in mass transfer via stellar wind primarily donate mass from the stellar envelope \citep[e.g.,][]{bondiaccretion}. Since we expect AGB stars to be the donors, barium stars should {primarily} show enhancements in elements that are synthesized during this evolutionary phase \citep{dawesreview}. Finding a star that is chemically normal except for $s$-process element enrichment could therefore be evidence of the mass transfer formation scenario. Figure \ref{fig:bigplot} presents the [X/Fe] vs. [Fe/H] distributions for barium stars in a subset of the elements reported by GALAH. Cool, RUWE $\ge 1.4$ barium stars have {[O/Fe], [Mg/Fe], [Si/Fe], [Ca/Fe], [Na/Fe], [Al/Fe], [Mn/Fe]}, and [Eu/Fe] abundance distributions similar to those of the background field. Their [Y/Fe], [La/Fe], and [Nd/Fe] ($s$-process elements) distributions, however, are systematically enhanced, as are their [C/Fe] abundances, though this is more visible in Figure \ref{fig:cspectra}. Hot, RUWE $<$ 1.2 barium stars show depletions in the $\alpha$-elements O, Mg, Si, and Ca as well as light element C, consistent with expectations from models of radiative acceleration \citep[e.g.,][]{2010Vick}. As in their cooler counterparts, these stars also show enhancements in other $s$-process elements. {We provide a complete quantitative investigation of the $\langle\rm[X/Fe]\rangle$ differences between the barium-rich and field populations in Appendix \ref{appendixmean}.}

\begin{figure}
    \centering
    \includegraphics[width=\linewidth]{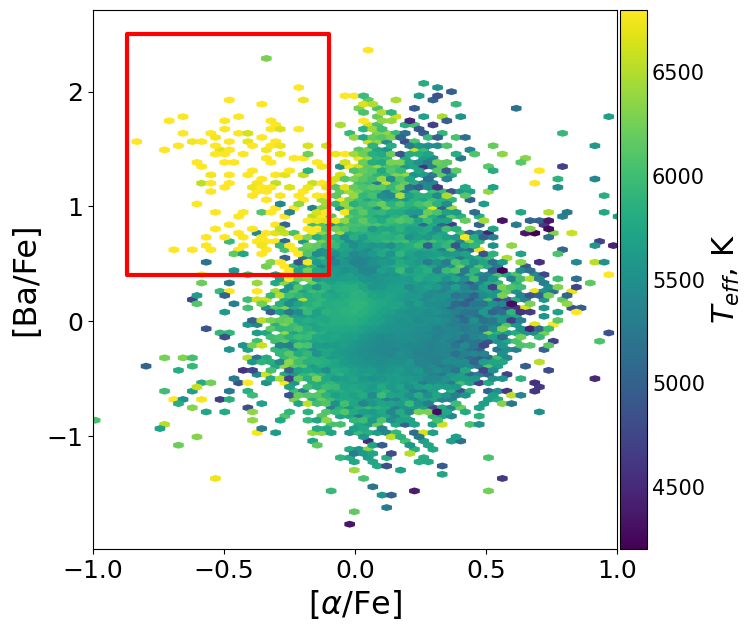}
    \caption{[Ba/Fe] vs. $\rm[\alpha/Fe]$ distribution for the nearby ($<1$ kpc) stars in our sample. We find a correlation between barium enhancement, $\alpha-$element depletion, and $T_{\rm eff}$ in our sample, something that is likely a result of radiative levitation in stellar envelopes.}
    \label{fig:alphabateff}
\end{figure}

{Unlike the cool barium stars, hot barium stars may also display enhanced Eu as a signature of signature of radiative levitation. {We see the strongest evidence for radiative levitation in Figure \ref{fig:alphabateff}. When restricting to nearby stars, we find that $\alpha-$element-poor, barium-enhanced stars are nearly all hotter than $T_{\rm eff}$ = 6000 K, exemplifying radiative levitation's temperature dependence and its role in barium star formation. We find two barium stars cooler than 6000 K: a metal-rich ([Fe/H]$=0.46\pm0.04$) giant  with $T_{\rm eff}=5397.51\pm69.64$ K and RUWE $=1.29$ (Gaia DR3 5921223471126699776), and Gaia DR3 4767775373462950016 which has RUWE $=1.17$ and ${T_{\rm eff}}=5876.11\pm137.35$ K. The $T_{\rm eff}$ uncertainties of the latter star could push its true $T_{\rm eff}$ closer to 6000 K. Importantly, barring the two aforementioned exceptions, none of the barium stars with $\alpha-$element depletions are cool stars, further motivating the mass transfer origins of our cooler ($T_{\rm eff}<6000$ K) subpopulation. Of the 98 barium stars that are in the boxed region of Figure \ref{fig:alphabateff}, 67 have RUWE $<1.2$, 21 have RUWE $\ge1.4$, and 10 have RUWE values between 1.2 and 1.4.  We posit that the hot, nearby, low-RUWE barium stars likely formed through radiative levitation while the hot, nearby, high-RUWE barium stars could have formed either through mass transfer or radiative levitation. This ability for the hot barium stars to form through either mass transfer or radiative levitation is likely one reason that hot stars are overrepresented amongst barium stars (Figure \ref{fig:temphist}).}}

\begin{figure*}
  \centering
  \includegraphics[width=\textwidth]{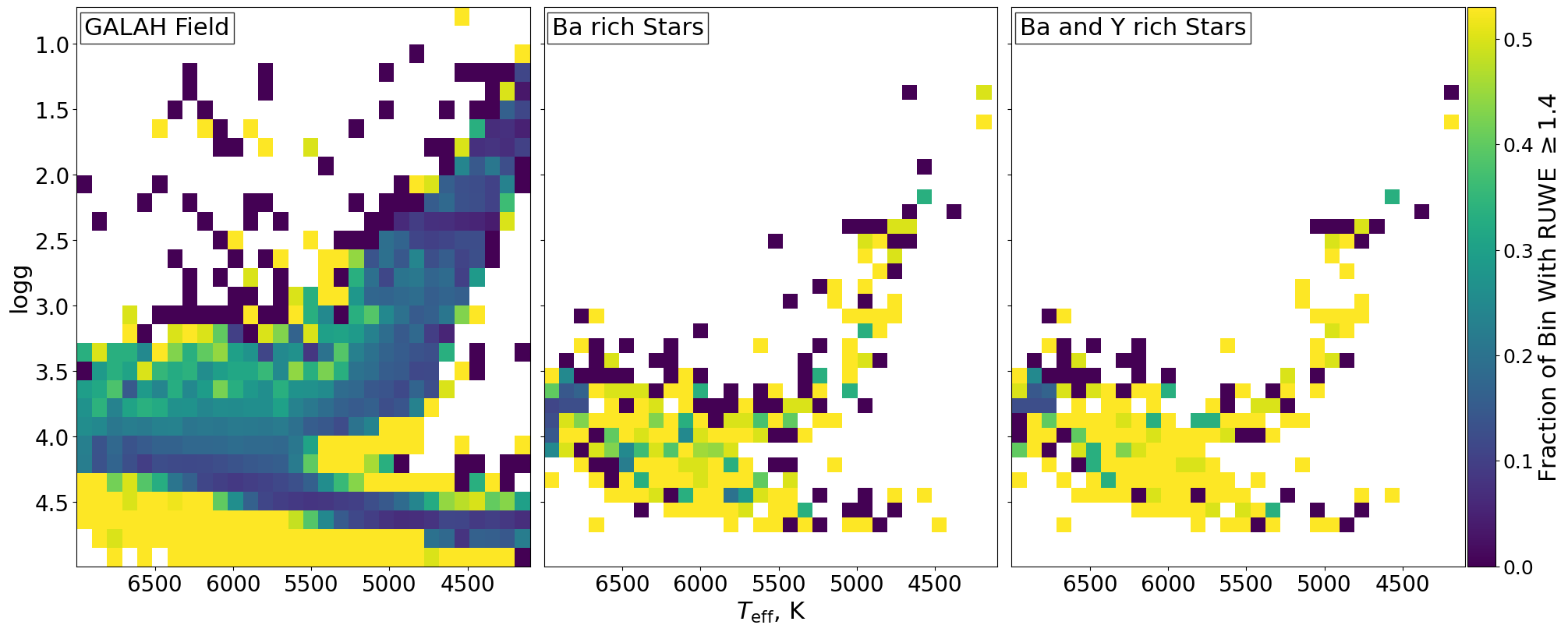}
  \caption{Kiel diagrams showing the fraction of a bin that has RUWE $\ge 1.4$ for nearby ($\le$ 1 kpc) stars in our sample of the GALAH field. We show the GALAH field without barium stars (left), barium-rich stars (middle), and Ba and Y-rich ([Ba/Fe] \& [Y/Fe] $\ge 1$) stars (right). The barium-rich sample displays high rates of RUWE $\ge 1.4$ compared to the GALAH field. This correlation is stronger in the Ba and Y-rich population. The barium-rich and the Ba-and-Y-rich populations support a mass transfer formation mechanism on the main sequence and into the giant branch}
  \label{fig:meshgrid}
\end{figure*}

In this work, we have investigated the GALAH barium star population and identified two distinct temperature-based subgroups (the cool $T_{\text{eff}} < 6000$ K, RUWE $\geq$ 1.4 group and the hot $T_{\text{eff}} > 6000$ K, RUWE $<$ 1.2 group) that may trace two possible formation scenarios for barium stars. The distribution of evolutionary states for each subgroup could explain the temperature distribution in Figure \ref{fig:temphist} and constrain the conditions and timing of the mass transfer process. If barium stars have a higher tendency to be dwarfs, it may imply that the mass transfer occurs in systems with {more extreme initial mass ratios}. Conversely, a higher relative incidence of giant barium stars could suggest that mass transfer happens in systems with mass ratios closer to unity \citep[e.g.][]{Escorza2019}. A higher incidence of giants may alternatively suggest these systems are older, {allowing} the barium star time to evolve into a giant. {We begin this investigation by using} surface gravity cuts to define dwarfs as log g $>3.7$ dex, subgiants as $3.5 \text{ dex} <\text{log g}<3.7 \text{ dex}$, and giants as $\text{log g}<3.5$ dex. Considering RUWE and thus limiting our sample to within 1 kpc, {we compare the 168 cool, nearby barium stars with RUWE $\ge 1.4$ to the non-barium enhanced GALAH field with {identical} constraints (32,961 stars), and find $f_{\rm excess}=-0.16$ for dwarfs, $f_{\rm excess}=0.24$ for subgiants, and $f_{\rm excess}=1.34$ for giants. Doing this comparison for hot, nearby, RUWE $<1.2$ stars, we find that the barium stars (216 stars) have fractional excesses: $f_{\rm excess}=-0.19$ for dwarfs, $f_{\rm excess}=2.45$ for subgiants, and $f_{\rm excess}=7.65$ for giants relative to the GALAH field (55,582 stars). Table \ref{luminosity_table} shows the respective evolutionary state composition of the above populations as a way to further visualize the trend we find: barium stars tend towards later stages of stellar evolution relative to the GALAH field. Returning to RUWE signatures, Figure \ref{fig:meshgrid} illustrates that wherever barium stars lie on the Kiel diagram, whether they are hot or cool, giants or dwarfs, they show increased rates of RUWE $\ge 1.4$, further establishing the correlation between elevated RUWE and barium abundance.}

{Recall from Section \ref{subsec:finding} that nearby (again limiting detection bias of giants) barium stars were hotter on average than the nearby GALAH field ($\rm \Delta$$\langle T_{\rm eff}\rangle$ = $+386\pm^{3.39}_{3.36}$ K). In the case of blue stragglers \citep[e.g.,][]{bluestragglers,Nine2024, Pal2024}, mass transfer has been shown to increase stellar $T_{\rm eff}$ in dwarf stars, implying that this trend in $T_{\rm eff}$ may be indicative of mass transfer. However, mass transfer most likely is not the sole cause of this temperature-related phenomenon, particularly with barium giants, whose surface temperature is relatively uncorrelated with mass transfer, exhibiting a similar tendency towards higher temperatures. Underlying metallicity trends can also have a role in this temperature trend.  Metallicity affects stellar opacity, resulting in smaller envelopes and therefore higher surface temperatures \citep[e.g.][]{Ramirez2005, Huang2015}. Nearby barium stars have an average difference in metallicity of $\Delta\langle\rm[Fe/H]\rangle=-0.24\pm0.004$, potentially explaining the temperature trend in giant stars. Finally, the trend toward higher $T_{\rm eff}$ could be the result of a selection effect toward barium stars created via radiative levitation. The effects of radiative levitation increase with increasing $T_{\rm eff}$  \citep[e.g.,][]{Richard2002}, meaning that by probing stars experiencing the strongest effects of radiative levitation (those capable of reaching the [Ba/Fe] $\ge1$ threshold), we are biasing a hotter selection of stars with the [Ba/Fe] cutoff. We save further explorations of the metallicity and $T_{\rm eff}$ trends of barium stars for future works as they do not represent the main focus of this manuscript.}

\begin{table*}[] 
\centering
 \caption{{Evolutionary state distributions for nearby barium-rich and non-barium-rich (GALAH field) populations as defined by log g (See Section \ref{subsec:masstransfer}). Values are derived from MC sampling with GALAH-provided log g uncertainties.}}
 \begin{tabular}{c|c|c|c|c}

 \multirow{ 2}{*}{Evolutionary State} & Cool RUWE $\ge1.4$  & Cool RUWE $\ge1.4$  & Hot RUWE $<1.2$  & Hot RUWE $<1.2$  \\
 & Barium Stars & GALAH Field & Barium Stars & GALAH Field \\
\hline
\hline
Dwarf & $72.62\pm^{1.19}_{1.79}$\% & $86.06\pm^{0.11}_{0.11}\%$ & $76.85\pm^{2.31}_{1.85}\%$ & $94.90\pm^{0.08}_{0.08}$\%  \\
Subgiant & $5.95\pm^{1.79}_{1.79}$\% & $4.78\pm^{0.11}_{0.11}\%$ & $13.89\pm^{1.86}_{2.31}\%$ & $4.03\pm^{0.07}_{0.07}\%$\\ 
Giant & $21.43\pm^{1.79}_{1.19}\%$ & $9.17\pm^{0.08}_{0.08}\%$ & $9.26\pm^{1.39}_{1.39}\%$ & $1.07\pm^{0.04}_{0.04}\%$ \\
\hline

 \end{tabular}
 \label{luminosity_table}
\end{table*}

\subsection{Galactic Membership of Barium Stars} \label{subsec:where}

\begin{figure*}
  \centering
  \includegraphics[width=.8\textwidth]{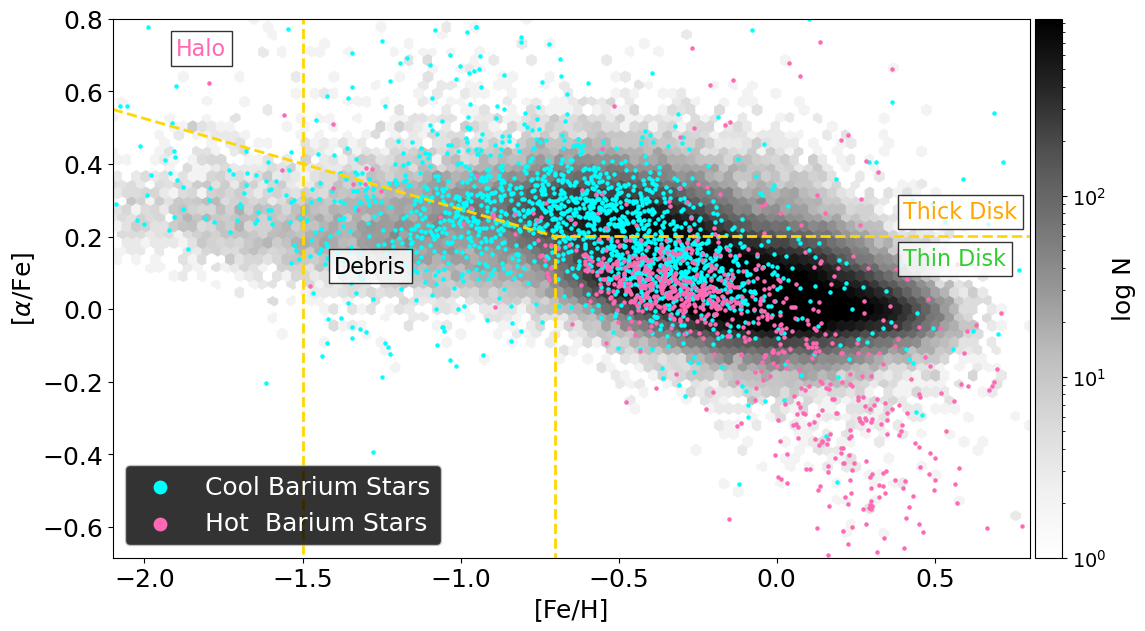}
  \caption{[$\alpha$/Fe] versus [Fe/H] for the GALAH field (background) with cool barium stars (blue) and hot barium stars (red) overplotted. The horizontal/sloped dashed yellow line marks our adopted chemical cutoff between the thin and thick disks while the vertical lines define the debris/accreted and halo populations, adopted from \citep{navarro2011}. Cool barium stars exist in all regions of the Galaxy. Hot barium stars are subject to $\alpha$-element depletion {(evidence of radiative levitation)} due to atomic diffusion and thus [$\alpha$/Fe] does not accurately trace their intrinsic Galactic memberships. This figure presents evidence that mass transfer (as traced by the cool, high RUWE barium stars) is not limited to a specific Galactic subpopulation.}
  \label{fig:tinsleywaller}
\end{figure*}

The occurrence of barium stars among different Galactic populations can add further context to the origins of these systems. In order to sort barium stars according to Galactic membership, we explore the application of both chemical and kinematic definitions of Galactic populations. Starting with chemical definitions, in Figure \ref{fig:tinsleywaller}, we present [$\alpha$/Fe] vs. [Fe/H] distributions for our cool and hot barium star subgroups to characterize these populations according to their Galactic membership \citep{Tinsley1980}. {Figure \ref{fig:tinsleywaller} and the following Galactic membership explorations omit RUWE and distance cuts to enable a larger investigation of barium star occurrence in the thick disk and halo.} Adopting the conventions of \citet{navarro2011}, we define the thin (low-$\alpha$) disk as stars with [$\alpha$/Fe] $<0.2$ for [Fe/H] $>-0.7$ and define the thick disk (high-$\alpha$) as stars with [$\alpha$/Fe] $>0.2-(\text{[Fe/H]/4}+0.7)$ for $-1.5<\text{[Fe/H]}<-0.7$ and [$\alpha$/Fe] $>0.2$ for [Fe/H] $>-0.7$. Stars with [Fe/H] $<-1.5$ tend to be associated with the halo and stars with $-1.5<\text{[Fe/H]}<-0.7$ are commonly classified as tidal ``debris'' (also termed accreted halo) stars \citep[e.g.][]{navarro2011, Buder2022}. Out of the {1883} cool barium stars $\alpha$-element measurements in our sample, we find {fractional excesses of $f_{\rm excess}=-0.54$ for the thin disk, $f_{\rm excess}=1.47$ for the thick disk, $f_{\rm excess}=6.30$ for debris stars, and $f_{\rm excess}=6.02$ for halo stars relative to the GALAH field without barium stars under the same constraints (352,617 stars). The memberships of hot barium stars should be interpreted with caution due to their suspected $\alpha$-element depletion. However, the following fractional excesses can be used qualitatively to compare with kinematic assignments of Galactic membership below. Repeating this examination for hot stars that have reported $\alpha$-element abundances, we find that the barium stars (786 stars) have $f_{\rm excess}=-0.09$ for the thin disk, $f_{\rm excess}=1.00$ for the thick disk, $f_{\rm excess}=2.28$ for debris stars, and $f_{\rm excess}=6.11$ for halo stars relative to the non-barium-enhanced GALAH field (54,552 stars). We present the equivalent membership distributions in terms of percentages in Table \ref{disk_table}. Overall, based on chemical assignments of Galactic membership, we find a higher fraction of barium stars existing in older, more metal poor regions of the Galaxy relative to the field rate of thick disk, debris, and halo membership.}

\begin{figure*}
  \centering
  \includegraphics[width=.8\textwidth]{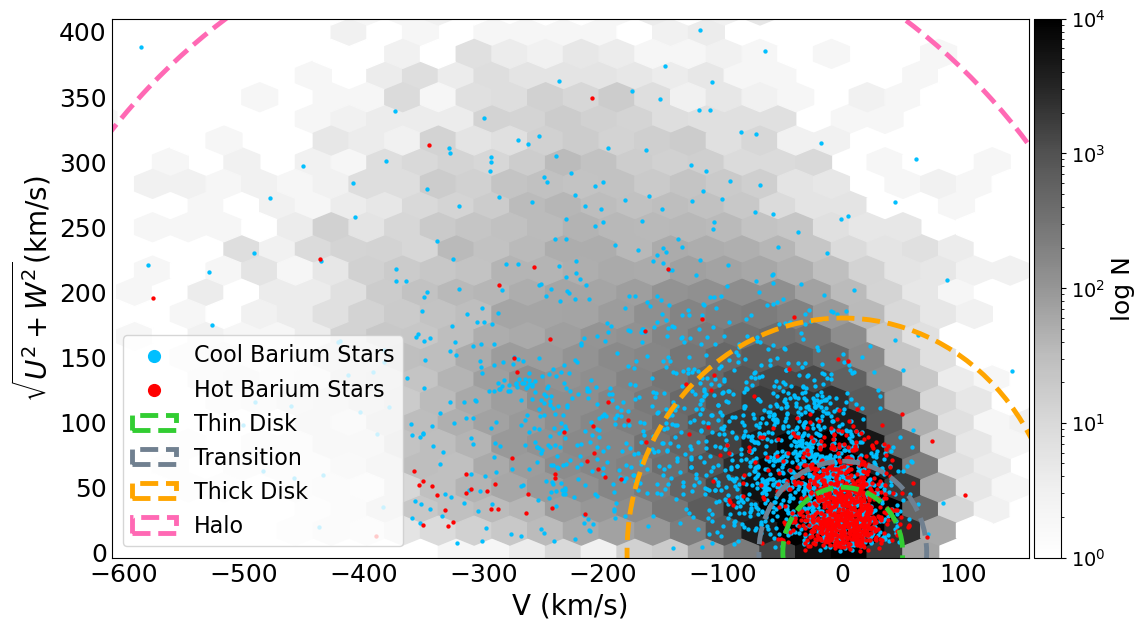}
  \caption{A Toomre diagram showing the LSR-corrected UVW velocities of the GALAH field (background), cool barium stars (blue), and hot barium stars (red). The dashed lines represent the kinematic definitions for each Galactic subpopulation. Stars interior to the green dashed line share the kinematics of stars in the thin disk, stars between the dashed gray and dashed green line have overlapping probabilities of being in the thin or thick disks, stars interior to the orange dashed line but exterior to the gray have thick disk kinematics, and those within the dashed pink line and exterior to the orange dashed line have the motions of typical halo stars. Stars exterior to the pink dashed line have speeds potentially exceeding the Galactic escape speed \citep{escapevelo} (see discussion in Section \ref{discuss:barlocation}). Hot barium stars are found primarily in the thin disk, but are also found in the thick disk and the halo. Cool barium stars are found in the thin disk, thick disk, and the halo. This is further evidence that barium star formation is not limited to a specific Galactic subpopulation.}
  \label{fig:toomre}
\end{figure*}

In addition to chemical definitions, we employ kinematic definitions to determine the Galactic membership of barium stars \citep[e.g.,][]{Toomre1964}. This gives us two independent methods of determining barium star membership and offers the chance to characterize the memberships of the hot barium stars in lieu of accurate intrinsic [$\alpha$/Fe] abundances. Using kinematic data from Gaia, we present the LSR-corrected UVW velocities for two barium star temperature subgroups in Figure \ref{fig:toomre}. Kinematically, we define thin disk as stars with $\nu_{\rm tot} \equiv \sqrt{U^2+V^2+W^2} < 50 \text{ km/s}$, the thick disk as stars with $70<\nu_{\rm tot}<180 \text{ km/s}$ \citep[e.g.,][]{Nissen2004,Bensby2014}, and the halo as stars with $\nu_{\rm tot}>180 \text{ km/s}$ \citep{Nissen2010}. We include the lower bound of the confidence interval for the Galactic escape speed ($\nu_{\rm tot}>498$ km/s in Galactic standard of rest) shown as the dashed pink semicircle in Figure \ref{fig:toomre} \citep{escapevelo}. In Figure \ref{fig:toomre}, for the 1919 cool barium stars in our sample {with available velocity information, we find $f_{\rm excess}=-0.56$ for the thin disk, $f_{\rm excess}=-0.38$ for the transition zone ($50<\nu_{\rm tot}<70$ km/s region where probabilities for thin and thick disk overlap), $f_{\rm excess}=0.33$ for the thick disk, and $f_{\rm excess}=6.74$ for the halo when compared to the cool, non-barium-enhanced GALAH field with velocity measurements (352,514 stars). When comparing the hot stars that have kinematic measurements, barium stars (890 stars) have $f_{\rm excess}=-0.17$ for the thin disk, $f_{\rm excess}=0.14$ for the transition zone, $f_{\rm excess}=0.57$ for the thick disk, and $f_{\rm excess}=9.56$ for the halo relative to the GALAH field without barium stars (94,503 stars). We further present the percentages of Galactic membership as determined by kinematics in Table \ref{disk_table}. We find the cool and hot barium stars show similar trends kinematically as they did chemically, with the barium stars as a whole showing an increased incidence of residing in the thick disk and halo when compared to the GALAH field.}

In general, our population of barium stars shows a preference towards old, metal poor populations compared to the barium-normal field. {This trend may reflect the dependence of $s$-process efficiency on metallicity when the $^{13}$C($\alpha$, n)$^{16}$O reaction serves as the dominant neutron source. In cases of lower metallicity, $s$-process yields, specifically barium, are shown to be significantly higher \citep{Clayton1961,Clayton1988}}. This trend may also reflect metallicity-dependent rates of multiplicity if mass transfer is the primary formation mechanism for barium stars \citep[e.g.,][]{El-Badry2019}. Similarly, mass transfer requires a time delay between system formation and mass transfer due to the necessary evolution of the higher mass companion into an AGB star. If 2-3 M$_{\odot}$ stars are the primary polluters of barium stars \citep[e.g.,][]{Cseh2022}, then there will be a time delay of 1-2 Gyr until the system forms a barium star. On average, this reliance on a companion star going through its stages of evolution could push the barium stars to be older, and thus having them be more likely to have low metallicities \citep[e.g,][]{carraro1998,doner2023}. Finally, this Galactic population distribution of barium stars could originate from our definition of a barium star. The more metal poor a star is, the less barium is required to reach the [Ba/Fe] $\ge 1.0 $. This is supported when looking at the [Fe/H] values of our barium star populations for each Galactic membership. Using chemical separations for the disks, barium stars in the thin disk have [Fe/H] ratios $-0.11\pm0.004$ dex lower than the field on average and those in the thick disk have [Fe/H] ratios $-0.17\pm0.004$ dex lower than the field on average. Using kinematic separations, barium stars in the thin disk and thick disk have [Fe/H] ratios $-0.11\pm0.003$ dex and $-0.30\pm0.004$ dex lower than the GALAH field {on average}. {We find that the tendency of barium stars to be more metal poor than their barium-normal counterparts persists across Galactic memberships}.

\begin{table*}[] 
\centering
 \caption{{Galactic membership distributions for cool and hot barium-rich and non-barium-rich (GALAH field) populations as defined by $\rm[\alpha/Fe]$ and [Fe/H] (top half) and as defined by LSR-corrected $U,~ V, ~W$ velocities (bottom half, See Section \ref{subsec:where}). Values are derived from MC sampling with GALAH-provided [$\alpha$/Fe] and [Fe/H] $1\sigma$ uncertainties and UVW 5th and 95th percentile uncertainties.}}
 \begin{tabular}{ccc|c|c|c|c}

 && Membership & Cool Barium Stars & Cool GALAH Field & Hot  Barium Stars & Hot GALAH Field \\ 
 
  \hline
 \hline
\parbox[t]{2mm}{\multirow{4}{*}{\rotatebox[origin=c]{90}{Chemical}}}&\parbox[t]{2mm}{\multirow{4}{*}{\rotatebox[origin=c]{90}{Definitions}}}&Thin Disk& $36.54\pm^{0.48}_{0.53}$\% & $80.00\pm^{0.03}_{0.03}\%$ & $85.24\pm^{0.76}_{0.64}\%$ & $93.52\pm^{0.06}_{0.05}$\%  \\
&&Thick Disk & $42.33\pm^{0.58}_{0.64}$\% & $17.08\pm^{0.03}_{0.04}\%$ & $10.69\pm^{0.64}_{0.64}\%$ & $5.34\pm^{0.05}_{0.05}\%$\\ 
&&Debris Stars & $16.78\pm^{0.48}_{0.53}\%$ & $2.30\pm^{0.02}_{0.02}\%$ & $3.44\pm^{0.51}_{0.51}\%$ & $1.05\pm^{0.03}_{0.02}\%$\\ 
&&Halo Stars & $4.35\pm^{0.21}_{0.21}\%$ & $0.62\pm^{0.01}_{0.01}\%$   & $0.64\pm^{0.13}_{0.13}\%$   & $0.09\pm^{0.01}_{0.01}\%$       \\
\hline
\hline

\parbox[t]{1mm}{\multirow{4}{*}{\rotatebox[origin=c]{90}{Kinematic}}}&\parbox[t]{1mm}{\multirow{4}{*}{\rotatebox[origin=c]{90}{Definitions}}}&Thin Disk & $20.58\pm^{0.16}_{0.10}$\% & $46.34\pm^{0.01}_{0.01}\%$ & $61.24\pm^{0.24}_{0.22}\%$ & $73.67\pm^{0.02}_{0.02}$\%  \\
&&Transition Zone & $13.13\pm^{0.21}_{0.21}\%$ & $21.18\pm^{0.02}_{0.02}\%$ & $18.88\pm^{0.34}_{0.34}\%$ & $16.58\pm^{0.03}_{0.03}\%$\\ 
&&Thick Disk & $38.51\pm^{0.26}_{0.26}$\% & $28.89\pm^{0.01}_{0.01}\%$ & $14.55\pm^{0.34}_{0.22}\%$ & $9.24\pm^{0.02}_{0.02}\%$\\ 
&&Halo Stars& $27.77\pm^{0.21}_{0.21}\%$ & $3.59\pm^{0.01}_{0.01}\%$   & $5.28\pm^{0.11}_{0.11}\%$   & $0.50\pm^{0.01}_{0.01}\%$       \\
 \hline
 \end{tabular}
 \label{disk_table}
\end{table*}

\section{Discussion} 
\label{sec:discussion}
{We have conducted one of the largest population-level studies of barium stars with high resolution spectra, identifying a sample of $\sim$3000 newfound barium stars.} This sample enables future work to better understand the physics of binary mass transfer and radiative levitation. Though our study confirms that barium stars are relatively rare ($0.62\%$ of our sample of the GALAH survey), they can serve as a laboratory within which to understand other flavors of mass transfer and internal chemical transport that do not obviously alter the surface compositions of stars. As mentioned in \cite{GCE}, barium stars are not predicted by current models of GCE as they do not typically account for stellar multiplicity or mass transfer. However, they may play a non-negligible role in the Galactic content and distribution of C and $s$-process elements. For example, \citet{Osborn2025} assess the impact of binarity on AGB star nucleosynthesis and evolution. They find that 20-25\% less C and $s$-process elements are ejected into the interstellar medium from AGB stars in a binary system when compared to single AGB stars. We find evidence for binarity playing an important role in the formation of barium stars, indicating that our sample is an ideal laboratory for testing theoretical yields and stellar wind models for AGB stars in binary systems \citep[e.g.,][]{Abate2013,saladino2019, Osborn2025, Kemp2025}. This is an important step toward improving {GCE models to account for AGB binarity and its impact on the C and $s$-process content of the Galaxy.}

\subsection{Implications for Barium Star Formation} \label{discuss:masstrans}
We find several pieces of evidence highlighting mass transfer as a dominant mechanism in the creation of barium stars in GALAH. Across both the cool ($T_{\text{eff}} < 6000$ K) and hot ($T_{\text{eff}} > 6000$ K) regimes, barium stars show enhanced rates of RUWE $\ge 1.4$ relative to the field (Figures \ref{fig:ruwehist}, \ref{fig:ba_y_ruwe}, and \ref{fig:meshgrid}), indicating that mass transfer {might have a significant} role in barium star formation. When isolating our analysis to cool barium stars with RUWE $\ge1.4$, stars with the strongest likelihood of past mass transfer, we find that barium stars tend to show enhancements in other $s$-process elemental abundance ratios, most strongly correlated [Y/Fe], but also [La, Ce, and Nd/Fe]. This is consistent with AGB star nucleosynthetic models which suggest that Y, Ba, La, Ce, and Nd are synthesized in comparable quantities ([X/Fe] within 0.3 dex of each other for solar metallicity AGB stars depending on mass, e.g., \citealp{Karakas2016}). However, using [Nd/Fe] $< 0.5$ to define a lack of Nd enhancement, {$18\pm^2_2$ of the} cool barium stars with RUWE $\ge1.4$ do not show notable enhancements in Nd. These cases may reflect mass transfer from an intermediate (M$\sim$3 M$_\odot$) mass AGB star \citep[e.g.,][]{Karakas2016}, though we leave the detailed investigation into the origins of this effect to future work. We find weaker, but still detectable, enhancements in [C/Fe] abundances in our barium star population, another element affected by AGB nucleosynthesis \citep[e.g.,][]{dawesreview, Karakas2016, Osborn2025, Roriz2025}. These enhancements are only detected among the barium dwarfs. Mixing along the giant branch that alters intrinsic [C/Fe] could explain the lack of C enhancement among barium giants \citep[e.g.][]{Motta2018, Spoo2022, Brady2023, Nine2024}. In addition to enhancements in $s$-process elements, barium stars with mass transfer origins should show otherwise typical abundances of elements unaffected by AGB star nucleosynthesis, something that we observe in our high RUWE, cool barium star sample (Figure \ref{fig:bigplot}). Furthermore, the {over-representation of early G- and F-type barium stars is} an additional piece of evidence supporting the mass transfer scenario. When {dwarf} stars experience mass transfer, the increase in mass causes an increase in surface temperature \citep[e.g.,][]{bluestragglers,Pal2024}. Typical barium stars that arise through mass transfer can be expected to accrete between $0.01-0.1 \text{ M}_{\odot}$ from a typical AGB companion \citep[e.g,][]{Theuns1996,Mohamed2012,Abate2013}. Using the mass-luminosity relation \citep[e.g,][]{Demircan1991,Eker2018} for barium dwarfs of mass $\text{M}\sim\text{0.5 - 1.5 M}_\odot$ , we {can expect} a $T_{\rm eff}$ increase of $\sim20-400$ K. This increase in $T_{\rm eff}$ from accretion could explain the higher incidence of early G and F barium stars compared to the GALAH field in Figure \ref{fig:temphist}. 

{The evidence presented here suggesting mass transfer as a primary reason for barium enhancement} directly supports the legacy of barium star studies that have {investigated binarity in barium stars} {\citep[e.g,][]{binarynature,lamost,Burbidge1957, Warner1965, McClure1980, Boffin1988, McClure1990, Jorissen1992, North1994, North2000, Jorissen1998, deCastro2016, Merle2016, Jorissen2019}}. {For example, recently,} \citet{Rekhi2024} identified main sequence white dwarf binary systems in Gaia DR3 and searched for signs of $s$-process enhancement using GALAH DR3-reported abundances. They indeed find enhanced $s$-process elemental abundances in their sample and demonstrate a dependence of $s$-process enhancement on progenitor white dwarf mass and metallicity. Our analysis takes the inverse approach, searching for signs of binarity in $s$-process enhanced stars. Our sample of GALAH barium stars is ideal for conducting studies similar to that of \citet{Rekhi2024} {and references throughout} to relate the orbital architectures of these systems with relative $s$-process enhancement. 

In addition to finding evidence supporting the mass transfer scenario, we find evidence for hot barium stars that arise from radiative levitation. Hot ($T_{\text{eff}}>6000$ K), RUWE $<1.2$, nearby ($<$ 1 kpc), and $\alpha$-element depleted barium stars are the subgroup that likely experiences this effect (Figures \ref{fig:bigplot} and \ref{fig:tinsleywaller}). These systems demonstrate that the observed surface composition of a star does not always reflect its natal composition and that radiative acceleration can alter stellar surface abundances even at temperatures as low as $\sim$ 6000 K.  

\subsection{Interpreting Barium Star Disk Memberships} \label{discuss:barlocation}
We found in Section \ref{subsec:where} that barium stars are pervasive across the thin disk, thick disk, and halo, though we find a slight over-representation of barium stars in the {more metal-poor} Galactic components (thick disk and halo) relative to the equivalent barium-normal field. This implies barium star formation is {generally} independent of disk membership, and Galactic membership is not a primary driver of formation. Following from Section \ref{subsec:masstransfer}, it can be posited that mass transfer is ubiquitous across Galactic space and time due to the existence of barium-rich giants and dwarfs {across} both disks and the halo.

Though most hot ($T_{\text{eff}}> 6000$ K) barium stars exist in the thin disk, we note the presence of a population of eight hot, dwarf barium stars {with both} chemical and kinematic signatures indicating halo membership. These systems represent another group of high-likelihood post-mass transfer systems. The youngest halo stars are $\sim$10 billion years old \citep[e.g.,][]{Jofre2011}.  At 10 billion years old, main sequence stars that are hotter than 6000 K should have already depleted their core H fuel and evolved off the main sequence \citep[e.g.,][]{Choi2016}. The presence of hot barium stars in the halo necessitates the transfer of additional mass to the star to prolong core fusion on the main sequence. This is the same phenomenon that could explain the existence of some young $\alpha$-rich stars \citep[e.g.,][]{Martig2015,Borisov2022, Lu2025, Maas2025}.

Finally, we identify a sample of eight barium stars with high total velocities in the Galactic reference frame ($\nu_{\rm tot}>395$ km/s).\footnote{We convert from LSR velocities into a total speed relative to the Galactic center by adopting $\nu_0=232.8$ km/s as the circular speed at the Sun \citep{McMillan2017}.} These stars are cool ($T_{\text{eff}}<$  6000 K), giant stars with the exception of one star at 6100 K (Gaia DR3 2884536461613379840). This hot ($T_{\text{eff}}>6000$ K) barium dwarf star has been identified in \citet{Giribaldi2023} as belonging to the Gaia-Sausage-Enceladus system, the most recent major merger experienced by the Milky Way. We argue that this is likely a post mass-transfer system that received mass from an AGB star. The cool barium giant Gaia DR3 5922493338335870080 occupies the ``chemically unevolved" quadrant of the [Mg/Mn] - [Al/Fe] plane, indicating a likely association with an accreted dwarf galaxy \citep[e.g.,][]{Hawkins2015, Horta2020, Fernandes2023}. The fastest moving star of this high velocity selection, Gaia DR3 6706018041696378752, has a total velocity of $\nu$=524.7 km/s with respect to the Galactic center. This {high velocity} exceeds the lower bound ($\nu_{\rm esc}> 498$ km/s) of the 90\% confidence interval in {Galactic escape velocity} from \cite{escapevelo}, {raising questions about its origins}. Each of these systems, alongside other halo barium stars in GALAH DR3, present an excellent sample for detailed spectroscopic follow-up to constrain AGB nucleosynthesis and mass transfer in metal-poor environments.

\subsection{{r-Process} Pollution as a Source of Barium Stars} \label{subsec:europiumdiscussion}
It is worth noting that \cite{EuContaminationPossiibility} found evidence for barium stars enriched by collapsars {that engaged in $r$-process nucleosynthesis} in dwarf galaxy NGC 1569. We explore this possible source of barium enrichment by looking at the [Eu/Fe] abundances of our barium star sample. \cite{EuContaminationPossiibility} measures a mean [Eu/Fe] $=$ 1.9 for the barium stars in the study. Our barium star sample generally does not display elevated levels of $\text{[Eu/Fe]}$, so {we conclude that} collapsars are not the primary formation mechanism for {our} barium-rich stars.  However, there are exceptions to this.  We find that 45 barium stars (5 of which are hot, the rest cool) have [Eu/Fe] greater than 1 {dex}, which represents a 10x [Eu/Fe] enrichment relative to Solar.  12 of the barium stars also show RUWE \textgreater 1.4. These Ba and Eu-enriched stars are potential candidates of enrichment from a collapsar or other $r-$process event (e.g., neutron star merger) and deserve dedicated follow-up \citep[e.g.,][]{Siegel2019, Brauer2021, EuContaminationPossiibility}. Alternate explanations for their enrichment include radiative levitation (for the hot stars) or enrichment from two separate events, one $r-$process event and a separate mass transfer event from an AGB star later on. Regardless, it remains that across our large population of barium stars, mass transfer and radiative levitation are the two primary formation mechanisms.

\section{Conclusions} \label{sec:conclusion}
{In this work, we identify} 2810 barium-enhanced ($\text{[Ba/Fe]} \ge 1$) stars in the GALAH survey, {which corresponds} to 0.62\% of the survey being barium-rich. Using chemical abundances from GALAH and Gaia astrometry, we find evidence supporting two {major} theories of barium star formation: 

\begin{enumerate}
  \item Stars become barium-enhanced via mass transfer from an AGB companion.
  \item Stars develop excess surface barium, though no wholistic enrichment, due to radiative levitation.
\end{enumerate} 

There is strong evidence for mass transfer being the dominant {source} of barium enhancement for cool ($T_{\text{eff}} < 6000$ K) stars, supporting a legacy of {existing literature that studied} these systems. This evidence consists of increased rates of RUWE $\ge 1.4$, enrichment in the $s$-process elements beyond barium, and, for dwarfs, enrichment in C {among our barium star sample}. We propose that the cool stars in our sample can be used to trace AGB star nucleosynthesis, constrain models of mass transfer, and identify certain types of unresolved binary systems.

No dominant formation mechanism {is} determined for the hot ($T_{\text{eff}} > 6000$ K) barium-enhanced stars, but evidence for both mechanisms {is presented.} We find that the hot barium stars also exhibit elevated RUWE, supporting mass transfer, but we {also} find cases where hot, low-RUWE stars {exhibit} $\alpha$-element depletions. Evidence for radiative levitation highlights that even in (likely) single main sequence stars as cool as $T_{\rm eff}$ = 6000 K, surface abundances can be altered by internal chemical transport and deviate from the stars' birth compositions.

Through kinematic and chemical analyses, we determine the memberships of barium stars to different Galactic subpopulations. Barium stars exist in the thin disk, thick disk, and halo, indicating that mass transfer is ubiquitous across Galactic environments and time. Although barium stars exist across all Galactic populations, we find that barium stars have a higher incidence of {residing} in low metallicity populations (thick disk and halo) when compared to the barium-normal GALAH field. We identify interesting barium star populations for potential follow-up. These populations include 8 hot ($T_{\rm eff} > 6000$ K), main sequence halo barium stars that have likely been rejuvenated by AGB mass transfer, akin to massive $\alpha-$enhanced stars. Furthermore, we identify one barium star with a total velocity that may {exceed} the Galactic escape speed at its position. Two additional noteworthy high velocity barium stars are identified, with one being a confirmed Gaia-Sausage-Enceladus member and the other a likely accreted halo member. Finally, we identify 43 barium stars that are also 10x enriched in [Eu/Fe] relative to the Sun.  These barium stars may have been enriched by $r-$process nucleosynthetic events such as neutron-star mergers or collapsars.  These unique populations of barium stars are excellent candidates for follow-up observations and/or analysis to constrain mass transfer in metal poor regimes and $r-$process nucleosynthesis.

Overall, we find that mass transfer plays a significant role in barium star formation. We find evidence for this mass transfer in hot and cool stars, giants and dwarfs, and thin and thick disk stars, implying that barium stars, and subsequently mass transfer, are both ancient and ongoing phenomena that are not limited to any one stellar type or Galactic component.

\section{Acknowledgments}
{We thank the referee for a careful review of the manuscript and the constructive suggestions that greatly improved the quality of this manuscript.}   CM thanks the Spring 2023 cohort of the Freshman Research Initiative (FRI) at the University of Texas at Austin for conducting an initial investigation into barium stars in GALAH under the guidance of CM to confirm the feasibility of this work.  CM also thanks Mike Montgomery (UT Austin), the leader of the FRI cohort, for preparing those students to engage in this project and supporting them through the process.  {We thank Adrian Price-Whelan for helpful discussions on RUWE as an indicator of binarity. We} thank Madeline Lucey for useful discussions regarding the connections between this sample of barium stars and carbon-enhanced metal poor stars. {We} thank {Borb\'{a}la} Cseh and {Madeleine} McKenzie for insightful conversations at the $s$-, $i-$, and $r-$ Element Nucleosynthesis (sirEN) conference\footnote{\url{https://indico.ict.inaf.it/event/2876/}} on the origins of barium stars, the state of the field, and GALAH DR3. {JL acknowledges support from the National Science Foundation (NSF) Research Experience for Undergraduates (REU) grant AST-2244278 (Principal investigator: Jogee) funded by the NSF and Department of Defense. CM is supported by the NSF Astronomy and Astrophysics Fellowship award number AST-2401638.}  ZM is partially supported by a NASA ROSES-2020 Exoplanet Research Program grant (20-XRP20 2-0125). KH is partially supported by NSF AST-2407975. KH acknowledge support from the Wootton Center for Astrophysical Plasma Properties, a U.S. Department of Energy NNSA Stewardship Science Academic Alliance Center of Excellence supported under award numbers DE-NA0003843 and DE-NA0004149, from the United States Department of Energy under grant DE-SC0010623. This work was performed in part at the Aspen Center for Physics, which is supported by National Science Foundation grant PHY-2210452.

\clearpage

\appendix

\section{RUWE Calibration} \label{appendix_ruwe}

\begin{figure}[h!]
    \centering
    \includegraphics[width=.7\linewidth]{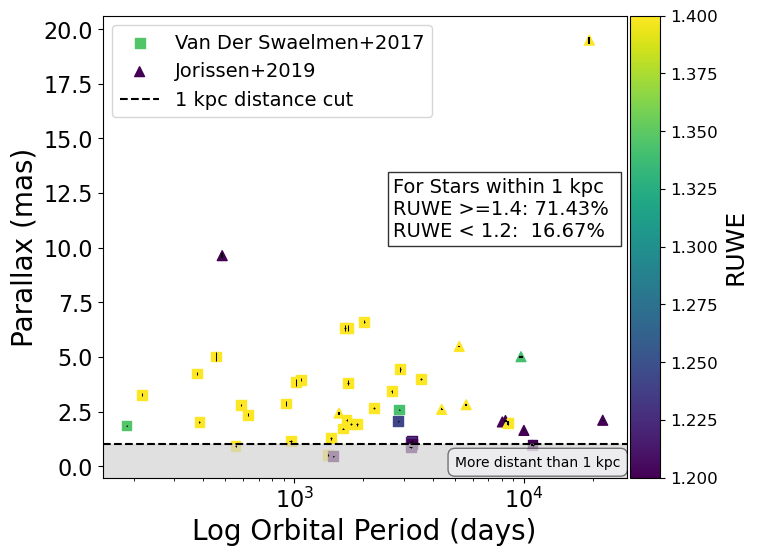}
    \caption{{Parallax vs. orbital period distributions for barium stars in known binary systems from \cite{VanderSwaelmen2017} and \cite{Jorissen2019}. Markers are colored by their RUWE value, with RUWE $>$ 1.4 (yellow) representing our adopted definition of a likely binary system and {RUWE $<$ 1.2 as likely single}.  For stars within 1 kpc, RUWE accurately indicates 30/42 of stars as binary systems while 7/42 are incorrectly labeled as single star systems.}}
    \label{fig:ruwecalibration}
\end{figure}

{Throughout this paper, we employ a strict RUWE cutoff for detecting binary systems (RUWE $>1.4$ \citealp[e.g.,][]{Lindegren2018,Fitton2022,Belokurov2020,Hernandez2023}). We acknowledge that a complete sample of binary stars is not possible using RUWE alone, so we instead focus on investigating and acknowledging the biases of RUWE. RUWE will bias intermediate mass intermediate separation binary systems. We perform a RUWE calibration experiment in the context of barium stars and present the results in Figure \ref{fig:ruwecalibration}. In this experiment, we examine the RUWE values of a sample of barium stars in known binaries from \citet{VanderSwaelmen2017, Jorissen2019b}.  We find that after applying the recommended 1 kpc distance cut, RUWE accurately identifies 30/42 (71.43\%) of stars as binary systems while incorrectly labeling 7/42 (16.67\%) as single star systems (RUWE $<1.2$). }

\section{Supplemental information of select barium stars} \label{appendixA}
{Table \ref{table} presents the full barium star sample investigated in this work. The ``Notes" column of the table highlights particularly noteworthy barium stars: the four stars used for spectral analysis in Figure \ref{fig:numbersubplot}, three high velocity stars nearing the Galactic escape velocity, eight hot halo dwarfs, 33 europium and barium rich stars, and 12 europium and barium rich stars that also have RUWE $\ge1.4$. Furthermore, we highlight stars that have both chemical and kinematic Galactic memberships that agree, labeling them as thin disk, thick disk, or halo stars. We also derive and include the LSR UVW velocities of the barium stars provided in Table \ref{table} by transforming GALAH's heliocentric velocities into the local standard of rest using $\rm(U, V, W)_\odot=(11.1, 12.24,7.25)$ km/s adopted from \citealt{schonrich2010}. This table should serve as a resource for prioritizing barium stars worthy of follow-up. }

\begin{table}[h!] 
\centering
 \caption{{Gaia DR3 source IDs and key characteristics for all barium stars identified in this work}. The barium stars used for spectral analysis in Figure \ref{fig:numbersubplot} and other noteworthy barium stars from Sections \ref{discuss:barlocation} and \ref{subsec:europiumdiscussion} {are designated in the ``Notes" column.} The full machine-readable version of this table {(which includes GALAH-reported $1\sigma$ uncertainties)} can be found in the journal-published version of this manuscript. }
 \begin{tabular}{c|c|c|c|c|c|c|c|c}

 Gaia DR3 ID & [Ba/Fe] & [Fe/H] & $T_{\rm eff}$ & log g & LSR U & LSR V & LSR W  & Notes\\ 
  &  &  & ($\mathrm{K}$) & log(cm s$^{-2}$) & ($\mathrm{km\,s^{-1}}$) & ($\mathrm{km\,s^{-1}}$) & ($\mathrm{km\,s^{-1}}$) &  \\
  \hline
 \hline
4422593353008932736 & 1.68 & -0.68 & 4700 & 2.2 & 8 & -13 & -34 & K Giant in Fig \ref{fig:numbersubplot} \\
6233918569409734400 & 1.18\ & -0.49 & 5878 & 4.1 & -68 & -103 & -125 & G Dwarf in Fig \ref{fig:numbersubplot} \\ 
3038914422505373824 & 1.602 & 0.022 & 5866 & 2.2 & -32 & -2 & 4 & G Giant in Fig \ref{fig:numbersubplot} \\ 
3790375232188743168 & 1.64 & -0.53 & 6100 & 4.1 & 45 & -27 & 10 & F Dwarf in Fig \ref{fig:numbersubplot} \\ 
 .. & .. & .. & .. & .. & .. & .. & .. & .. \\
2884536461613379840 & 1.24 & -1.48 & 6114 & 4.0 & -196 & -576 & -9 & Member of\\ 
 &  & &  & &  &  &  & Gaia-Sausage-Enceladus\\
 5922493338335870080 & 1.63 & -1.07 & 4339 & 1.1 & -181 & -580 & -126 & Likely member of \\
  &  & &  & &  &  &  & accreted halo\\
 6706018041696378752 & 1.54 & -0.78 & 4554 & 1.3 & -117 & -586 & -370 & May exceed $\nu_{\rm esc}$ \\
 \hline
 \end{tabular}
 \label{table}
\end{table}

\clearpage

\section{Comparison of Mean Abundances} \label{appendixmean}
{Investigating the average abundance trends of barium stars may reveal additional chemical peculiarities and provide insights into their origin. Table \ref{chemical_mean_table} presents the average abundance ratios for barium stars compared to those of the GALAH field. As mentioned in Sections \ref{subsec:masstransfer} and \ref{subsec:where}, barium stars are systematically more metal poor than the field. Beside the $s-$process elements, which are linked to mass transfer and radiative levitation, [Fe/H] has the largest mean difference between populations. The magnitude in this difference is largely carried by the cool barium stars, which have $\Delta\langle\rm[Fe/H]\rangle=-0.47\pm0.002$ relative to the cool GALAH field compared to $\Delta\langle\rm [Fe/H]\rangle=-0.03\pm0.005$ for the hot barium stars relative to the hot field. Many of the remaining elements show statistically significant chemical differences from the barium-normal GALAH field, but it is difficult to speculate the causes of these differences given that we are not making any distance, evolutionary, or temperature cuts in this table.  A quick investigation into applying a distance cut here (d $<1$ kpc) returns that the mean differences for [Fe/H] and the lighter elements (except carbon and sodium) decrease in magnitude by $\leq0.01$ dex while the mean differences in C, Na, Y, Nd, and Eu increase by $\sim0.1$ dex. The distance cut enhances chemical differences that arise from mass transfer while limiting differences caused by birth environment.}

\begin{table}[h!] 
\centering
 \caption{{Average [X/Fe] abundance ratios of barium stars, the non-barium-enhanced GALAH field, and the differences between the two derived with MC sampling. Uncertainties for the barium star means and the mean differences are symmetrical within $10^{-4}$.  Due to the large sample size of the field, the [X/Fe] uncertainties are $<10^{-4}$ dex and thus omitted from the table.  Note that we do not apply any distance, temperature, or RUWE cuts here.}}
 \begin{tabular}{c|c|c|c}

 & Barium Stars & GALAH Field & Mean Difference \\
\hline
\hline

$\langle\rm[Fe/H]\rangle$ & $-0.506\pm{0.002}$ & -0.175 & -$0.331\pm0.002$ \\
$\langle\rm[Ba/Fe]\rangle$ & $1.341\pm{0.002}$ & 0.035 & $+1.306\pm{0.002}$ \\
$\langle\rm[O/Fe]\rangle$ & $0.396\pm{0.003}$ & 0.175 & $+0.221\pm0.003$ \\
$\langle\rm[Mg/Fe]\rangle$ & $0.134\pm{0.002}$ & 0.075 & $+0.060\pm0.002$ \\
$\langle\rm[Si/Fe]\rangle$ & $0.138\pm{0.001}$ & 0.078 & $+0.061\pm0.002$ \\
$\langle\rm[Ca/Fe]\rangle$ & $0.188\pm{0.002}$ & 0.096 & $+0.092\pm0.002$ \\
$\langle\rm[C/Fe]\rangle$ & $0.263\pm{0.005}$ & 0.072 & $+0.191\pm0.005$ \\
$\langle\rm[Na/Fe]\rangle$ & $0.150\pm{0.002}$ & 0.075& $+0.075\pm0.002$\\
$\langle\rm[Al/Fe]\rangle$ & $0.164\pm{0.002}$ & 0.134& $+0.029\pm0.002$\\
$\langle\rm[Mn/Fe]\rangle$ & $-0.117\pm{0.002}$ &-0.029 & $-0.089\pm0.002$\\
$\langle\rm[Y/Fe]\rangle$ & $1.055\pm{0.004}$ & -0.021& $+1.076\pm0.004$\\
$\langle\rm[La/Fe]\rangle$ & $1.008\pm{0.002}$ & 0.284& $+0.723\pm0.002$\\
$\langle\rm[Ce/Fe]\rangle$ & $0.549\pm0.004$ & 0.035 & $+0.514\pm0.004$ \\
$\langle\rm[Nd/Fe]\rangle$ & $1.081\pm{0.003}$ & 0.308& $+0.773\pm0.003$\\
$\langle\rm[Eu/Fe]\rangle$ & $0.460\pm0.003$ & 0.170& $+0.289\pm0.003$\\

 \hline
 \end{tabular}
 \label{chemical_mean_table}
\end{table}



\bibliography{sample631}{}

@ARTICLE{Ramirez2011,
       author = {{Ram{\'\i}rez}, I. and {Allende Prieto}, C.},
        title = "{Fundamental Parameters and Chemical Composition of Arcturus}",
      journal = {\apj},
         year = 2011,
        month = dec,
       volume = {743},
       number = {2},
          eid = {135},
        pages = {135},
          doi = {10.1088/0004-637X/743/2/135},
archivePrefix = {arXiv},
       eprint = {1109.4425},
 primaryClass = {astro-ph.SR},
       adsurl = {https://ui.adsabs.harvard.edu/abs/2011ApJ...743..135R}
}

@ARTICLE{Brauer2021,
       author = {{Brauer}, Kaley and {Ji}, Alexander P. and {Drout}, Maria R. and {Frebel}, Anna},
        title = "{Collapsar R-process Yields Can Reproduce [Eu/Fe] Abundance Scatter in Metal-poor Stars}",
      journal = {\apj},
         year = 2021,
        month = jul,
       volume = {915},
       number = {2},
          eid = {81},
        pages = {81},
          doi = {10.3847/1538-4357/ac00b2},
archivePrefix = {arXiv},
       eprint = {2010.15837},
 primaryClass = {astro-ph.HE},
       adsurl = {https://ui.adsabs.harvard.edu/abs/2021ApJ...915...81B}
}

@ARTICLE{Jorissen2019b,
       author = {{Jorissen}, A.},
        title = "{Impact of binaries on stellar evolution in the Gaia era}",
      journal = {\memsai},
         year = 2019,
        month = jan,
       volume = {90},
        pages = {395},
       adsurl = {https://ui.adsabs.harvard.edu/abs/2019MmSAI..90..395J}
}

@ARTICLE{Elbadry2025,
       author = {{El-Badry}, Kareem},
        title = "{How to use Gaia parallaxes for stars with poor astrometric fits}",
      journal = {The Open Journal of Astrophysics},
         year = 2025,
        month = may,
       volume = {8},
          eid = {62},
        pages = {62},
          doi = {10.33232/001c.138448},
archivePrefix = {arXiv},
       eprint = {2504.11528},
 primaryClass = {astro-ph.SR},
       adsurl = {https://ui.adsabs.harvard.edu/abs/2025OJAp....8E..62E}
}

@ARTICLE{Hegedus2023,
       author = {{Heged{\H{u}}s}, Viola and {M{\'e}sz{\'a}ros}, Szabolcs and {Jofr{\'e}}, Paula and {Stringfellow}, Guy S. and {Feuillet}, Diane and {Garc{\'\i}a-Hern{\'a}ndez}, Domingo An{\'\i}bal and {Nitschelm}, Christian and {Zamora}, Olga},
        title = "{Comparative analysis of atmospheric parameters from high-resolution spectroscopic sky surveys: APOGEE, GALAH, Gaia-ESO}",
      journal = {\aap},
         year = 2023,
        month = feb,
       volume = {670},
          eid = {A107},
        pages = {A107},
          doi = {10.1051/0004-6361/202244813},
archivePrefix = {arXiv},
       eprint = {2211.03416},
 primaryClass = {astro-ph.SR},
       adsurl = {https://ui.adsabs.harvard.edu/abs/2023A&A...670A.107H}
}

@dataset{NSS,
       author = {{Gaia Collaboration}},
        title = "{VizieR Online Data Catalog: Gaia DR3 Part 3. Non-single stars (Gaia Collaboration, 2022)}",
 howpublished = {VizieR On-line Data Catalog: I/357.  Originally published in: 2023A\&A...674A...1G},
         year = 2022,
        month = may,
          eid = {I/357},
       adsurl = {https://ui.adsabs.harvard.edu/abs/2022yCat.1357....0G}
}

@ARTICLE{Lu2025,
       author = {{Lu}, Yuxi and {Manea}, Catherine and {Sayeed}, Maryum and {Douglas}, Stephanie T. and {McKenzie}, Madeleine and {Rowan}, Dominick and {Ilyin}, Ilya and {Wheeler}, Adam and {Buder}, Sven and {Amard}, Louis and {Pinsonneault}, Marc H. and {Johnson}, Jennifer A.},
        title = "{Spectroscopic Follow-up of Young High-$α$ Dwarf Star Candidates: Still Likely Genuinely Young}",
      journal = {arXiv e-prints},
         year = 2025,
        month = oct,
          eid = {arXiv:2510.15654},
        pages = {arXiv:2510.15654},
          doi = {10.48550/arXiv.2510.15654},
archivePrefix = {arXiv},
       eprint = {2510.15654},
 primaryClass = {astro-ph.GA},
       adsurl = {https://ui.adsabs.harvard.edu/abs/2025arXiv251015654L}
}

@ARTICLE{Aguado2021,
       author = {{Aguado}, David S. and {Belokurov}, Vasily and {Myeong}, G.~C. and {Evans}, N. Wyn and {Kobayashi}, Chiaki and {Sbordone}, Luca and {Chanam{\'e}}, Julio and {Navarrete}, Camila and {Koposov}, Sergey E.},
        title = "{Elevated r-process Enrichment in Gaia Sausage and Sequoia}",
      journal = {\apjl},
         year = 2021,
        month = feb,
       volume = {908},
       number = {1},
          eid = {L8},
        pages = {L8},
          doi = {10.3847/2041-8213/abdbb8},
archivePrefix = {arXiv},
       eprint = {2012.01430},
 primaryClass = {astro-ph.GA},
       adsurl = {https://ui.adsabs.harvard.edu/abs/2021ApJ...908L...8A}
}

@ARTICLE{Kemp2025,
       author = {{Kemp}, Alex and {Kaur}, Tejpreet},
        title = "{Binary stellar evolution yields in galactic chemical evolution calculations}",
      journal = {\aap},
         year = 2025,
        month = sep,
       volume = {701},
          eid = {A177},
        pages = {A177},
          doi = {10.1051/0004-6361/202554912},
archivePrefix = {arXiv},
       eprint = {2508.01717},
 primaryClass = {astro-ph.SR},
       adsurl = {https://ui.adsabs.harvard.edu/abs/2025A&A...701A.177K}
}

@ARTICLE{gbs,
       author = {{Jofr{\'e}}, Paula and {Heiter}, Ulrike and {Tucci Maia}, Marcelo and {Soubiran}, Caroline and {Worley}, C. Clare and {Hawkins}, Keith and {Blanco-Cuaresma}, Sergi and {Rodrigo}, Carlos},
        title = "{The Gaia FGK Benchmark Stars Version 2.1}",
      journal = {Research Notes of the American Astronomical Society},
         year = 2018,
        month = aug,
       volume = {2},
       number = {3},
          eid = {152},
        pages = {152},
          doi = {10.3847/2515-5172/aadc61},
archivePrefix = {arXiv},
       eprint = {1808.09778},
 primaryClass = {astro-ph.SR},
       adsurl = {https://ui.adsabs.harvard.edu/abs/2018RNAAS...2..152J}
}

@ARTICLE{Maas2025,
       author = {{Maas}, Zachary G. and {Hawkins}, Keith and {Gerber}, Jeffrey M. and {Hackshaw}, Zoe and {Manea}, Catherine},
        title = "{Exploring the History of Stellar Mergers with Chemistry: Examining the Origins of Massive {\ensuremath{\alpha}}-enriched Stars Using Carbon Isotope Ratios}",
      journal = {\apj},
         year = 2025,
        month = nov,
       volume = {994},
       number = {1},
          eid = {16},
        pages = {16},
          doi = {10.3847/1538-4357/ae0a2d},
archivePrefix = {arXiv},
       eprint = {2509.08900},
 primaryClass = {astro-ph.SR},
       adsurl = {https://ui.adsabs.harvard.edu/abs/2025ApJ...994...16M}
}

@ARTICLE{Katz2023,
       author = {{Katz}, D. and {Sartoretti}, P. and {Guerrier}, A. and {Panuzzo}, P. and {Seabroke}, G.~M. and {Th{\'e}venin}, F. and {Cropper}, M. and {Benson}, K. and {Blomme}, R. and {Haigron}, R. and {Marchal}, O. and {Smith}, M. and {Baker}, S. and {Chemin}, L. and {Damerdji}, Y. and {David}, M. and {Dolding}, C. and {Fr{\'e}mat}, Y. and {Gosset}, E. and {Jan{\ss}en}, K. and {Jasniewicz}, G. and {Lobel}, A. and {Plum}, G. and {Samaras}, N. and {Snaith}, O. and {Soubiran}, C. and {Vanel}, O. and {Zwitter}, T. and {Antoja}, T. and {Arenou}, F. and {Babusiaux}, C. and {Brouillet}, N. and {Caffau}, E. and {Di Matteo}, P. and {Fabre}, C. and {Fabricius}, C. and {Fragkoudi}, F. and {Haywood}, M. and {Huckle}, H.~E. and {Hottier}, C. and {Lasne}, Y. and {Leclerc}, N. and {Mastrobuono-Battisti}, A. and {Royer}, F. and {Teyssier}, D. and {Zorec}, J. and {Crifo}, F. and {Jean-Antoine Piccolo}, A. and {Turon}, C. and {Viala}, Y.},
        title = "{Gaia Data Release 3. Properties and validation of the radial velocities}",
      journal = {\aap},
         year = 2023,
        month = jun,
       volume = {674},
          eid = {A5},
        pages = {A5},
          doi = {10.1051/0004-6361/202244220},
archivePrefix = {arXiv},
       eprint = {2206.05902},
 primaryClass = {astro-ph.GA},
       adsurl = {https://ui.adsabs.harvard.edu/abs/2023A&A...674A...5K}
}

@ARTICLE{Chance2025,
       author = {{Chance}, Quadry and {Foreman-Mackey}, Daniel and {Ballard}, Sarah and {Casey}, Andrew R. and {David}, Trevor J. and {Price-Whelan}, Adrian M.},
        title = "{paired: A Statistical Framework for Detecting Stellar Binarity with Gaia RVs. I. Sensitivity to Unresolved Binaries}",
      journal = {\apj},
         year = 2025,
        month = oct,
       volume = {992},
       number = {1},
          eid = {131},
        pages = {131},
          doi = {10.3847/1538-4357/adfb68},
archivePrefix = {arXiv},
       eprint = {2206.11275},
 primaryClass = {astro-ph.EP},
       adsurl = {https://ui.adsabs.harvard.edu/abs/2025ApJ...992..131C}
}

@ARTICLE{Pourbaix2019,
       author = {{Pourbaix}, D.},
        title = "{Binaries in Gaia perspective. DR3 teaser}",
      journal = {\memsai},
         year = 2019,
        month = jan,
       volume = {90},
        pages = {318},
       adsurl = {https://ui.adsabs.harvard.edu/abs/2019MmSAI..90..318P}
}

@ARTICLE{Yamaguchi2025,
       author = {{Yamaguchi}, Natsuko and {El-Badry}, Kareem and {Reggiani}, Henrique and {Andrae}, Ren{\'e} and {Shahaf}, Sahar},
        title = "{Chemical Signatures of AGB Mass Transfer in Gaia White Dwarf Companions}",
      journal = {arXiv e-prints},
         year = 2025,
        month = dec,
          eid = {arXiv:2512.07972},
        pages = {arXiv:2512.07972},
          doi = {10.48550/arXiv.2512.07972},
archivePrefix = {arXiv},
       eprint = {2512.07972},
 primaryClass = {astro-ph.SR},
       adsurl = {https://ui.adsabs.harvard.edu/abs/2025arXiv251207972Y}
}

@ARTICLE{Burbidge1957,
       author = {{Burbidge}, E. Margaret and {Burbidge}, G.~R. and {Fowler}, William A. and {Hoyle}, F.},
        title = "{Synthesis of the Elements in Stars}",
      journal = {Reviews of Modern Physics},
         year = 1957,
        month = oct,
       volume = {29},
       number = {4},
        pages = {547-650},
          doi = {10.1103/RevModPhys.29.547},
       adsurl = {https://ui.adsabs.harvard.edu/abs/1957RvMP...29..547B}
}

@ARTICLE{Buder2022,
       author = {{Buder}, Sven and {Lind}, Karin and {Ness}, Melissa K. and {Feuillet}, Diane K. and {Horta}, Danny and {Monty}, Stephanie and {Buck}, Tobias and {Nordlander}, Thomas and {Bland-Hawthorn}, Joss and {Casey}, Andrew R. and {de Silva}, Gayandhi M. and {D'Orazi}, Valentina and {Freeman}, Ken C. and {Hayden}, Michael R. and {Kos}, Janez and {Martell}, Sarah L. and {Lewis}, Geraint F. and {Lin}, Jane and {Schlesinger}, Katharine J. and {Sharma}, Sanjib and {Simpson}, Jeffrey D. and {Stello}, Dennis and {Zucker}, Daniel B. and {Zwitter}, Toma{\v{z}} and {Ciuc{\u{a}}}, Ioana and {Horner}, Jonathan and {Kobayashi}, Chiaki and {Ting}, Yuan-Sen and {Wyse}, Rosemary F.~G. and {Wyse}, Galah Collaboration},
        title = "{The GALAH Survey: chemical tagging and chrono-chemodynamics of accreted halo stars with GALAH+ DR3 and Gaia eDR3}",
      journal = {\mnras},
         year = 2022,
        month = feb,
       volume = {510},
       number = {2},
        pages = {2407-2436},
          doi = {10.1093/mnras/stab3504},
archivePrefix = {arXiv},
       eprint = {2109.04059},
 primaryClass = {astro-ph.GA},
       adsurl = {https://ui.adsabs.harvard.edu/abs/2022MNRAS.510.2407B}
}

@ARTICLE{Spoo2022,
       author = {{Spoo}, Taylor and {Tayar}, Jamie and {Frinchaboy}, Peter M. and {Cunha}, Katia and {Myers}, Natalie and {Donor}, John and {Majewski}, Steven R. and {Bizyaev}, Dmitry and {Garc{\'\i}a-Hern{\'a}ndez}, D.~A. and {J{\"o}nsson}, Henrik and {Lane}, Richard R. and {Pan}, Kaike and {Longa-Pe{\~n}a}, Pen{\'e}lope and {Roman-Lopes}, A.},
        title = "{The Open Cluster Chemical Abundances and Mapping Survey. VII. APOGEE DR17 [C/N]-Age Calibration}",
      journal = {\aj},
         year = 2022,
        month = may,
       volume = {163},
       number = {5},
          eid = {229},
        pages = {229},
          doi = {10.3847/1538-3881/ac5d53},
archivePrefix = {arXiv},
       eprint = {2203.05463},
 primaryClass = {astro-ph.GA},
       adsurl = {https://ui.adsabs.harvard.edu/abs/2022AJ....163..229S}
}

@ARTICLE{Roriz2025,
       author = {{Roriz}, M.~P. and {Drake}, N.~A. and {Holanda}, N. and {Lugaro}, M. and {Cseh}, B. and {Junqueira}, S. and {Pereira}, C.~B.},
        title = "{Exploring giant barium stars: $^{12}\rm{C}/^{13}\rm{C}$ ratio and elemental abundances of carbon, nitrogen, and oxygen}",
      journal = {arXiv e-prints},
         year = 2025,
        month = sep,
          eid = {arXiv:2509.02441},
        pages = {arXiv:2509.02441},
          doi = {10.48550/arXiv.2509.02441},
archivePrefix = {arXiv},
       eprint = {2509.02441},
 primaryClass = {astro-ph.SR},
       adsurl = {https://ui.adsabs.harvard.edu/abs/2025arXiv250902441R}
}

@ARTICLE{Brady2023,
       author = {{Brady}, K.~E. and {Sneden}, C. and {Pilachowski}, C.~A. and {Af{\c{s}}ar}, Melike and {Mace}, G.~N. and {Jaffe}, D.~T. and {Gosnell}, N.~M. and {Seifert}, R.},
        title = "{M67 Blue Stragglers with High-resolution Infrared Spectroscopy}",
      journal = {\aj},
         year = 2023,
        month = oct,
       volume = {166},
       number = {4},
          eid = {154},
        pages = {154},
          doi = {10.3847/1538-3881/acf2f3},
archivePrefix = {arXiv},
       eprint = {2309.16863},
 primaryClass = {astro-ph.SR},
       adsurl = {https://ui.adsabs.harvard.edu/abs/2023AJ....166..154B}
}

@ARTICLE{Motta2018,
       author = {{Bertelli Motta}, Clio and {Pasquali}, Anna and {Caffau}, Elisabetta and {Grebel}, Eva K.},
        title = "{A chemical study of M67 candidate blue stragglers and evolved blue stragglers observed with APOGEE DR14}",
      journal = {\mnras},
         year = 2018,
        month = nov,
       volume = {480},
       number = {4},
        pages = {4314-4326},
          doi = {10.1093/mnras/sty2147},
archivePrefix = {arXiv},
       eprint = {1808.04601},
 primaryClass = {astro-ph.SR},
       adsurl = {https://ui.adsabs.harvard.edu/abs/2018MNRAS.480.4314B}
}

@ARTICLE{Cristallo2009,
       author = {{Cristallo}, S. and {Straniero}, O. and {Gallino}, R. and {Piersanti}, L. and {Dom{\'\i}nguez}, I. and {Lederer}, M.~T.},
        title = "{Evolution, Nucleosynthesis, and Yields of Low-Mass Asymptotic Giant Branch Stars at Different Metallicities}",
      journal = {\apj},
         year = 2009,
        month = may,
       volume = {696},
       number = {1},
        pages = {797-820},
          doi = {10.1088/0004-637X/696/1/797},
archivePrefix = {arXiv},
       eprint = {0902.0243},
 primaryClass = {astro-ph.SR},
       adsurl = {https://ui.adsabs.harvard.edu/abs/2009ApJ...696..797C}
}

@BOOK{Michaud2015,
       author = {{Michaud}, Georges and {Alecian}, Georges and {Richer}, Jacques},
        title = "{Atomic Diffusion in Stars}",
         year = 2015,
          doi = {10.1007/978-3-319-19854-5},
       adsurl = {https://ui.adsabs.harvard.edu/abs/2015ads..book.....M},
    publisher = "Springer Cham"
}

@ARTICLE{Nine2024,
       author = {{Nine}, Andrew C. and {Mathieu}, Robert D. and {Schuler}, Simon C. and {Milliman}, Katelyn E.},
        title = "{WIYN Open Cluster Study. XC. Barium Surface Abundances of Blue Straggler Stars in the Open Clusters NGC 7789 and M67}",
      journal = {\apj},
         year = 2024,
        month = aug,
       volume = {970},
       number = {2},
          eid = {187},
        pages = {187},
          doi = {10.3847/1538-4357/ad534b},
archivePrefix = {arXiv},
       eprint = {2405.20242},
 primaryClass = {astro-ph.SR},
       adsurl = {https://ui.adsabs.harvard.edu/abs/2024ApJ...970..187N}
}

@ARTICLE{Nine2023,
       author = {{Nine}, Andrew C. and {Mathieu}, Robert D. and {Gosnell}, Natalie M. and {Leiner}, Emily M.},
        title = "{WIYN Open Cluster Study. LXXXVII. Hubble Space Telescope Ultraviolet Detection of Hot White Dwarf Companions to Blue Lurkers in M67}",
      journal = {\apj},
         year = 2023,
        month = feb,
       volume = {944},
       number = {2},
          eid = {145},
        pages = {145},
          doi = {10.3847/1538-4357/acb046},
archivePrefix = {arXiv},
       eprint = {2301.02303},
 primaryClass = {astro-ph.SR},
       adsurl = {https://ui.adsabs.harvard.edu/abs/2023ApJ...944..145N}
}

@ARTICLE{Pal2024,
       author = {{Pal}, Harshit and {Subramaniam}, Annapurni and {Reddy}, Arumalla B.~S. and {Jadhav}, Vikrant V.},
        title = "{Discovery of a Barium Blue Straggler Star in M67 and ``Sighting'' of Its White Dwarf Companion}",
      journal = {\apjl},
         year = 2024,
        month = aug,
       volume = {970},
       number = {2},
          eid = {L39},
        pages = {L39},
          doi = {10.3847/2041-8213/ad6316},
archivePrefix = {arXiv},
       eprint = {2407.06897},
 primaryClass = {astro-ph.SR},
       adsurl = {https://ui.adsabs.harvard.edu/abs/2024ApJ...970L..39P}
}

@ARTICLE{Kong2018,
       author = {{Kong}, X.~M. and {Bharat Kumar}, Y. and {Zhao}, G. and {Zhao}, J.~K. and {Fang}, X.~S. and {Shi}, J.~R. and {Wang}, L. and {Zhang}, J.~B. and {Yan}, H.~L.},
        title = "{Three new barium dwarfs with white dwarf companions: BD+68{\textdegree}1027, RE J0702+129 and BD+80{\textdegree}670}",
      journal = {\mnras},
         year = 2018,
        month = feb,
       volume = {474},
       number = {2},
        pages = {2129-2141},
          doi = {10.1093/mnras/stx2809},
archivePrefix = {arXiv},
       eprint = {1710.10750},
 primaryClass = {astro-ph.SR},
       adsurl = {https://ui.adsabs.harvard.edu/abs/2018MNRAS.474.2129K}
}

@ARTICLE{Gray2011,
       author = {{Gray}, R.~O. and {McGahee}, C.~E. and {Griffin}, R.~E.~M. and {Corbally}, C.~J.},
        title = "{First Direct Evidence That Barium Dwarfs Have White Dwarf Companions}",
      journal = {\aj},
         year = 2011,
        month = may,
       volume = {141},
       number = {5},
          eid = {160},
        pages = {160},
          doi = {10.1088/0004-6256/141/5/160},
       adsurl = {https://ui.adsabs.harvard.edu/abs/2011AJ....141..160G}
}

@ARTICLE{Escorza2019,
       author = {{Escorza}, A. and {Karinkuzhi}, D. and {Jorissen}, A. and {Siess}, L. and {Van Winckel}, H. and {Pourbaix}, D. and {Johnston}, C. and {Miszalski}, B. and {Oomen}, G. -M. and {Abdul-Masih}, M. and {Boffin}, H.~M.~J. and {North}, P. and {Manick}, R. and {Shetye}, S. and {Miko{\l}ajewska}, J.},
        title = "{Barium and related stars, and their white-dwarf companions. II. Main-sequence and subgiant starss}",
      journal = {\aap},
         year = 2019,
        month = jun,
       volume = {626},
          eid = {A128},
        pages = {A128},
          doi = {10.1051/0004-6361/201935390},
archivePrefix = {arXiv},
       eprint = {1904.04095},
 primaryClass = {astro-ph.SR},
       adsurl = {https://ui.adsabs.harvard.edu/abs/2019A&A...626A.128E}
}

@ARTICLE{Buntain2017,
       author = {{Buntain}, J.~F. and {Doherty}, C.~L. and {Lugaro}, M. and {Lattanzio}, J.~C. and {Stancliffe}, R.~J. and {Karakas}, A.~I.},
        title = "{Partial mixing and the formation of $^{13}$C pockets in AGB stars: effects on the s-process elements}",
      journal = {\mnras},
         year = 2017,
        month = oct,
       volume = {471},
       number = {1},
        pages = {824-838},
          doi = {10.1093/mnras/stx1502},
archivePrefix = {arXiv},
       eprint = {1706.05802},
 primaryClass = {astro-ph.SR},
       adsurl = {https://ui.adsabs.harvard.edu/abs/2017MNRAS.471..824B}
}

@ARTICLE{Cseh2022,
       author = {{Cseh}, B. and {Vil{\'a}gos}, B. and {Roriz}, M.~P. and {Pereira}, C.~B. and {D'Orazi}, V. and {Karakas}, A.~I. and {So{\'o}s}, B. and {Drake}, N.~A. and {Junqueira}, S. and {Lugaro}, M.},
        title = "{Barium stars as tracers of s-process nucleosynthesis in AGB stars. I. 28 stars with independently derived AGB mass}",
      journal = {\aap},
         year = 2022,
        month = apr,
       volume = {660},
          eid = {A128},
        pages = {A128},
          doi = {10.1051/0004-6361/202142468},
archivePrefix = {arXiv},
       eprint = {2201.13379},
 primaryClass = {astro-ph.SR},
       adsurl = {https://ui.adsabs.harvard.edu/abs/2022A&A...660A.128C}
}

@ARTICLE{Kappeler2011,
       author = {{K{\"a}ppeler}, F. and {Gallino}, R. and {Bisterzo}, S. and {Aoki}, Wako},
        title = "{The s process: Nuclear physics, stellar models, and observations}",
      journal = {Reviews of Modern Physics},
         year = 2011,
        month = jan,
       volume = {83},
       number = {1},
        pages = {157-194},
          doi = {10.1103/RevModPhys.83.157},
archivePrefix = {arXiv},
       eprint = {1012.5218},
 primaryClass = {astro-ph.SR},
       adsurl = {https://ui.adsabs.harvard.edu/abs/2011RvMP...83..157K}
}

@ARTICLE{Karakas2006,
       author = {{Karakas}, A.~I. and {Lugaro}, M.~A. and {Wiescher}, M. and {G{\"o}rres}, J. and {Ugalde}, C.},
        title = "{The Uncertainties in the $^{22}$Ne+{\ensuremath{\alpha}}-Capture Reaction Rates and the Production of the Heavy Magnesium Isotopes in Asymptotic Giant Branch Stars of Intermediate Mass}",
      journal = {\apj},
         year = 2006,
        month = may,
       volume = {643},
       number = {1},
        pages = {471-483},
          doi = {10.1086/502793},
archivePrefix = {arXiv},
       eprint = {astro-ph/0601645},
 primaryClass = {astro-ph},
       adsurl = {https://ui.adsabs.harvard.edu/abs/2006ApJ...643..471K}
}

@ARTICLE{Lugaro2003,
       author = {{Lugaro}, Maria and {Herwig}, Falk and {Lattanzio}, John C. and {Gallino}, Roberto and {Straniero}, Oscar},
        title = "{s-Process Nucleosynthesis in Asymptotic Giant Branch Stars: A Test for Stellar Evolution}",
      journal = {\apj},
         year = 2003,
        month = apr,
       volume = {586},
       number = {2},
        pages = {1305-1319},
          doi = {10.1086/367887},
archivePrefix = {arXiv},
       eprint = {astro-ph/0212364},
 primaryClass = {astro-ph},
       adsurl = {https://ui.adsabs.harvard.edu/abs/2003ApJ...586.1305L}
}

@ARTICLE{Vilagos2024,
       author = {{Vil{\'a}gos}, B. and {Cseh}, B. and {Yag{\"u}e L{\'o}pez}, A. and {Joyce}, M. and {Karakas}, A. and {Tagliente}, G. and {Lugaro}, M.},
        title = "{Barium stars as tracers of s-process nucleosynthesis in AGB stars. III. Systematic deviations from the AGB models}",
      journal = {\aap},
         year = 2024,
        month = aug,
       volume = {688},
          eid = {A164},
        pages = {A164},
          doi = {10.1051/0004-6361/202450084},
archivePrefix = {arXiv},
       eprint = {2405.19330},
 primaryClass = {astro-ph.SR},
       adsurl = {https://ui.adsabs.harvard.edu/abs/2024A&A...688A.164V}
}

@ARTICLE{GaiaESO,
       author = {{Gilmore}, G. and {Randich}, S. and {Asplund}, M. and {Binney}, J. and {Bonifacio}, P. and {Drew}, J. and {Feltzing}, S. and {Ferguson}, A. and {Jeffries}, R. and {Micela}, G. and {Negueruela}, I. and {Prusti}, T. and {Rix}, H. -W. and {Vallenari}, A. and {Alfaro}, E. and {Allende-Prieto}, C. and {Babusiaux}, C. and {Bensby}, T. and {Blomme}, R. and {Bragaglia}, A. and {Flaccomio}, E. and {Fran{\c{c}}ois}, P. and {Irwin}, M. and {Koposov}, S. and {Korn}, A. and {Lanzafame}, A. and {Pancino}, E. and {Paunzen}, E. and {Recio-Blanco}, A. and {Sacco}, G. and {Smiljanic}, R. and {Van Eck}, S. and {Walton}, N. and {Aden}, D. and {Aerts}, C. and {Affer}, L. and {Alcala}, J. -M. and {Altavilla}, G. and {Alves}, J. and {Antoja}, T. and {Arenou}, F. and {Argiroffi}, C. and {Asensio Ramos}, A. and {Bailer-Jones}, C. and {Balaguer-Nunez}, L. and {Bayo}, A. and {Barbuy}, B. and {Barisevicius}, G. and {Barrado y Navascues}, D. and {Battistini}, C. and {Bellas Velidis}, I. and {Bellazzini}, M. and {Belokurov}, V. and {Bergemann}, M. and {Bertelli}, G. and {Biazzo}, K. and {Bienayme}, O. and {Bland-Hawthorn}, J. and {Boeche}, C. and {Bonito}, S. and {Boudreault}, S. and {Bouvier}, J. and {Brandao}, I. and {Brown}, A. and {de Bruijne}, J. and {Burleigh}, M. and {Caballero}, J. and {Caffau}, E. and {Calura}, F. and {Capuzzo-Dolcetta}, R. and {Caramazza}, M. and {Carraro}, G. and {Casagrande}, L. and {Casewell}, S. and {Chapman}, S. and {Chiappini}, C. and {Chorniy}, Y. and {Christlieb}, N. and {Cignoni}, M. and {Cocozza}, G. and {Colless}, M. and {Collet}, R. and {Collins}, M. and {Correnti}, M. and {Covino}, E. and {Crnojevic}, D. and {Cropper}, M. and {Cunha}, M. and {Damiani}, F. and {David}, M. and {Delgado}, A. and {Duffau}, S. and {Edvardsson}, B. and {Eldridge}, J. and {Enke}, H. and {Eriksson}, K. and {Evans}, N.~W. and {Eyer}, L. and {Famaey}, B. and {Fellhauer}, M. and {Ferreras}, I. and {Figueras}, F. and {Fiorentino}, G. and {Flynn}, C. and {Folha}, D. and {Franciosini}, E. and {Frasca}, A. and {Freeman}, K. and {Fremat}, Y. and {Friel}, E. and {Gaensicke}, B. and {Gameiro}, J. and {Garzon}, F. and {Geier}, S. and {Geisler}, D. and {Gerhard}, O. and {Gibson}, B. and {Gomboc}, A. and {Gomez}, A. and {Gonzalez-Fernandez}, C. and {Gonzalez Hernandez}, J. and {Gosset}, E. and {Grebel}, E. and {Greimel}, R. and {Groenewegen}, M. and {Grundahl}, F. and {Guarcello}, M. and {Gustafsson}, B. and {Hadrava}, P. and {Hatzidimitriou}, D. and {Hambly}, N. and {Hammersley}, P. and {Hansen}, C. and {Haywood}, M. and {Heber}, U. and {Heiter}, U. and {Held}, E. and {Helmi}, A. and {Hensler}, G. and {Herrero}, A. and {Hill}, V. and {Hodgkin}, S. and {Huelamo}, N. and {Huxor}, A. and {Ibata}, R. and {Jackson}, R. and {de Jong}, R. and {Jonker}, P. and {Jordan}, S. and {Jordi}, C. and {Jorissen}, A. and {Katz}, D. and {Kawata}, D. and {Keller}, S. and {Kharchenko}, N. and {Klement}, R. and {Klutsch}, A. and {Knude}, J. and {Koch}, A. and {Kochukhov}, O. and {Kontizas}, M. and {Koubsky}, P. and {Lallement}, R. and {de Laverny}, P. and {van Leeuwen}, F. and {Lemasle}, B. and {Lewis}, G. and {Lind}, K. and {Lindstrom}, H.~P.~E. and {Lobel}, A. and {Lopez Santiago}, J. and {Lucas}, P. and {Ludwig}, H. and {Lueftinger}, T. and {Magrini}, L. and {Maiz Apellaniz}, J. and {Maldonado}, J. and {Marconi}, G. and {Marino}, A. and {Martayan}, C. and {Martinez-Valpuesta}, I. and {Matijevic}, G. and {McMahon}, R. and {Messina}, S. and {Meyer}, M. and {Miglio}, A. and {Mikolaitis}, S. and {Minchev}, I. and {Minniti}, D. and {Moitinho}, A. and {Momany}, Y. and {Monaco}, L. and {Montalto}, M. and {Monteiro}, M.~J. and {Monier}, R. and {Montes}, D. and {Mora}, A. and {Moraux}, E. and {Morel}, T. and {Mowlavi}, N.},
        title = "{The Gaia-ESO Public Spectroscopic Survey}",
      journal = {The Messenger},
         year = 2012,
        month = mar,
       volume = {147},
        pages = {25-31},
       adsurl = {https://ui.adsabs.harvard.edu/abs/2012Msngr.147...25G}
}

@ARTICLE{Delgado2017,
       author = {{Delgado Mena}, E. and {Tsantaki}, M. and {Adibekyan}, V. Zh. and {Sousa}, S.~G. and {Santos}, N.~C. and {Gonz{\'a}lez Hern{\'a}ndez}, J.~I. and {Israelian}, G.},
        title = "{Chemical abundances of 1111 FGK stars from the HARPS GTO planet search program. II. Cu, Zn, Sr, Y, Zr, Ba, Ce, Nd, and Eu}",
      journal = {\aap},
         year = 2017,
        month = oct,
       volume = {606},
          eid = {A94},
        pages = {A94},
          doi = {10.1051/0004-6361/201730535},
archivePrefix = {arXiv},
       eprint = {1705.04349},
 primaryClass = {astro-ph.SR},
       adsurl = {https://ui.adsabs.harvard.edu/abs/2017A&A...606A..94D}
}

@ARTICLE{Bensby2014,
       author = {{Bensby}, T. and {Feltzing}, S. and {Oey}, M.~S.},
        title = "{Exploring the Milky Way stellar disk. A detailed elemental abundance study of 714 F and G dwarf stars in the solar neighbourhood}",
      journal = {\aap},
         year = 2014,
        month = feb,
       volume = {562},
          eid = {A71},
        pages = {A71},
          doi = {10.1051/0004-6361/201322631},
archivePrefix = {arXiv},
       eprint = {1309.2631},
 primaryClass = {astro-ph.GA},
       adsurl = {https://ui.adsabs.harvard.edu/abs/2014A&A...562A..71B}
}

@ARTICLE{Storm2025,
       author = {{Storm}, Nicholas and {Bergemann}, Maria and {Eitner}, Philipp and {Hoppe}, Richard and {Kemp}, Alex J. and {Ruiter}, Ashley J. and {Janka}, Hans-Thomas and {Sieverding}, Andre and {de Mink}, Selma E. and {Seitenzahl}, Ivo R. and {Owusu}, Evans K.},
        title = "{Observational constraints on the origin of the elements. IX. 3D NLTE abundances of metals in the context of Galactic Chemical Evolution models and 4MOST}",
      journal = {\mnras},
         year = 2025,
        month = apr,
       volume = {538},
       number = {4},
        pages = {3284-3313},
          doi = {10.1093/mnras/staf472},
archivePrefix = {arXiv},
       eprint = {2503.16946},
 primaryClass = {astro-ph.SR},
       adsurl = {https://ui.adsabs.harvard.edu/abs/2025MNRAS.538.3284S}
}

@ARTICLE{Bedell2018,
       author = {{Bedell}, Megan and {Bean}, Jacob L. and {Mel{\'e}ndez}, Jorge and {Spina}, Lorenzo and {Ram{\'\i}rez}, Ivan and {Asplund}, Martin and {Alves-Brito}, Alan and {dos Santos}, Leonardo and {Dreizler}, Stefan and {Yong}, David and {Monroe}, TalaWanda and {Casagrande}, Luca},
        title = "{The Chemical Homogeneity of Sun-like Stars in the Solar Neighborhood}",
      journal = {\apj},
         year = 2018,
        month = sep,
       volume = {865},
       number = {1},
          eid = {68},
        pages = {68},
          doi = {10.3847/1538-4357/aad908},
archivePrefix = {arXiv},
       eprint = {1802.02576},
 primaryClass = {astro-ph.SR},
       adsurl = {https://ui.adsabs.harvard.edu/abs/2018ApJ...865...68B}
}

@ARTICLE{LAMOSTsurvey,
       author = {{Cui}, Xiang-Qun and {Zhao}, Yong-Heng and {Chu}, Yao-Quan and {Li}, Guo-Ping and {Li}, Qi and {Zhang}, Li-Ping and {Su}, Hong-Jun and {Yao}, Zheng-Qiu and {Wang}, Ya-Nan and {Xing}, Xiao-Zheng and {Li}, Xin-Nan and {Zhu}, Yong-Tian and {Wang}, Gang and {Gu}, Bo-Zhong and {Luo}, A. -Li and {Xu}, Xin-Qi and {Zhang}, Zhen-Chao and {Liu}, Gen-Rong and {Zhang}, Hao-Tong and {Yang}, De-Hua and {Cao}, Shu-Yun and {Chen}, Hai-Yuan and {Chen}, Jian-Jun and {Chen}, Kun-Xin and {Chen}, Ying and {Chu}, Jia-Ru and {Feng}, Lei and {Gong}, Xue-Fei and {Hou}, Yong-Hui and {Hu}, Hong-Zhuan and {Hu}, Ning-Sheng and {Hu}, Zhong-Wen and {Jia}, Lei and {Jiang}, Fang-Hua and {Jiang}, Xiang and {Jiang}, Zi-Bo and {Jin}, Ge and {Li}, Ai-Hua and {Li}, Yan and {Li}, Ye-Ping and {Liu}, Guan-Qun and {Liu}, Zhi-Gang and {Lu}, Wen-Zhi and {Mao}, Yin-Dun and {Men}, Li and {Qi}, Yong-Jun and {Qi}, Zhao-Xiang and {Shi}, Huo-Ming and {Tang}, Zheng-Hong and {Tao}, Qing-Sheng and {Wang}, Da-Qi and {Wang}, Dan and {Wang}, Guo-Min and {Wang}, Hai and {Wang}, Jia-Ning and {Wang}, Jian and {Wang}, Jian-Ling and {Wang}, Jian-Ping and {Wang}, Lei and {Wang}, Shu-Qing and {Wang}, You and {Wang}, Yue-Fei and {Xu}, Ling-Zhe and {Xu}, Yan and {Yang}, Shi-Hai and {Yu}, Yong and {Yuan}, Hui and {Yuan}, Xiang-Yan and {Zhai}, Chao and {Zhang}, Jing and {Zhang}, Yan-Xia and {Zhang}, Yong and {Zhao}, Ming and {Zhou}, Fang and {Zhou}, Guo-Hua and {Zhu}, Jie and {Zou}, Si-Cheng},
        title = "{The Large Sky Area Multi-Object Fiber Spectroscopic Telescope (LAMOST)}",
      journal = {Research in Astronomy and Astrophysics},
         year = 2012,
        month = sep,
       volume = {12},
       number = {9},
        pages = {1197-1242},
          doi = {10.1088/1674-4527/12/9/003},
       adsurl = {https://ui.adsabs.harvard.edu/abs/2012RAA....12.1197C}
}

@ARTICLE{Jorissen2019,
       author = {{Jorissen}, A. and {Boffin}, H.~M.~J. and {Karinkuzhi}, D. and {Van Eck}, S. and {Escorza}, A. and {Shetye}, S. and {Van Winckel}, H.},
        title = "{Barium and related stars, and their white-dwarf companions. I. Giant stars}",
      journal = {\aap},
         year = 2019,
        month = jun,
       volume = {626},
          eid = {A127},
        pages = {A127},
          doi = {10.1051/0004-6361/201834630},
archivePrefix = {arXiv},
       eprint = {1904.03975},
 primaryClass = {astro-ph.SR},
       adsurl = {https://ui.adsabs.harvard.edu/abs/2019A&A...626A.127J}
}

@ARTICLE{Merle2016,
       author = {{Merle}, T. and {Jorissen}, A. and {Van Eck}, S. and {Masseron}, T. and {Van Winckel}, H.},
        title = "{To Ba or not to Ba: Enrichment in s-process elements in binary systems with WD companions of various masses}",
      journal = {\aap},
         year = 2016,
        month = feb,
       volume = {586},
          eid = {A151},
        pages = {A151},
          doi = {10.1051/0004-6361/201526944},
archivePrefix = {arXiv},
       eprint = {1510.05908},
 primaryClass = {astro-ph.SR},
       adsurl = {https://ui.adsabs.harvard.edu/abs/2016A&A...586A.151M}
}

@ARTICLE{deCastro2016,
       author = {{de Castro}, D.~B. and {Pereira}, C.~B. and {Roig}, F. and {Jilinski}, E. and {Drake}, N.~A. and {Chavero}, C. and {Sales Silva}, J.~V.},
        title = "{Chemical abundances and kinematics of barium stars}",
      journal = {\mnras},
         year = 2016,
        month = jul,
       volume = {459},
       number = {4},
        pages = {4299-4324},
          doi = {10.1093/mnras/stw815},
archivePrefix = {arXiv},
       eprint = {1604.03031},
 primaryClass = {astro-ph.SR},
       adsurl = {https://ui.adsabs.harvard.edu/abs/2016MNRAS.459.4299D}
}

@ARTICLE{Jorissen1998,
       author = {{Jorissen}, A. and {Van Eck}, S. and {Mayor}, M. and {Udry}, S.},
        title = "{Insights into the formation of barium and Tc-poor S stars from an extended sample of orbital elements}",
      journal = {\aap},
         year = 1998,
        month = apr,
       volume = {332},
        pages = {877-903},
          doi = {10.48550/arXiv.astro-ph/9801272},
archivePrefix = {arXiv},
       eprint = {astro-ph/9801272},
 primaryClass = {astro-ph},
       adsurl = {https://ui.adsabs.harvard.edu/abs/1998A&A...332..877J}
}

@INPROCEEDINGS{North2000,
       author = {{North}, Pierre and {Jorissen}, Alain and {Mayor}, Michel},
        title = "{Binarity among Barium Dwarfs and CH Subgiants: Will They Become Barium Giants?}",
    booktitle = {The Carbon Star Phenomenon},
         year = 2000,
       editor = {{Wing}, Robert F.},
       series = {IAU Symposium},
       volume = {177},
        month = jun,
        pages = {269},
       adsurl = {https://ui.adsabs.harvard.edu/abs/2000IAUS..177..269N}
}

@ARTICLE{North1994,
       author = {{North}, P. and {Berthet}, S. and {Lanz}, T.},
        title = "{The nature of the F STR lambda 4077 stars. III. Spectroscopy of the barium dwarfs and other CP stars.}",
      journal = {\aap},
         year = 1994,
        month = jan,
       volume = {281},
        pages = {775-796},
       adsurl = {https://ui.adsabs.harvard.edu/abs/1994A&A...281..775N}
}

@ARTICLE{Jorissen1992,
       author = {{Jorissen}, A. and {Manfroid}, J. and {Sterken}, C.},
        title = "{Photometric variability of barium stars.}",
      journal = {\aap},
         year = 1992,
        month = jan,
       volume = {253},
        pages = {407-424},
       adsurl = {https://ui.adsabs.harvard.edu/abs/1992A&A...253..407J}
}

@ARTICLE{Siegel2019,
       author = {{Siegel}, Daniel M. and {Barnes}, Jennifer and {Metzger}, Brian D.},
        title = "{Collapsars as a major source of r-process elements}",
      journal = {\nat},
         year = 2019,
        month = may,
       volume = {569},
       number = {7755},
        pages = {241-244},
          doi = {10.1038/s41586-019-1136-0},
archivePrefix = {arXiv},
       eprint = {1810.00098},
 primaryClass = {astro-ph.HE},
       adsurl = {https://ui.adsabs.harvard.edu/abs/2019Natur.569..241S}
}

@ARTICLE{Goldman2017,
       author = {{Goldman}, Steven R. and {van Loon}, Jacco Th. and {Zijlstra}, Albert A. and {Green}, James A. and {Wood}, Peter R. and {Nanni}, Ambra and {Imai}, Hiroshi and {Whitelock}, Patricia A. and {Matsuura}, Mikako and {Groenewegen}, Martin A.~T. and {G{\'o}mez}, Jos{\'e} F.},
        title = "{The wind speeds, dust content, and mass-loss rates of evolved AGB and RSG stars at varying metallicity}",
      journal = {\mnras},
         year = 2017,
        month = feb,
       volume = {465},
       number = {1},
        pages = {403-433},
          doi = {10.1093/mnras/stw2708},
archivePrefix = {arXiv},
       eprint = {1610.05761},
 primaryClass = {astro-ph.SR},
       adsurl = {https://ui.adsabs.harvard.edu/abs/2017MNRAS.465..403G}
}

@ARTICLE{El-Badry2019,
       author = {{El-Badry}, Kareem and {Rix}, Hans-Walter},
        title = "{The wide binary fraction of solar-type stars: emergence of metallicity dependence at a < 200 au}",
      journal = {\mnras},
         year = 2019,
        month = jan,
       volume = {482},
       number = {1},
        pages = {L139-L144},
          doi = {10.1093/mnrasl/sly206},
archivePrefix = {arXiv},
       eprint = {1809.06860},
 primaryClass = {astro-ph.SR},
       adsurl = {https://ui.adsabs.harvard.edu/abs/2019MNRAS.482L.139E}
}

@ARTICLE{Michaud1970,
       author = {{Michaud}, Georges},
        title = "{Diffusion Processes in Peculiar a Stars}",
      journal = {\apj},
         year = 1970,
        month = may,
       volume = {160},
        pages = {641},
          doi = {10.1086/150459},
       adsurl = {https://ui.adsabs.harvard.edu/abs/1970ApJ...160..641M}
}

@ARTICLE{McClure1990,
       author = {{McClure}, Robert D. and {Woodsworth}, A.~W.},
        title = "{The Binary Nature of the Barium and CH Stars. III. Orbital Parameters}",
      journal = {\apj},
         year = 1990,
        month = apr,
       volume = {352},
        pages = {709},
          doi = {10.1086/168573},
       adsurl = {https://ui.adsabs.harvard.edu/abs/1990ApJ...352..709M}
}

@ARTICLE{Boffin1988,
       author = {{Boffin}, H.~M.~J. and {Jorissen}, A.},
        title = "{Can a barium star be produced by wind accretion in a detached binary ?}",
      journal = {\aap},
         year = 1988,
        month = oct,
       volume = {205},
        pages = {155-163},
       adsurl = {https://ui.adsabs.harvard.edu/abs/1988A&A...205..155B}
}

@ARTICLE{Warner1965,
       author = {{Warner}, B.},
        title = "{The barium stars}",
      journal = {\mnras},
         year = 1965,
        month = jan,
       volume = {129},
        pages = {263},
          doi = {10.1093/mnras/129.3.263},
       adsurl = {https://ui.adsabs.harvard.edu/abs/1965MNRAS.129..263W}
}

@ARTICLE{McClure1980,
       author = {{McClure}, R.~D. and {Fletcher}, J.~M. and {Nemec}, J.~M.},
        title = "{The binary nature of the barium stars.}",
      journal = {\apjl},
         year = 1980,
        month = may,
       volume = {238},
        pages = {L35-L38},
          doi = {10.1086/183252},
       adsurl = {https://ui.adsabs.harvard.edu/abs/1980ApJ...238L..35M}
}

@ARTICLE{GCE,
       author = {{Lian}, Jianhui and {Storm}, Nicholas and {Guiglion}, Guillaume and {Serenelli}, Aldo and {Cote}, Benoit and {Karakas}, Amanda I. and {Boardman}, Nicholas and {Bergemann}, Maria},
        title = "{Observational constraints on the origin of the elements - VI. Origin and evolution of neutron-capture elements as probed by the Gaia-ESO survey}",
      journal = {\mnras},
         year = 2023,
        month = oct,
       volume = {525},
       number = {1},
        pages = {1329-1341},
          doi = {10.1093/mnras/stad2390},
archivePrefix = {arXiv},
       eprint = {2308.01111},
 primaryClass = {astro-ph.SR},
       adsurl = {https://ui.adsabs.harvard.edu/abs/2023MNRAS.525.1329L}
}

@ARTICLE{Belokurov2020,
       author = {{Belokurov}, Vasily and {Penoyre}, Zephyr and {Oh}, Semyeong and {Iorio}, Giuliano and {Hodgkin}, Simon and {Evans}, N. Wyn and {Everall}, Andrew and {Koposov}, Sergey E. and {Tout}, Christopher A. and {Izzard}, Robert and {Clarke}, Cathie J. and {Brown}, Anthony G.~A.},
        title = "{Unresolved stellar companions with Gaia DR2 astrometry}",
      journal = {\mnras},
         year = 2020,
        month = aug,
       volume = {496},
       number = {2},
        pages = {1922-1940},
          doi = {10.1093/mnras/staa1522},
archivePrefix = {arXiv},
       eprint = {2003.05467},
 primaryClass = {astro-ph.SR},
       adsurl = {https://ui.adsabs.harvard.edu/abs/2020MNRAS.496.1922B}
}

@ARTICLE{Clayton1988,
       author = {{Clayton}, Donald D.},
        title = "{Nuclear cosmochronology within analytic models of the chemical evolution of the solar neighbourhood.}",
      journal = {\mnras},
         year = 1988,
        month = sep,
       volume = {234},
        pages = {1-36},
          doi = {10.1093/mnras/234.1.1},
       adsurl = {https://ui.adsabs.harvard.edu/abs/1988MNRAS.234....1C}
}

@ARTICLE{Clayton1961,
       author = {{Clayton}, D.~D. and {Fowler}, W.~A. and {Hull}, T.~E. and {Zimmerman}, B.~A.},
        title = "{Neutron capture chains in heavy element synthesis}",
      journal = {Annals of Physics},
         year = 1961,
        month = mar,
       volume = {12},
       number = {3},
        pages = {331-408},
          doi = {10.1016/0003-4916(61)90067-7},
       adsurl = {https://ui.adsabs.harvard.edu/abs/1961AnPhy..12..331C}
}

@ARTICLE{Lindegren2018,
       author = {{Lindegren}, L. and {Hern{\'a}ndez}, J. and {Bombrun}, A. and {Klioner}, S. and {Bastian}, U. and {Ramos-Lerate}, M. and {de Torres}, A. and {Steidelm{\"u}ller}, H. and {Stephenson}, C. and {Hobbs}, D. and {Lammers}, U. and {Biermann}, M. and {Geyer}, R. and {Hilger}, T. and {Michalik}, D. and {Stampa}, U. and {McMillan}, P.~J. and {Casta{\~n}eda}, J. and {Clotet}, M. and {Comoretto}, G. and {Davidson}, M. and {Fabricius}, C. and {Gracia}, G. and {Hambly}, N.~C. and {Hutton}, A. and {Mora}, A. and {Portell}, J. and {van Leeuwen}, F. and {Abbas}, U. and {Abreu}, A. and {Altmann}, M. and {Andrei}, A. and {Anglada}, E. and {Balaguer-N{\'u}{\~n}ez}, L. and {Barache}, C. and {Becciani}, U. and {Bertone}, S. and {Bianchi}, L. and {Bouquillon}, S. and {Bourda}, G. and {Br{\"u}semeister}, T. and {Bucciarelli}, B. and {Busonero}, D. and {Buzzi}, R. and {Cancelliere}, R. and {Carlucci}, T. and {Charlot}, P. and {Cheek}, N. and {Crosta}, M. and {Crowley}, C. and {de Bruijne}, J. and {de Felice}, F. and {Drimmel}, R. and {Esquej}, P. and {Fienga}, A. and {Fraile}, E. and {Gai}, M. and {Garralda}, N. and {Gonz{\'a}lez-Vidal}, J.~J. and {Guerra}, R. and {Hauser}, M. and {Hofmann}, W. and {Holl}, B. and {Jordan}, S. and {Lattanzi}, M.~G. and {Lenhardt}, H. and {Liao}, S. and {Licata}, E. and {Lister}, T. and {L{\"o}ffler}, W. and {Marchant}, J. and {Martin-Fleitas}, J. -M. and {Messineo}, R. and {Mignard}, F. and {Morbidelli}, R. and {Poggio}, E. and {Riva}, A. and {Rowell}, N. and {Salguero}, E. and {Sarasso}, M. and {Sciacca}, E. and {Siddiqui}, H. and {Smart}, R.~L. and {Spagna}, A. and {Steele}, I. and {Taris}, F. and {Torra}, J. and {van Elteren}, A. and {van Reeven}, W. and {Vecchiato}, A.},
        title = "{Gaia Data Release 2. The astrometric solution}",
      journal = {\aap},
         year = 2018,
        month = aug,
       volume = {616},
          eid = {A2},
        pages = {A2},
          doi = {10.1051/0004-6361/201832727},
archivePrefix = {arXiv},
       eprint = {1804.09366},
 primaryClass = {astro-ph.IM},
       adsurl = {https://ui.adsabs.harvard.edu/abs/2018A&A...616A...2L}
}

@ARTICLE{Giribaldi2023,
       author = {{Giribaldi}, R.~E. and {Smiljanic}, R.},
        title = "{Chronology of the chemical enrichment of the old Galactic stellar populations}",
      journal = {\aap},
         year = 2023,
        month = may,
       volume = {673},
          eid = {A18},
        pages = {A18},
          doi = {10.1051/0004-6361/202245404},
archivePrefix = {arXiv},
       eprint = {2302.09640},
 primaryClass = {astro-ph.GA},
       adsurl = {https://ui.adsabs.harvard.edu/abs/2023A&A...673A..18G}
}

@ARTICLE{Choi2016,
       author = {{Choi}, Jieun and {Dotter}, Aaron and {Conroy}, Charlie and {Cantiello}, Matteo and {Paxton}, Bill and {Johnson}, Benjamin D.},
        title = "{Mesa Isochrones and Stellar Tracks (MIST). I. Solar-scaled Models}",
      journal = {\apj},
         year = 2016,
        month = jun,
       volume = {823},
       number = {2},
          eid = {102},
        pages = {102},
          doi = {10.3847/0004-637X/823/2/102},
archivePrefix = {arXiv},
       eprint = {1604.08592},
 primaryClass = {astro-ph.SR},
       adsurl = {https://ui.adsabs.harvard.edu/abs/2016ApJ...823..102C}
}

@ARTICLE{Jofre2011,
       author = {{Jofr{\'e}}, P. and {Weiss}, A.},
        title = "{The age of the Milky Way halo stars from the Sloan Digital Sky Survey}",
      journal = {\aap},
         year = 2011,
        month = sep,
       volume = {533},
          eid = {A59},
        pages = {A59},
          doi = {10.1051/0004-6361/201117131},
archivePrefix = {arXiv},
       eprint = {1105.2022},
 primaryClass = {astro-ph.GA},
       adsurl = {https://ui.adsabs.harvard.edu/abs/2011A&A...533A..59J}
}

@ARTICLE{Horta2020,
       author = {{Horta}, Danny and {Schiavon}, Ricardo P. and {Mackereth}, J. Ted and {Beers}, Timothy C. and {Fern{\'a}ndez-Trincado}, Jos{\'e} G. and {Frinchaboy}, Peter M. and {Garc{\'\i}a-Hern{\'a}ndez}, D.~A. and {Geisler}, Doug and {Hasselquist}, Sten and {J{\"o}nsson}, Henrik and {Lane}, Richard R. and {Majewski}, Steven R. and {M{\'e}sz{\'a}ros}, Szabolcs and {Bidin}, Christian Moni and {Nataf}, David M. and {Roman-Lopes}, Alexandre and {Nitschelm}, Christian and {Vargas-Gonz{\'a}lez}, J. and {Zasowski}, Gail},
        title = "{The chemical compositions of accreted and in situ galactic globular clusters according to SDSS/APOGEE}",
      journal = {\mnras},
         year = 2020,
        month = apr,
       volume = {493},
       number = {3},
        pages = {3363-3378},
          doi = {10.1093/mnras/staa478},
archivePrefix = {arXiv},
       eprint = {2001.03177},
 primaryClass = {astro-ph.GA},
       adsurl = {https://ui.adsabs.harvard.edu/abs/2020MNRAS.493.3363H}
}

@ARTICLE{Fernandes2023,
       author = {{Fernandes}, Laura and {Mason}, Andrew C. and {Horta}, Danny and {Schiavon}, Ricardo P. and {Hayes}, Christian and {Hasselquist}, Sten and {Feuillet}, Diane and {Beaton}, Rachael L. and {J{\"o}nsson}, Henrik and {Kisku}, Shobhit and {Lacerna}, Ivan and {Lian}, Jianhui and {Minniti}, Dante and {Villanova}, Sandro},
        title = "{A comparative analysis of the chemical compositions of Gaia-Enceladus/Sausage and Milky Way satellites using APOGEE}",
      journal = {\mnras},
         year = 2023,
        month = mar,
       volume = {519},
       number = {3},
        pages = {3611-3622},
          doi = {10.1093/mnras/stac3543},
archivePrefix = {arXiv},
       eprint = {2301.01302},
 primaryClass = {astro-ph.GA},
       adsurl = {https://ui.adsabs.harvard.edu/abs/2023MNRAS.519.3611F}
}

@ARTICLE{Borisov2022,
       author = {{Borisov}, S. and {Prantzos}, N. and {Charbonnel}, C.},
        title = "{Lithium, masses, and kinematics of young Galactic dwarf and giant stars with extreme [{\ensuremath{\alpha}}/Fe] ratios}",
      journal = {\aap},
         year = 2022,
        month = dec,
       volume = {668},
          eid = {A181},
        pages = {A181},
          doi = {10.1051/0004-6361/202244468},
archivePrefix = {arXiv},
       eprint = {2209.11915},
 primaryClass = {astro-ph.SR},
       adsurl = {https://ui.adsabs.harvard.edu/abs/2022A&A...668A.181B}
}

@ARTICLE{Lu2025wCatherine,
       author = {{Lu}, Yuxi(Lucy) and {Colman}, Isabel L. and {Sayeed}, Maryum and {Amard}, Louis and {Buder}, Sven and {Manea}, Catherine and {Hattori}, Soichiro and {Pinsonneault}, Marc H. and {Price-Whelan}, Adrian M. and {Bedell}, Megan and {Nidever}, David and {Johnson}, Jennifer A. and {Ness}, Melissa and {Angus}, Ruth and {Claytor}, Zachary R. and {Horta}, Danny and {Behmard}, Aida},
        title = "{Evidence of Truly Young High-{\ensuremath{\alpha}} Dwarf Stars}",
      journal = {\aj},
         year = 2025,
        month = mar,
       volume = {169},
       number = {3},
          eid = {168},
        pages = {168},
          doi = {10.3847/1538-3881/ada9e0},
archivePrefix = {arXiv},
       eprint = {2410.02962},
 primaryClass = {astro-ph.SR},
       adsurl = {https://ui.adsabs.harvard.edu/abs/2025AJ....169..168L}
}

@ARTICLE{Rekhi2024,
       author = {{Rekhi}, Param and {Ben-Ami}, Sagi and {Hallakoun}, Na'ama and {Shahaf}, Sahar and {Toonen}, Silvia and {Rix}, Hans-Walter},
        title = "{Ba Enrichment in Gaia MS+WD Binaries: Tracing s-process Element Production}",
      journal = {\apjl},
         year = 2024,
        month = oct,
       volume = {973},
       number = {2},
          eid = {L56},
        pages = {L56},
          doi = {10.3847/2041-8213/ad77b9},
archivePrefix = {arXiv},
       eprint = {2407.07048},
 primaryClass = {astro-ph.SR},
       adsurl = {https://ui.adsabs.harvard.edu/abs/2024ApJ...973L..56R}
}

@ARTICLE{Karakas2016,
       author = {{Karakas}, Amanda I. and {Lugaro}, Maria},
        title = "{Stellar Yields from Metal-rich Asymptotic Giant Branch Models}",
      journal = {\apj},
         year = 2016,
        month = jul,
       volume = {825},
       number = {1},
          eid = {26},
        pages = {26},
          doi = {10.3847/0004-637X/825/1/26},
archivePrefix = {arXiv},
       eprint = {1604.02178},
 primaryClass = {astro-ph.SR},
       adsurl = {https://ui.adsabs.harvard.edu/abs/2016ApJ...825...26K}
}

@ARTICLE{2010Vick,
       author = {{Vick}, M. and {Michaud}, G. and {Richer}, J. and {Richard}, O.},
        title = "{AmFm and lithium gap stars. Stellar evolution models with mass loss}",
      journal = {\aap},
         year = 2010,
        month = oct,
       volume = {521},
          eid = {A62},
        pages = {A62},
          doi = {10.1051/0004-6361/201014307},
archivePrefix = {arXiv},
       eprint = {1006.5711},
 primaryClass = {astro-ph.SR},
       adsurl = {https://ui.adsabs.harvard.edu/abs/2010A&A...521A..62V}
}

@ARTICLE{Fitton2022,
       author = {{Fitton}, Shannon and {Tofflemire}, Benjamin M. and {Kraus}, Adam L.},
        title = "{Disk Material Inflates Gaia RUWE Values in Single Stars}",
      journal = {Research Notes of the American Astronomical Society},
         year = 2022,
        month = jan,
       volume = {6},
       number = {1},
          eid = {18},
        pages = {18},
          doi = {10.3847/2515-5172/ac4bb7},
archivePrefix = {arXiv},
       eprint = {2206.02695},
 primaryClass = {astro-ph.SR},
       adsurl = {https://ui.adsabs.harvard.edu/abs/2022RNAAS...6...18F}
}

@ARTICLE{Amarsi2020,
       author = {{Amarsi}, A.~M. and {Lind}, K. and {Osorio}, Y. and {Nordlander}, T. and {Bergemann}, M. and {Reggiani}, H. and {Wang}, E.~X. and {Buder}, S. and {Asplund}, M. and {Barklem}, P.~S. and {Wehrhahn}, A. and {Sk{\'u}lad{\'o}ttir}, {\'A}. and {Kobayashi}, C. and {Karakas}, A.~I. and {Gao}, X.~D. and {Bland-Hawthorn}, J. and {de Silva}, G.~M. and {Kos}, J. and {Lewis}, G.~F. and {Martell}, S.~L. and {Sharma}, S. and {Simpson}, J.~D. and {Zucker}, D.~B. and {{\v{C}}otar}, K. and {Horner}, J. and {GALAH Collaboration}},
        title = "{The GALAH Survey: non-LTE departure coefficients for large spectroscopic surveys}",
      journal = {\aap},
         year = 2020,
        month = oct,
       volume = {642},
          eid = {A62},
        pages = {A62},
          doi = {10.1051/0004-6361/202038650},
archivePrefix = {arXiv},
       eprint = {2008.09582},
 primaryClass = {astro-ph.SR},
       adsurl = {https://ui.adsabs.harvard.edu/abs/2020A&A...642A..62A}
}

@ARTICLE{GALAH,
       author = {{Buder}, Sven and {Sharma}, Sanjib and {Kos}, Janez and {Amarsi}, Anish M. and {Nordlander}, Thomas and {Lind}, Karin and {Martell}, Sarah L. and {Asplund}, Martin and {Bland-Hawthorn}, Joss and {Casey}, Andrew R. and {de Silva}, Gayandhi M. and {D'Orazi}, Valentina and {Freeman}, Ken C. and {Hayden}, Michael R. and {Lewis}, Geraint F. and {Lin}, Jane and {Schlesinger}, Katharine J. and {Simpson}, Jeffrey D. and {Stello}, Dennis and {Zucker}, Daniel B. and {Zwitter}, Toma{\v{z}} and {Beeson}, Kevin L. and {Buck}, Tobias and {Casagrande}, Luca and {Clark}, Jake T. and {{\v{C}}otar}, Klemen and {da Costa}, Gary S. and {de Grijs}, Richard and {Feuillet}, Diane and {Horner}, Jonathan and {Kafle}, Prajwal R. and {Khanna}, Shourya and {Kobayashi}, Chiaki and {Liu}, Fan and {Montet}, Benjamin T. and {Nandakumar}, Govind and {Nataf}, David M. and {Ness}, Melissa K. and {Spina}, Lorenzo and {Tepper-Garc{\'\i}a}, Thor and {Ting}, Yuan-Sen and {Traven}, Gregor and {Vogrin{\v{c}}i{\v{c}}}, Rok and {Wittenmyer}, Robert A. and {Wyse}, Rosemary F.~G. and {{\v{Z}}erjal}, Maru{\v{s}}a and {Galah Collaboration}},
        title = "{The GALAH+ survey: Third data release}",
      journal = {\mnras},
         year = 2021,
        month = sep,
       volume = {506},
       number = {1},
        pages = {150-201},
          doi = {10.1093/mnras/stab1242},
archivePrefix = {arXiv},
       eprint = {2011.02505},
 primaryClass = {astro-ph.GA},
       adsurl = {https://ui.adsabs.harvard.edu/abs/2021MNRAS.506..150B}
}

@ARTICLE{Demircan1991,
       author = {{Demircan}, Osman and {Kahraman}, Goksel},
        title = "{Stellar Mass / Luminosity and Mass / Radius Relations}",
      journal = {\apss},
         year = 1991,
        month = jul,
       volume = {181},
       number = {2},
        pages = {313-322},
          doi = {10.1007/BF00639097},
       adsurl = {https://ui.adsabs.harvard.edu/abs/1991Ap&SS.181..313D}
}

@ARTICLE{Eker2018,
       author = {{Eker}, Z. and {Bak{\i}{\c{s}}}, V. and {Bilir}, S. and {Soydugan}, F. and {Steer}, I. and {Soydugan}, E. and {Bak{\i}{\c{s}}}, H. and {Ali{\c{c}}avu{\c{s}}}, F. and {Aslan}, G. and {Alpsoy}, M.},
        title = "{Interrelated main-sequence mass-luminosity, mass-radius, and mass-effective temperature relations}",
      journal = {\mnras},
         year = 2018,
        month = oct,
       volume = {479},
       number = {4},
        pages = {5491-5511},
          doi = {10.1093/mnras/sty1834},
archivePrefix = {arXiv},
       eprint = {1807.02568},
 primaryClass = {astro-ph.SR},
       adsurl = {https://ui.adsabs.harvard.edu/abs/2018MNRAS.479.5491E}
}

@ARTICLE{Mohamed2012,
       author = {{Mohamed}, S. and {Podsiadlowski}, Ph.},
        title = "{Mass Transfer in Mira-type Binaries}",
      journal = {Baltic Astronomy},
         year = 2012,
        month = jan,
       volume = {21},
        pages = {88-96},
          doi = {10.1515/astro-2017-0362},
       adsurl = {https://ui.adsabs.harvard.edu/abs/2012BaltA..21...88M}
}

@ARTICLE{GaiaDR3,
       author = {{Gaia Collaboration} and {Vallenari}, A. and {Brown}, A.~G.~A. and {Prusti}, T. and {de Bruijne}, J.~H.~J. and {Arenou}, F. and {Babusiaux}, C. and {Biermann}, M. and {Creevey}, O.~L. and {Ducourant}, C. and {Evans}, D.~W. and {Eyer}, L. and {Guerra}, R. and {Hutton}, A. and {Jordi}, C. and {Klioner}, S.~A. and {Lammers}, U.~L. and {Lindegren}, L. and {Luri}, X. and {Mignard}, F. and {Panem}, C. and {Pourbaix}, D. and {Randich}, S. and {Sartoretti}, P. and {Soubiran}, C. and {Tanga}, P. and {Walton}, N.~A. and {Bailer-Jones}, C.~A.~L. and {Bastian}, U. and {Drimmel}, R. and {Jansen}, F. and {Katz}, D. and {Lattanzi}, M.~G. and {van Leeuwen}, F. and {Bakker}, J. and {Cacciari}, C. and {Casta{\~n}eda}, J. and {De Angeli}, F. and {Fabricius}, C. and {Fouesneau}, M. and {Fr{\'e}mat}, Y. and {Galluccio}, L. and {Guerrier}, A. and {Heiter}, U. and {Masana}, E. and {Messineo}, R. and {Mowlavi}, N. and {Nicolas}, C. and {Nienartowicz}, K. and {Pailler}, F. and {Panuzzo}, P. and {Riclet}, F. and {Roux}, W. and {Seabroke}, G.~M. and {Sordo}, R. and {Th{\'e}venin}, F. and {Gracia-Abril}, G. and {Portell}, J. and {Teyssier}, D. and {Altmann}, M. and {Andrae}, R. and {Audard}, M. and {Bellas-Velidis}, I. and {Benson}, K. and {Berthier}, J. and {Blomme}, R. and {Burgess}, P.~W. and {Busonero}, D. and {Busso}, G. and {C{\'a}novas}, H. and {Carry}, B. and {Cellino}, A. and {Cheek}, N. and {Clementini}, G. and {Damerdji}, Y. and {Davidson}, M. and {de Teodoro}, P. and {Nu{\~n}ez Campos}, M. and {Delchambre}, L. and {Dell'Oro}, A. and {Esquej}, P. and {Fern{\'a}ndez-Hern{\'a}ndez}, J. and {Fraile}, E. and {Garabato}, D. and {Garc{\'\i}a-Lario}, P. and {Gosset}, E. and {Haigron}, R. and {Halbwachs}, J. -L. and {Hambly}, N.~C. and {Harrison}, D.~L. and {Hern{\'a}ndez}, J. and {Hestroffer}, D. and {Hodgkin}, S.~T. and {Holl}, B. and {Jan{\ss}en}, K. and {Jevardat de Fombelle}, G. and {Jordan}, S. and {Krone-Martins}, A. and {Lanzafame}, A.~C. and {L{\"o}ffler}, W. and {Marchal}, O. and {Marrese}, P.~M. and {Moitinho}, A. and {Muinonen}, K. and {Osborne}, P. and {Pancino}, E. and {Pauwels}, T. and {Recio-Blanco}, A. and {Reyl{\'e}}, C. and {Riello}, M. and {Rimoldini}, L. and {Roegiers}, T. and {Rybizki}, J. and {Sarro}, L.~M. and {Siopis}, C. and {Smith}, M. and {Sozzetti}, A. and {Utrilla}, E. and {van Leeuwen}, M. and {Abbas}, U. and {{\'A}brah{\'a}m}, P. and {Abreu Aramburu}, A. and {Aerts}, C. and {Aguado}, J.~J. and {Ajaj}, M. and {Aldea-Montero}, F. and {Altavilla}, G. and {{\'A}lvarez}, M.~A. and {Alves}, J. and {Anders}, F. and {Anderson}, R.~I. and {Anglada Varela}, E. and {Antoja}, T. and {Baines}, D. and {Baker}, S.~G. and {Balaguer-N{\'u}{\~n}ez}, L. and {Balbinot}, E. and {Balog}, Z. and {Barache}, C. and {Barbato}, D. and {Barros}, M. and {Barstow}, M.~A. and {Bartolom{\'e}}, S. and {Bassilana}, J. -L. and {Bauchet}, N. and {Becciani}, U. and {Bellazzini}, M. and {Berihuete}, A. and {Bernet}, M. and {Bertone}, S. and {Bianchi}, L. and {Binnenfeld}, A. and {Blanco-Cuaresma}, S. and {Blazere}, A. and {Boch}, T. and {Bombrun}, A. and {Bossini}, D. and {Bouquillon}, S. and {Bragaglia}, A. and {Bramante}, L. and {Breedt}, E. and {Bressan}, A. and {Brouillet}, N. and {Brugaletta}, E. and {Bucciarelli}, B. and {Burlacu}, A. and {Butkevich}, A.~G. and {Buzzi}, R. and {Caffau}, E. and {Cancelliere}, R. and {Cantat-Gaudin}, T. and {Carballo}, R. and {Carlucci}, T. and {Carnerero}, M.~I. and {Carrasco}, J.~M. and {Casamiquela}, L. and {Castellani}, M. and {Castro-Ginard}, A. and {Chaoul}, L. and {Charlot}, P. and {Chemin}, L. and {Chiaramida}, V. and {Chiavassa}, A. and {Chornay}, N. and {Comoretto}, G. and {Contursi}, G. and {Cooper}, W.~J. and {Cornez}, T. and {Cowell}, S. and {Crifo}, F. and {Cropper}, M. and {Crosta}, M. and {Crowley}, C. and {Dafonte}, C. and {Dapergolas}, A. and {David}, M. and {David}, P. and {de Laverny}, P. and {De Luise}, F. and {De March}, R. and {De Ridder}, J. and {de Souza}, R. and {de Torres}, A. and {del Peloso}, E.~F. and {del Pozo}, E. and {Delbo}, M. and {Delgado}, A. and {Delisle}, J. -B. and {Demouchy}, C. and {Dharmawardena}, T.~E. and {Di Matteo}, P. and {Diakite}, S. and {Diener}, C. and {Distefano}, E. and {Dolding}, C. and {Edvardsson}, B. and {Enke}, H. and {Fabre}, C. and {Fabrizio}, M. and {Faigler}, S. and {Fedorets}, G. and {Fernique}, P. and {Fienga}, A. and {Figueras}, F. and {Fournier}, Y. and {Fouron}, C. and {Fragkoudi}, F. and {Gai}, M. and {Garcia-Gutierrez}, A. and {Garcia-Reinaldos}, M. and {Garc{\'\i}a-Torres}, M. and {Garofalo}, A. and {Gavel}, A. and {Gavras}, P. and {Gerlach}, E. and {Geyer}, R. and {Giacobbe}, P. and {Gilmore}, G. and {Girona}, S. and {Giuffrida}, G. and {Gomel}, R. and {Gomez}, A. and {Gonz{\'a}lez-N{\'u}{\~n}ez}, J. and {Gonz{\'a}lez-Santamar{\'\i}a}, I. and {Gonz{\'a}lez-Vidal}, J.~J. and {Granvik}, M. and {Guillout}, P. and {Guiraud}, J. and {Guti{\'e}rrez-S{\'a}nchez}, R. and {Guy}, L.~P. and {Hatzidimitriou}, D. and {Hauser}, M. and {Haywood}, M. and {Helmer}, A. and {Helmi}, A. and {Sarmiento}, M.~H. and {Hidalgo}, S.~L. and {Hilger}, T. and {H{\l}adczuk}, N. and {Hobbs}, D. and {Holland}, G. and {Huckle}, H.~E. and {Jardine}, K. and {Jasniewicz}, G. and {Jean-Antoine Piccolo}, A. and {Jim{\'e}nez-Arranz}, {\'O}. and {Jorissen}, A. and {Juaristi Campillo}, J. and {Julbe}, F. and {Karbevska}, L. and {Kervella}, P. and {Khanna}, S. and {Kontizas}, M. and {Kordopatis}, G. and {Korn}, A.~J. and {K{\'o}sp{\'a}l}, {\'A}. and {Kostrzewa-Rutkowska}, Z. and {Kruszy{\'n}ska}, K. and {Kun}, M. and {Laizeau}, P. and {Lambert}, S. and {Lanza}, A.~F. and {Lasne}, Y. and {Le Campion}, J. -F. and {Lebreton}, Y. and {Lebzelter}, T. and {Leccia}, S. and {Leclerc}, N. and {Lecoeur-Taibi}, I. and {Liao}, S. and {Licata}, E.~L. and {Lindstr{\o}m}, H.~E.~P. and {Lister}, T.~A. and {Livanou}, E. and {Lobel}, A. and {Lorca}, A. and {Loup}, C. and {Madrero Pardo}, P. and {Magdaleno Romeo}, A. and {Managau}, S. and {Mann}, R.~G. and {Manteiga}, M. and {Marchant}, J.~M. and {Marconi}, M. and {Marcos}, J. and {Marcos Santos}, M.~M.~S. and {Mar{\'\i}n Pina}, D. and {Marinoni}, S. and {Marocco}, F. and {Marshall}, D.~J. and {Martin Polo}, L. and {Mart{\'\i}n-Fleitas}, J.~M. and {Marton}, G. and {Mary}, N. and {Masip}, A. and {Massari}, D. and {Mastrobuono-Battisti}, A. and {Mazeh}, T. and {McMillan}, P.~J. and {Messina}, S. and {Michalik}, D. and {Millar}, N.~R. and {Mints}, A. and {Molina}, D. and {Molinaro}, R. and {Moln{\'a}r}, L. and {Monari}, G. and {Mongui{\'o}}, M. and {Montegriffo}, P. and {Montero}, A. and {Mor}, R. and {Mora}, A. and {Morbidelli}, R. and {Morel}, T. and {Morris}, D. and {Muraveva}, T. and {Murphy}, C.~P. and {Musella}, I. and {Nagy}, Z. and {Noval}, L. and {Oca{\~n}a}, F. and {Ogden}, A. and {Ordenovic}, C. and {Osinde}, J.~O. and {Pagani}, C. and {Pagano}, I. and {Palaversa}, L. and {Palicio}, P.~A. and {Pallas-Quintela}, L. and {Panahi}, A. and {Payne-Wardenaar}, S. and {Pe{\~n}alosa Esteller}, X. and {Penttil{\"a}}, A. and {Pichon}, B. and {Piersimoni}, A.~M. and {Pineau}, F. -X. and {Plachy}, E. and {Plum}, G. and {Poggio}, E. and {Pr{\v{s}}a}, A. and {Pulone}, L. and {Racero}, E. and {Ragaini}, S. and {Rainer}, M. and {Raiteri}, C.~M. and {Rambaux}, N. and {Ramos}, P. and {Ramos-Lerate}, M. and {Re Fiorentin}, P. and {Regibo}, S. and {Richards}, P.~J. and {Rios Diaz}, C. and {Ripepi}, V. and {Riva}, A. and {Rix}, H. -W. and {Rixon}, G. and {Robichon}, N. and {Robin}, A.~C. and {Robin}, C. and {Roelens}, M. and {Rogues}, H.~R.~O. and {Rohrbasser}, L. and {Romero-G{\'o}mez}, M. and {Rowell}, N. and {Royer}, F. and {Ruz Mieres}, D. and {Rybicki}, K.~A. and {Sadowski}, G. and {S{\'a}ez N{\'u}{\~n}ez}, A. and {Sagrist{\`a} Sell{\'e}s}, A. and {Sahlmann}, J. and {Salguero}, E. and {Samaras}, N. and {Sanchez Gimenez}, V. and {Sanna}, N. and {Santove{\~n}a}, R. and {Sarasso}, M. and {Schultheis}, M. and {Sciacca}, E. and {Segol}, M. and {Segovia}, J.~C. and {S{\'e}gransan}, D. and {Semeux}, D. and {Shahaf}, S. and {Siddiqui}, H.~I. and {Siebert}, A. and {Siltala}, L. and {Silvelo}, A. and {Slezak}, E. and {Slezak}, I. and {Smart}, R.~L. and {Snaith}, O.~N. and {Solano}, E. and {Solitro}, F. and {Souami}, D. and {Souchay}, J. and {Spagna}, A. and {Spina}, L. and {Spoto}, F. and {Steele}, I.~A. and {Steidelm{\"u}ller}, H. and {Stephenson}, C.~A. and {S{\"u}veges}, M. and {Surdej}, J. and {Szabados}, L. and {Szegedi-Elek}, E. and {Taris}, F. and {Taylor}, M.~B. and {Teixeira}, R. and {Tolomei}, L. and {Tonello}, N. and {Torra}, F. and {Torra}, J. and {Torralba Elipe}, G. and {Trabucchi}, M. and {Tsounis}, A.~T. and {Turon}, C. and {Ulla}, A. and {Unger}, N. and {Vaillant}, M.~V. and {van Dillen}, E. and {van Reeven}, W. and {Vanel}, O. and {Vecchiato}, A. and {Viala}, Y. and {Vicente}, D. and {Voutsinas}, S. and {Weiler}, M. and {Wevers}, T. and {Wyrzykowski}, {\L}. and {Yoldas}, A. and {Yvard}, P. and {Zhao}, H. and {Zorec}, J. and {Zucker}, S. and {Zwitter}, T.},
        title = "{Gaia Data Release 3. Summary of the content and survey properties}",
      journal = {\aap},
         year = 2023,
        month = jun,
       volume = {674},
          eid = {A1},
        pages = {A1},
          doi = {10.1051/0004-6361/202243940},
archivePrefix = {arXiv},
       eprint = {2208.00211},
 primaryClass = {astro-ph.GA},
       adsurl = {https://ui.adsabs.harvard.edu/abs/2023A&A...674A...1G}
}

@ARTICLE{VanderSwaelmen2017,
       author = {{Van der Swaelmen}, M. and {Boffin}, H.~M.~J. and {Jorissen}, A. and {Van Eck}, S.},
        title = "{The mass-ratio and eccentricity distributions of barium and S stars, and red giants in open clusters}",
      journal = {\aap},
         year = 2017,
        month = jan,
       volume = {597},
          eid = {A68},
        pages = {A68},
          doi = {10.1051/0004-6361/201628867},
archivePrefix = {arXiv},
       eprint = {1608.04949},
 primaryClass = {astro-ph.SR},
       adsurl = {https://ui.adsabs.harvard.edu/abs/2017A&A...597A..68V}
}

@ARTICLE{lamost,
       author = {{Zhang}, Meng and {Xiang}, Maosheng and {Zhang}, Hua-Wei and {Ting}, Yuan-Sen and {Wu}, Ya-Qian and {Liu}, Xiao-Wei},
        title = "{Ba-enhanced Dwarf and Subgiant Stars in the LAMOST Galactic Surveys}",
      journal = {\apj},
         year = 2023,
        month = apr,
       volume = {946},
       number = {2},
          eid = {110},
        pages = {110},
          doi = {10.3847/1538-4357/acbcc4},
archivePrefix = {arXiv},
       eprint = {2302.10504},
 primaryClass = {astro-ph.SR},
       adsurl = {https://ui.adsabs.harvard.edu/abs/2023ApJ...946..110Z}
}

@ARTICLE{Tinsley1980,
       author = {{Tinsley}, B.~M.},
        title = "{Evolution of the Stars and Gas in Galaxies}",
      journal = {\fcp},
         year = 1980,
        month = jan,
       volume = {5},
        pages = {287-388},
          doi = {10.48550/arXiv.2203.02041},
archivePrefix = {arXiv},
       eprint = {2203.02041},
 primaryClass = {astro-ph.GA},
       adsurl = {https://ui.adsabs.harvard.edu/abs/1980FCPh....5..287T}
}

@ARTICLE{Huang2015,
       author = {{Huang}, Y. and {Liu}, X.-W. and {Yuan}, H.-B. and {Xiang}, M.-S. and {Chen}, B.-Q. and {Zhang}, H.-W.},
        title = "{Empirical metallicity-dependent calibrations of effective temperature against colours for dwarfs and giants based on interferometric data}",
      journal = {\mnras},
         year = 2015,
        month = dec,
       volume = {454},
       number = {3},
        pages = {2863-2889},
          doi = {10.1093/mnras/stv1991},
archivePrefix = {arXiv},
       eprint = {1508.06080},
 primaryClass = {astro-ph.SR},
       adsurl = {https://ui.adsabs.harvard.edu/abs/2015MNRAS.454.2863H}
}

@ARTICLE{Ramirez2005,
       author = {{Ram{\'\i}rez}, Iv{\'a}n and {Mel{\'e}ndez}, Jorge},
        title = "{The Effective Temperature Scale of FGK Stars. II. T$_{eff}$:Color:[Fe/H] Calibrations}",
      journal = {\apj},
         year = 2005,
        month = jun,
       volume = {626},
       number = {1},
        pages = {465-485},
          doi = {10.1086/430102},
archivePrefix = {arXiv},
       eprint = {astro-ph/0503110},
 primaryClass = {astro-ph},
       adsurl = {https://ui.adsabs.harvard.edu/abs/2005ApJ...626..465R}
}

@ARTICLE{Toomre1964,
       author = {{Toomre}, A.},
        title = "{On the gravitational stability of a disk of stars.}",
      journal = {\apj},
         year = 1964,
        month = may,
       volume = {139},
        pages = {1217-1238},
          doi = {10.1086/147861},
       adsurl = {https://ui.adsabs.harvard.edu/abs/1964ApJ...139.1217T}
}

@ARTICLE{Richard2002,
       author = {{Richard}, O. and {Michaud}, G. and {Richer}, J.},
        title = "{Models of Metal-poor Stars with Gravitational Settling and Radiative Accelerations. III. Metallicity Dependence}",
      journal = {\apj},
         year = 2002,
        month = dec,
       volume = {580},
       number = {2},
        pages = {1100-1117},
          doi = {10.1086/343733},
archivePrefix = {arXiv},
       eprint = {astro-ph/0112113},
 primaryClass = {astro-ph},
       adsurl = {https://ui.adsabs.harvard.edu/abs/2002ApJ...580.1100R}
}

@ARTICLE{radiativelev,
       author = {{Preston}, G.~W.},
        title = "{The chemically peculiar stars of the upper main sequence.}",
      journal = {\araa},
         year = 1974,
        month = jan,
       volume = {12},
        pages = {257-277},
          doi = {10.1146/annurev.aa.12.090174.001353},
       adsurl = {https://ui.adsabs.harvard.edu/abs/1974ARA&A..12..257P}
}

@ARTICLE{bluestragglers,
       author = {{Leonard}, Peter J.~T.},
        title = "{Stellar Collisions in Globular Clusters and the Blue Straggler Problem}",
      journal = {\aj},
         year = 1989,
        month = jul,
       volume = {98},
        pages = {217},
          doi = {10.1086/115138},
       adsurl = {https://ui.adsabs.harvard.edu/abs/1989AJ.....98..217L}
}

@ARTICLE{firstba,
       author = {{Bidelman}, William P. and {Keenan}, Philip C.},
        title = "{The Ba II Stars.}",
      journal = {\apj},
         year = 1951,
        month = nov,
       volume = {114},
        pages = {473},
          doi = {10.1086/145488},
       adsurl = {https://ui.adsabs.harvard.edu/abs/1951ApJ...114..473B}
}

@ARTICLE{binarynature,
       author = {{McClure}, R.~D.},
        title = "{The binary nature of the barium stars. II. Velocities, binary frequency, and preliminary orbits.}",
      journal = {\apj},
         year = 1983,
        month = may,
       volume = {268},
        pages = {264-273},
          doi = {10.1086/160951},
       adsurl = {https://ui.adsabs.harvard.edu/abs/1983ApJ...268..264M}
}

@ARTICLE{Bailerjones,
       author = {{Bailer-Jones}, C.~A.~L. and {Rybizki}, J. and {Fouesneau}, M. and {Mantelet}, G. and {Andrae}, R.},
        title = "{Estimating Distance from Parallaxes. IV. Distances to 1.33 Billion Stars in Gaia Data Release 2}",
      journal = {\aj},
         year = 2018,
        month = aug,
       volume = {156},
       number = {2},
          eid = {58},
        pages = {58},
          doi = {10.3847/1538-3881/aacb21},
archivePrefix = {arXiv},
       eprint = {1804.10121},
 primaryClass = {astro-ph.SR},
       adsurl = {https://ui.adsabs.harvard.edu/abs/2018AJ....156...58B}
}

@ARTICLE{oldsme,
       author = {{Valenti}, J.~A. and {Piskunov}, N.},
        title = "{Spectroscopy made easy: A new tool for fitting observations with synthetic spectra.}",
      journal = {\aaps},
         year = 1996,
        month = sep,
       volume = {118},
        pages = {595-603},
       adsurl = {https://ui.adsabs.harvard.edu/abs/1996A&AS..118..595V}
}

@ARTICLE{newsme,
       author = {{Piskunov}, Nikolai and {Valenti}, Jeff A.},
        title = "{Spectroscopy Made Easy: Evolution}",
      journal = {\aap},
         year = 2017,
        month = jan,
       volume = {597},
          eid = {A16},
        pages = {A16},
          doi = {10.1051/0004-6361/201629124},
archivePrefix = {arXiv},
       eprint = {1606.06073},
 primaryClass = {astro-ph.IM},
       adsurl = {https://ui.adsabs.harvard.edu/abs/2017A&A...597A..16P}
}

@ARTICLE{1976MARCS,
       author = {{Bell}, R.~A. and {Eriksson}, K. and {Gustafsson}, B. and {Nordlund}, A.},
        title = "{A grid of model atmospheres for metal-deficient giant stars. II.}",
      journal = {\aaps},
         year = 1976,
        month = jan,
       volume = {23},
        pages = {37-95},
       adsurl = {https://ui.adsabs.harvard.edu/abs/1976A&AS...23...37B}
}

@ARTICLE{1975MARCS,
       author = {{Gustafsson}, B. and {Bell}, R.~A. and {Eriksson}, K. and {Nordlund}, A.},
        title = "{A grid of model atmospheres for metal-deficient giant stars. I.}",
      journal = {\aap},
         year = 1975,
        month = sep,
       volume = {42},
        pages = {407-432},
       adsurl = {https://ui.adsabs.harvard.edu/abs/1975A&A....42..407G}
}

@ARTICLE{2008MARCS,
       author = {{Gustafsson}, B. and {Edvardsson}, B. and {Eriksson}, K. and {J{\o}rgensen}, U.~G. and {Nordlund}, {\r{A}}. and {Plez}, B.},
        title = "{A grid of MARCS model atmospheres for late-type stars. I. Methods and general properties}",
      journal = {\aap},
         year = 2008,
        month = aug,
       volume = {486},
       number = {3},
        pages = {951-970},
          doi = {10.1051/0004-6361:200809724},
archivePrefix = {arXiv},
       eprint = {0805.0554},
 primaryClass = {astro-ph},
       adsurl = {https://ui.adsabs.harvard.edu/abs/2008A&A...486..951G}
}

@ARTICLE{dawesreview,
       author = {{Karakas}, Amanda I. and {Lattanzio}, John C.},
        title = "{The Dawes Review 2: Nucleosynthesis and Stellar Yields of Low- and Intermediate-Mass Single Stars}",
      journal = {\pasa},
         year = 2014,
        month = jul,
       volume = {31},
          eid = {e030},
        pages = {e030},
          doi = {10.1017/pasa.2014.21},
archivePrefix = {arXiv},
       eprint = {1405.0062},
 primaryClass = {astro-ph.SR},
       adsurl = {https://ui.adsabs.harvard.edu/abs/2014PASA...31...30K}
}

@ARTICLE{parallaxfordistancebad,
       author = {{Luri}, X. and {Brown}, A.~G.~A. and {Sarro}, L.~M. and {Arenou}, F. and {Bailer-Jones}, C.~A.~L. and {Castro-Ginard}, A. and {de Bruijne}, J. and {Prusti}, T. and {Babusiaux}, C. and {Delgado}, H.~E.},
        title = "{Gaia Data Release 2. Using Gaia parallaxes}",
      journal = {\aap},
         year = 2018,
        month = aug,
       volume = {616},
          eid = {A9},
        pages = {A9},
          doi = {10.1051/0004-6361/201832964},
archivePrefix = {arXiv},
       eprint = {1804.09376},
 primaryClass = {astro-ph.IM},
       adsurl = {https://ui.adsabs.harvard.edu/abs/2018A&A...616A...9L}
}

@ARTICLE{Molero2023,
       author = {{Molero}, Marta and {Magrini}, Laura and {Matteucci}, Francesca and {Romano}, Donatella and {Palla}, Marco and {Cescutti}, Gabriele and {Viscasillas V{\'a}zquez}, Carlos and {Spitoni}, Emanuele},
        title = "{Origin of neutron-capture elements with the Gaia-ESO survey: the evolution of s- and r-process elements across the Milky Way}",
      journal = {\mnras},
         year = 2023,
        month = aug,
       volume = {523},
       number = {2},
        pages = {2974-2989},
          doi = {10.1093/mnras/stad1577},
archivePrefix = {arXiv},
       eprint = {2304.06452},
 primaryClass = {astro-ph.GA},
       adsurl = {https://ui.adsabs.harvard.edu/abs/2023MNRAS.523.2974M}
}

@ARTICLE{Osborn2025,
       author = {{Osborn}, Zara and {Karakas}, Amanda and {Kemp}, Alex and {Izzard}, Robert and {Kamath}, Devika and {Lugaro}, Maria},
        title = "{Using binary population synthesis to examine the impact of binary evolution on the C, N, O, and S-process yields of solar-metallicity low- and intermediate-mass stars}",
      journal = {\pasa},
         year = 2025,
        month = feb,
       volume = {42},
          eid = {e020},
        pages = {e020},
          doi = {10.1017/pasa.2024.124},
archivePrefix = {arXiv},
       eprint = {2412.01025},
 primaryClass = {astro-ph.SR},
       adsurl = {https://ui.adsabs.harvard.edu/abs/2025PASA...42...20O}
}

@ARTICLE{EuContaminationPossiibility,
       author = {{Leicester}, Brayden and {Bekki}, Kenji and {Tsujimoto}, Takuji},
        title = "{Chemical enrichment by collapsars as the origin of the unusually high [Ba/Fe] in a massive star cluster of the dwarf galaxy NGC 1569}",
      journal = {\mnras},
         year = 2025,
        month = feb,
       volume = {537},
       number = {2},
        pages = {1889-1903},
          doi = {10.1093/mnras/staf142},
archivePrefix = {arXiv},
       eprint = {2410.19257},
 primaryClass = {astro-ph.GA},
       adsurl = {https://ui.adsabs.harvard.edu/abs/2025MNRAS.537.1889L}
}

@ARTICLE{bondiaccretion,
       author = {{Comerford}, T.~A.~F. and {Izzard}, R.~G. and {Booth}, R.~A. and {Rosotti}, G.},
        title = "{Bondi-Hoyle-Lyttleton accretion by binary stars}",
      journal = {\mnras},
         year = 2019,
        month = dec,
       volume = {490},
       number = {4},
        pages = {5196-5209},
          doi = {10.1093/mnras/stz2977},
archivePrefix = {arXiv},
       eprint = {1910.13353},
 primaryClass = {astro-ph.SR},
       adsurl = {https://ui.adsabs.harvard.edu/abs/2019MNRAS.490.5196C}
}

@ARTICLE{atomicdifussionandturbulentmixing,
       author = {{Moedas}, Nuno and {Deal}, Morgan and {Bossini}, Diego and {Campilho}, Bernardo},
        title = "{Atomic diffusion and turbulent mixing in solar-like stars: Impact on the fundamental properties of FG-type stars}",
      journal = {\aap},
         year = 2022,
        month = oct,
       volume = {666},
          eid = {A43},
        pages = {A43},
          doi = {10.1051/0004-6361/202243210},
archivePrefix = {arXiv},
       eprint = {2207.02779},
 primaryClass = {astro-ph.SR},
       adsurl = {https://ui.adsabs.harvard.edu/abs/2022A&A...666A..43M}
}

@ARTICLE{Alecian2020,
       author = {{Alecian}, G. and {LeBlanc}, F.},
        title = "{An improved parametric method for evaluating radiative accelerations in stellar interiors}",
      journal = {\mnras},
         year = 2020,
        month = nov,
       volume = {498},
       number = {3},
        pages = {3420-3428},
          doi = {10.1093/mnras/staa2584},
archivePrefix = {arXiv},
       eprint = {2008.10954},
 primaryClass = {astro-ph.SR},
       adsurl = {https://ui.adsabs.harvard.edu/abs/2020MNRAS.498.3420A}
}

@ARTICLE{radiativelevmodeling,
       author = {{Hui-Bon-Hoa}, A. and {LeBlanc}, F. and {Hauschildt}, P.~H. and {Baron}, E.},
        title = "{Radiative accelerations in stellar atmospheres}",
      journal = {\aap},
         year = 2002,
        month = jan,
       volume = {381},
        pages = {197-208},
          doi = {10.1051/0004-6361:20011494},
       adsurl = {https://ui.adsabs.harvard.edu/abs/2002A&A...381..197Hmn}
}

@ARTICLE{Sullivan2024,
       author = {{Sullivan}, Kendall and {Kraus}, Adam L. and {Berger}, Travis A. and {Huber}, Daniel},
        title = "{Quantifying the Contamination from nearby Stellar Companions in Gaia DR3 Photometry}",
      journal = {\aj},
         year = 2025,
        month = jan,
       volume = {169},
       number = {1},
          eid = {29},
        pages = {29},
          doi = {10.3847/1538-3881/ad9330},
archivePrefix = {arXiv},
       eprint = {2411.04196},
 primaryClass = {astro-ph.IM},
       adsurl = {https://ui.adsabs.harvard.edu/abs/2025AJ....169...29S}
}

@ARTICLE{radlevalphaelement,
       author = {{Deal}, M. and {Alecian}, G. and {Lebreton}, Y. and {Goupil}, M.~J. and {Marques}, J.~P. and {LeBlanc}, F. and {Morel}, P. and {Pichon}, B.},
        title = "{Impacts of radiative accelerations on solar-like oscillating main-sequence stars}",
      journal = {\aap},
         year = 2018,
        month = oct,
       volume = {618},
          eid = {A10},
        pages = {A10},
          doi = {10.1051/0004-6361/201833361},
archivePrefix = {arXiv},
       eprint = {1806.10533},
 primaryClass = {astro-ph.SR},
       adsurl = {https://ui.adsabs.harvard.edu/abs/2018A&A...618A..10D}
}

@ARTICLE{navarro2011,
       author = {{Navarro}, Julio F. and {Abadi}, Mario G. and {Venn}, Kim A. and {Freeman}, K.~C. and {Anguiano}, Borja},
        title = "{Through thick and thin: kinematic and chemical components in the solar neighbourhood}",
      journal = {\mnras},
         year = 2011,
        month = apr,
       volume = {412},
       number = {2},
        pages = {1203-1209},
          doi = {10.1111/j.1365-2966.2010.17975.x},
archivePrefix = {arXiv},
       eprint = {1009.0020},
 primaryClass = {astro-ph.GA},
       adsurl = {https://ui.adsabs.harvard.edu/abs/2011MNRAS.412.1203N}
}

@INPROCEEDINGS{Nissen2004,
       author = {{Nissen}, Poul E.},
        title = "{Thin and Thick Galactic Disks}",
    booktitle = {Origin and Evolution of the Elements},
         year = 2004,
       editor = {{McWilliam}, Andrew and {Rauch}, Michael},
        month = jan,
        pages = {154},
          doi = {10.48550/arXiv.astro-ph/0310326},
archivePrefix = {arXiv},
       eprint = {astro-ph/0310326},
 primaryClass = {astro-ph},
       adsurl = {https://ui.adsabs.harvard.edu/abs/2004oee..symp..154N}
}

@ARTICLE{Nissen2010,
       author = {{Nissen}, P.~E. and {Schuster}, W.~J.},
        title = "{Two distinct halo populations in the solar neighborhood. Evidence from stellar abundance ratios and kinematics}",
      journal = {\aap},
         year = 2010,
        month = feb,
       volume = {511},
          eid = {L10},
        pages = {L10},
          doi = {10.1051/0004-6361/200913877},
archivePrefix = {arXiv},
       eprint = {1002.4514},
 primaryClass = {astro-ph.GA},
       adsurl = {https://ui.adsabs.harvard.edu/abs/2010A&A...511L..10N}
}

@ARTICLE{escapevelo,
       author = {{Smith}, Martin C. and {Ruchti}, Gregory R. and {Helmi}, Amina and {Wyse}, Rosemary F.~G. and {Fulbright}, J.~P. and {Freeman}, K.~C. and {Navarro}, J.~F. and {Seabroke}, G.~M. and {Steinmetz}, M. and {Williams}, M. and {Bienaym{\'e}}, O. and {Binney}, J. and {Bland-Hawthorn}, J. and {Dehnen}, W. and {Gibson}, B.~K. and {Gilmore}, G. and {Grebel}, E.~K. and {Munari}, U. and {Parker}, Q.~A. and {Scholz}, R. -D. and {Siebert}, A. and {Watson}, F.~G. and {Zwitter}, T.},
        title = "{The RAVE survey: constraining the local Galactic escape speed}",
      journal = {\mnras},
         year = 2007,
        month = aug,
       volume = {379},
       number = {2},
        pages = {755-772},
          doi = {10.1111/j.1365-2966.2007.11964.x},
archivePrefix = {arXiv},
       eprint = {astro-ph/0611671},
 primaryClass = {astro-ph},
       adsurl = {https://ui.adsabs.harvard.edu/abs/2007MNRAS.379..755S}
}

@ARTICLE{Pecaut2013,
       author = {{Pecaut}, Mark J. and {Mamajek}, Eric E.},
        title = "{Intrinsic Colors, Temperatures, and Bolometric Corrections of Pre-main-sequence Stars}",
      journal = {\apjs},
         year = 2013,
        month = sep,
       volume = {208},
       number = {1},
          eid = {9},
        pages = {9},
          doi = {10.1088/0067-0049/208/1/9},
archivePrefix = {arXiv},
       eprint = {1307.2657},
 primaryClass = {astro-ph.SR},
       adsurl = {https://ui.adsabs.harvard.edu/abs/2013ApJS..208....9P}
}

@ARTICLE{schonrich2010,
       author = {{Sch{\"o}nrich}, Ralph and {Binney}, James and {Dehnen}, Walter},
        title = "{Local kinematics and the local standard of rest}",
      journal = {\mnras},
         year = 2010,
        month = apr,
       volume = {403},
       number = {4},
        pages = {1829-1833},
          doi = {10.1111/j.1365-2966.2010.16253.x},
archivePrefix = {arXiv},
       eprint = {0912.3693},
 primaryClass = {astro-ph.GA},
       adsurl = {https://ui.adsabs.harvard.edu/abs/2010MNRAS.403.1829S}
}

@ARTICLE{doner2023,
       author = {{D{\"o}ner}, Sibel and {Ak}, Serap and {{\"O}nal Ta{\c{s}}}, {\"O}zgecan and {Plevne}, Olcay},
        title = "{The Age-Metallicity Relation in the Solar Neighbourhood}",
      journal = {Physics and Astronomy Reports},
         year = 2023,
        month = may,
       volume = {1},
       number = {1},
        pages = {11-26},
          doi = {10.26650/PAR.2023.00002},
archivePrefix = {arXiv},
       eprint = {2304.14747},
 primaryClass = {astro-ph.GA},
       adsurl = {https://ui.adsabs.harvard.edu/abs/2023PARep...1...11D}
}

@ARTICLE{carraro1998,
       author = {{Carraro}, Giovanni and {Ng}, Yuen Keong and {Portinari}, Laura},
        title = "{On the Galactic disc age-metallicity relation}",
      journal = {\mnras},
         year = 1998,
        month = jun,
       volume = {296},
       number = {4},
        pages = {1045-1056},
          doi = {10.1046/j.1365-8711.1998.01460.x},
archivePrefix = {arXiv},
       eprint = {astro-ph/9707185},
 primaryClass = {astro-ph},
       adsurl = {https://ui.adsabs.harvard.edu/abs/1998MNRAS.296.1045C}
}

@ARTICLE{Theuns1996,
       author = {{Theuns}, Tom and {Boffin}, Henri M.~J. and {Jorissen}, Alain},
        title = "{Wind accretion in binary stars - II. Accretion rates}",
      journal = {\mnras},
         year = 1996,
        month = jun,
       volume = {280},
       number = {4},
        pages = {1264-1276},
          doi = {10.1093/mnras/280.4.1264},
archivePrefix = {arXiv},
       eprint = {astro-ph/9602089},
 primaryClass = {astro-ph},
       adsurl = {https://ui.adsabs.harvard.edu/abs/1996MNRAS.280.1264T}
}

@ARTICLE{Martig2015,
       author = {{Martig}, Marie and {Rix}, Hans-Walter and {Silva Aguirre}, Victor and {Hekker}, Saskia and {Mosser}, Benoit and {Elsworth}, Yvonne and {Bovy}, Jo and {Stello}, Dennis and {Anders}, Friedrich and {Garc{\'\i}a}, Rafael A. and {Tayar}, Jamie and {Rodrigues}, Tha{\'\i}se S. and {Basu}, Sarbani and {Carrera}, Ricardo and {Ceillier}, Tugdual and {Chaplin}, William J. and {Chiappini}, Cristina and {Frinchaboy}, Peter M. and {Garc{\'\i}a-Hern{\'a}ndez}, D.~A. and {Hearty}, Fred R. and {Holtzman}, Jon and {Johnson}, Jennifer A. and {Majewski}, Steven R. and {Mathur}, Savita and {M{\'e}sz{\'a}ros}, Szabolcs and {Miglio}, Andrea and {Nidever}, David and {Pan}, Kaike and {Pinsonneault}, Marc and {Schiavon}, Ricardo P. and {Schneider}, Donald P. and {Serenelli}, Aldo and {Shetrone}, Matthew and {Zamora}, Olga},
        title = "{Young {\ensuremath{\alpha}}-enriched giant stars in the solar neighbourhood}",
      journal = {\mnras},
         year = 2015,
        month = aug,
       volume = {451},
       number = {2},
        pages = {2230-2243},
          doi = {10.1093/mnras/stv1071},
archivePrefix = {arXiv},
       eprint = {1412.3453},
 primaryClass = {astro-ph.GA},
       adsurl = {https://ui.adsabs.harvard.edu/abs/2015MNRAS.451.2230M}
}

@ARTICLE{McMillan2017,
       author = {{McMillan}, Paul J.},
        title = "{The mass distribution and gravitational potential of the Milky Way}",
      journal = {\mnras},
         year = 2017,
        month = feb,
       volume = {465},
       number = {1},
        pages = {76-94},
          doi = {10.1093/mnras/stw2759},
archivePrefix = {arXiv},
       eprint = {1608.00971},
 primaryClass = {astro-ph.GA},
       adsurl = {https://ui.adsabs.harvard.edu/abs/2017MNRAS.465...76M}
}
\bibliographystyle{aasjournal}



\end{document}